\documentclass[12pt]{article}
\usepackage{amsmath, amssymb,graphicx,psfrag}
\usepackage{setspace}

\makeatletter\def\input@path{{pics/}}\makeatother\graphicspath{{pics/}}

\newwrite\ffile\global\newcount\figno \global\figno=1

\def\writedef#1{}

\input epsf
\def\figin{\epsfcheck\figin}\def\figins{\epsfcheck\figins}
\def\epsfcheck{\ifx\epsfbox\UnDeFiNeD
\message{(NO epsf.tex, FIGURES WILL BE IGNORED)}
\gdef\figin##1{\vskip2in}\gdef\figins##1{\hskip.5in}
\else\message{(FIGURES WILL BE INCLUDED)}%
\gdef\figin##1{##1}\gdef\figins##1{##1}\fi}

\def\figinsert{}
\def\ifig#1#2#3{\xdef#1{fig.~\the\figno}
\writedef{#1\leftbracket fig.\noexpand~\the\figno}%
\figinsert\figin{\centerline{#3}}\medskip\centerline{\vbox{\baselineskip12pt
\advance\hsize by -1truein\center\footnotesize{  Fig.~\the\figno.} #2}}
\bigskip\endinsert\global\advance\figno by1}
\def\endinsert{}

\usepackage{amssymb}
\usepackage{graphics}

\DeclareMathOperator{\diag}{diag}

\begin{document}
\baselineskip 18pt
\newcommand{\Tr}{\mbox{Tr\,}}
\renewcommand{\Re}{\mbox{Re}\,}
\renewcommand{\Im}{\mbox{Im}\,}

\newcommand{\ZZ}{\mathbb{Z}}
\newcommand{\RR}{\mathbb{R}}
\newcommand{\CC}{\mathbb{C}}
\newcommand{\w}{w}
\def\Ukk{U_{\rm KK}}
\def\nc{N_c}
\def\sac{\, \qquad}
\def\mkk{M_{\rm KK}}
\def\ut{U_{\rm KK}}
\def\l{\lambda}
\def\ua{U(1)_A}
\def\dd{d}
\def\p{\partial}
\def\half{\frac{1}{2}}

\newcommand{\beq}{\begin{equation}}
\newcommand{\eeq}{\end{equation}}
\newcommand{\bea}{\begin{eqnarray}}
\newcommand{\eea}[1]{\label{#1}\end{eqnarray}}
\newcommand{\acro}[1]{#1\@} 
\newcommand{\N}{{\cal N}}
\def\O{{\cal O}}
\def\pr{{\partial}}
\newcommand{\vr}{\varrho}
\newcommand{\df}{d}
\def\ap {\alpha'}
\def\vp {\varphi}
\newcommand{\cL}{{\cal L}}
\newcommand{\cY}{\ensuremath{{\cal Y}}}


\thispagestyle{empty}
\renewcommand{\thefootnote}{\fnsymbol{footnote}}

{\hfill \parbox{4cm}{
    MPP-2007-168 \\    SHEP-07-45 \\ NI-07-071
}}

\bigskip

\begin{center} \noindent \Large \bf
Mesons in Gauge/Gravity Duals

A Review

\end{center}

\bigskip\bigskip\bigskip

\centerline{ \normalsize \bf Johanna Erdmenger $^a$, Nick Evans
$^{bc}$, Ingo Kirsch $^d$  and Ed Threlfall $^b$ \footnote[1]{\noindent \tt
 jke@mppmu.mpg.de, evans@phys.soton.ac.uk, kirsch@phys.ethz.ch, ejt@phys.soton.ac.uk} }

\bigskip

\centerline{ \it $^a$ Max-Planck-Institut f\"ur Physik
(Werner-Heisenberg-Institut)} \centerline{ \it F\"ohringer Ring 6,
80805 M\"unchen, Germany}
\bigskip

\bigskip

\centerline{ \it $^b$ School of Physics \& Astronomy,
Southampton University} \centerline{\it  Southampton, S017 1BJ,
United Kingdom}
\bigskip

\bigskip

\centerline{ \it $^c$ Isaac Newton Institute for Mathematical Sciences,}
\centerline{\it 20 Clarkson Road, Cambridge, CB3 0EH, United Kingdom}

\bigskip

\bigskip

\centerline{\it ${}^d$ Institut f\"ur Theoretische Physik, ETH
  Z\"urich}
\centerline{\it
CH-8093 Z\"urich, Switzerland}

\bigskip\bigskip

\bigskip\bigskip

\renewcommand{\thefootnote}{\arabic{footnote}}

\centerline{\bf \small Abstract}
\medskip

{\small \noindent We review recent progress in studying mesons within
  gauge/gravity duality, in the context of adding flavour degrees of
  freedom to generalizations of the AdS/CFT correspondence.  Our main
  focus is on the `top-down approach' of considering models
  constructed within string theory. We explain the string-theoretical
  constructions in detail, aiming at non-specialists.  These give rise
  to a new way of describing strongly coupled confining large $N$
  gauge gauge theories similar to large $N$ QCD.  In particular, we
  consider gravity dual descriptions of spontaneous chiral symmetry
  breaking, and compare with lattice results.  A further topic covered
  is the behaviour of flavour bound states in finite temperature field
  theories dual to a gravity background involving a black hole.  We
  also describe the `bottom up' phenomenological approach to mesons
  within AdS/QCD.  -- Some previously unpublished results are also
  included. }

\newpage


\tableofcontents

\newpage

\section{Introduction}
\setcounter{equation}{0}\setcounter{figure}{0}\setcounter{table}{0}

String theory\footnote{Some introductory texts on string theory are
  listed in \cite{Zwiebach:2004tj, Kiritsis:2007zz, Becker:2007zj,
    Green:1987sp, Polchinski:1998rq}.}  originated as a theory of
hadrons in the 1960's, when it was noticed that the hadron spectra
contains Regge trajectories that can be reproduced by the
properties of a rotating relativistic string.  However, it was
subsequently realized that four-dimensional string theories
contain unphysical modes such as tachyons and a massless vector
particle. At that time string theory was abandoned as a theory of
the strong interactions, and took a rather different route as a
promising candidate for a unified theory of all four fundamental
interactions including gravity, due to the fact that it contains a
graviton in its spectrum.  It was realized that a fully consistent
string theory must contain supersymmetry and live in ten
space-time dimensions.  Frustratingly, since gravity is so weak,
none of the novel physics of string theory need appear
experimentally below energies close to the Planck scale (10$^{19}$
GeV), making the ideas of string theory difficult to test.

Since the 1970's though, our understanding of the strong interactions
has developed greatly.  Quantum Chromodynamics (QCD) has established
itself as a very successful quantum field theory description of the
strong interactions, and is by now very well tested experimentally.
The matter degrees of freedom in QCD consist of quarks transforming in
the fundamental representation of a non-abelian $SU(3)$ gauge theory.
Interactions are mediated by gauge bosons, the gluon fields, in the
adjoint of $SU(3)$. The theory has been shown to be asymptotically
free \cite{Gross:1973id,Politzer:1973fx}.  This means that at
arbitrarily large energy scales, or equivalently at very short
distances, the quarks become weakly interacting, whilst at long
distances the force becomes ferociously strong. The result of the
strong interaction regime is that quarks are confined into bound
states, the hadrons. In addition, the dynamics generate a large
constituent mass for the quarks which mixes left and right handed
quarks, and hence breaks their chiral symmetries.

QCD does provide a heuristic understanding for why the hadron spectrum
looks like a string spectrum. An excited meson may be thought of as a
quark and an anti-quark connected by a tube of strong interaction
flux.  Such a configuration indeed resembles a string. 't~Hooft made an
additional step towards making the connection more concrete when he
noticed that $SU(N)$ gauge theories with a large number of colours $N$
simplify \cite{tHooft:1973jz}. The leading Feynman
diagrams in an expansion in $N$ are planar
diagrams. The description of a meson in this limit has two quark lines
propagating in time connected by a dense ``sheet'' of gluons - it
suggests the world-sheet swept out by a string through time. An
explicit understanding of the relation remained mysterious though.

Despite the successes of both string theory and QCD, a number of
unsolved issues remain in both areas. On the one hand, it would be
desirable to find closer links between string theory and
experimentally testable theories.  On the other hand, there are
properties of QCD which are still poorly understood. Despite the
tremendous successes of large scale computer simulations (lattice
gauge theory - for introductory texts see
\cite{latticeintro1,latticeintro2,latticeintro3}), in particular the
low-energy mechanisms in QCD for confinement and chiral symmetry
breaking remain unclear conceptually.  New theoretical input in
addition to lattice gauge theory appears to be desirable.  String
theory always seemed like a potential candidate to provide new
insights.

These questions have recently led to new relations between modern
superstring theory and QCD.  These new relations have been made
possible by the second superstring revolution in 1995, introducing the
concept of {\it D~branes}
\cite{Polchinski:1995mt,Polchinski:1996na,Johnson:2003gi}.  D~branes
arose on the one hand as solitonic solutions of ten-dimensional
supergravity (the low energy effective theory of superstring theory at
scales lower than the string scale), and on the other hand as
hypersurfaces in the fundamental string theory on which open strings
can end.  In the first of these pictures, the excitations are
gravitational closed-string modes sourced by the tension of the brane.
In the second, where the charged endpoints of open strings move on the
D~branes, the low-energy limit of the lightest string is a gauge
theory.

This ``dual'' interpretation of D~branes is at the heart of the
AdS/CFT correspondence (AdS: Anti-de Sitter space, CFT: conformal
field theory) put forward by Maldacena in 1997
\cite{Maldacena:1997re}.  In its original form, this correspondence
provides a map between a highly symmetric, strongly coupled large $N$
gauge theory and a weakly coupled supergravity theory. The gauge
theory is just the simplest 3+1 dimensional theory to emerge on the
world-volume of the most basic D3 brane configuration. A number of $N$
coincident D3 branes generates an $SU(N)$ gauge theory in the
low-energy limit.  $N$ must be large since very many D3 branes are
required in order to ensure that the dual supergravity background is
weakly coupled.  Gauge invariant composite operators of the quantum
field theory are mapped to supergravity fields in the same
representation of the large symmetry group present.  For this original
case, many non-trivial tests have been found. The field theory is
$\N=4$ large $N$ $SU(N)$ gauge theory (in addition to the usual gauge
fields there are 4 two-component gauginos and 6 real adjoint scalars),
whose $\beta$~function has been shown to vanish to all orders in
perturbation theory, and thus is conformal even when quantized. More
precisely, the correlation functions of the quantum field theory -
which, since they involve expectation values, are classical functions
although involving Hilbert space operators - are mapped to classical
correlation functions in supergravity.

An obvious question after the discovery of this duality was whether it
could teach us about QCD, a different strongly coupled gauge
theory. To achieve a description of QCD-like theories, it is necessary
to break supersymmetry and to remove conformal invariance, so as to
obtain a running coupling, as well as to introduce quark
fields. Technologies have been developed that allow all of these required
features at least to some degree. This review will discuss these
technologies, their implications for QCD and their limitations. A main
feature is that so far, gauge/gravity dualities describe large $N$
field theories only.

It was suggested very shortly after Maldacena's original paper to find
gravity duals of less symmetric large $N$ gauge theories, in
particular of confining theories. A number of examples of gravity
duals of quantum field theories with less supersymmetry and running
couplings have been found.  Examples include renormalization group
flows obtained by adding relevant operators, for instance mass terms
for the adjoint fermions and scalars present in $\N=4$ theory. These
perturbations can be chosen to maintain some or none of the
supersymmetries of the original model.  The common feature is that a
strongly coupled gauge theory is mapped to a weakly coupled -
{\em i.e.}\ solvable - classical gravity theory.  Whilst this is considerable
progress, it must be noted that the relevant operators are
essentially perturbations - since the gauge dynamics is strongly
coupled at all energy scales, one cannot completely decouple massive
fields from the dynamics. The far ultraviolet region
of these theories generically
displays a large degree of supersymmetry. On the other hand, a mass
perturbation in a conformal field theory fundamentally changes the
dynamics, and the resulting behaviour of these theories is very
different from the conformal $\N=4$ theory.

Further progress has been made by adding flavour degrees of freedom in
the fundamental representation of the gauge group to the gravity dual
description. The original Maldacena set-up contains $N$
$3+1$-dimensional D3 branes, on which open strings, which have charged
endpoints, may end. This corresponds to $\N=4$ $SU(N)$ gauge theory
which has only adjoint degrees of freedom. The addition of different
types of branes into the set up introduces strings stretched between
the new brane and the D3 branes - these strings have only one charge
under the SU($N$) group on the D3 branes and are therefore quark
fields. The best understood example consists of a small number, $N_f$,
of $7+1$-dimensional D7 probe branes \cite{Karch:2002sh}. Treating them as
a probe means they do not change the background geometry or, in the
gauge theory language, that quark loop effects are suppressed in the gauge
background - this corresponds to the quenched approximation
which is formally valid when $N_f \ll
N$. \footnote{However, the effect of quark degrees of freedom on flavour
  physics may indeed be described in the gauge/gravity dual approximation.
The prime example for this is the study of the condensate phase diagram in
presence of a quark chemical potential.}
In the supergravity picture, one has $AdS_5 \times S^5$ generated
by the D3 branes with the D7 brane probe wrapping - for massless
quarks - an $AdS_5 \times S^3$ subspace.  This corresponds to a
four-dimensional $\N=2$ supersymmetric large $N$ gauge theory with the
field content of $\N=4$ plus a small number of fundamental
hypermultiplets.  It is the remaining supersymmetry of this theory
that makes clear analysis possible.

Strings with both ends on the flavour brane are dual to
quark-antiquark operators (they are in the adjoint of SU($N_f$)).  On
the gravity side of the correspondence, these strings describe the
vacuum position of the brane and its fluctuations if perturbed. The
embedding of a brane in a geometry dual to a gauge background
therefore encodes the mass and quark bilinear condensate in the
theory. Linearized fluctuations are dual to mesonic excitations in the
gauge theory. It is possible to extract the bound state masses
\cite{Kruczenski:2003be}.  In the supersymmetric theory of D3 and D7
branes, supersymmetry forbids a quark condensate. The meson
spectrum consists of tightly bound states of a quark and its antiquark
- the mass of the bound state is smaller than the mass of the
constituent quarks' mass by a factor of the square root of the
't~Hooft coupling $ \lambda \equiv g_{YM}^2 N$.
On the other hand, mesons made of
two quarks with distinct masses are heavier with mass of order the
heavier quark mass. The suppression of some meson masses relative to
others is rather different from what is observed in QCD. This
suppression is a result of the very strong coupling present in the
models across a large range of energy scales.

The next step towards QCD is to combine supersymmetry breaking
deformations of the original $AdS_5 \times S^5$ background and the
adding of D7 brane probes to include quarks.  In the UV, the field
theory returns to the four-dimensional $\N=2$ theory of
\cite{Karch:2002sh}, but the IR is QCD-like with a running gauge
coupling. This combination has been used to obtain a gravity dual
description of dynamical chiral symmetry breaking by a quark
condensate \cite{Babington:2003vm}.
Moreover the associated Goldstone boson has
been identified: It is obtained from the fluctuations of the probe D7
brane around its minimum energy configuration. Since in this set-up
the spontaneously broken symmetry is $U(1)_A$, which is non-anomalous
in the limit $N\rightarrow \infty$, the Goldstone boson corresponds to
the $\eta '$. The $\rho$ mass, as well as
interaction terms involving both the Goldstone field and the $\rho$ can
also be computed. Comparison with recent lattice results
\cite{Bali:2007kt,lattice} for $m_\rho$ and $m_\pi$ at
large $N$ is possible and shows good agreement, at least for small quark
mass.

Several similar scenarios in which supersymmetry breaking leads to
chiral symmetry breaking have also been found - for example a
set-up of D4 and D6 branes \cite{Kruczenski:2003uq}, by placing the
gauge theory on an anti-de-Sitter space \cite{Ghoroku:2006nh},
or introducing a background magnetic field \cite{Filev:2007gb}.

More recently a string theory model of D4, D8 and $\rm
\overline{D8}$ branes has been constructed in
\cite{Sakai:2004cn,Sakai:2005yt} which realizes the larger
non-abelian $SU(N_f) \times SU(N_f)$ chiral symmetry of QCD and
its spontaneous breaking to $SU(N_f)$. The symmetry is broken when
the D8 and $\overline{\rm D8}$ brane probes join to form a
continuous object. In this approach, meson masses such as for
instance of the $\rho$ and $a_1$ have been calculated, with
results surprisingly close to experimental measurements.  However,
as in the D4/D6 case, in the far UV  the corresponding gauge
theory runs to a five-dimensional non-renormalizable theory.

These string theory models have inspired phenomenological approaches
to QCD dubbed AdS/QCD. AdS/QCD are a group of models that are
essentially a distillation of the key elements of the string models
above relevant to QCD phenomenology.  Parameters such as the 't~Hooft
coupling and the quark mass are fitted to the QCD data, and
predictions result for the meson masses and couplings. The agreement
is surprisingly good (typically of order 10$\%$) although systematic
errors are uncalculable. Again one should stress that one would expect
the results to suffer from being at large $N$, from near-conformality,
from the presence of (broken) super-partners and from the quenched
approximation.

Some progress has been made towards addressing the quenching issue. In
the supersymmetric D3/D7 model, it is possible to investigate also the
case of a large number of flavours, of the same order as the number of
colours \cite{Aharony:1998xz, Grana:2001xn,Bertolini:2001qa, Burrington:2004id,
  Kirsch:2005uy}.  In this case, even in the presence of supersymmetry
the beta function is no longer zero, and there is a Landau pole in the
UV. On the gravity side, the backreaction has to be taken into
account, and the gravity dual of the Landau pole is identified as a
certain singularity.

A very fruitful area for extended gauge/gravity dualities is the case
of finite temperature field theories, whose gravity dual is given by a
black hole background \cite{Witten:1998qj, Witten:1998zw}.  In this
case, gauge/gravity duality is ideally suited for describing dynamical
and non-equilibrium processes. This is considered to be of particular
importance for the physics of the quark-gluon plasma as studied at the
RHIC accelerator. At high temperature or density, mesons become
unstable and melt into the quark-gluon plasma.  This phenomenon is
obtained in the gauge/gravity dual description, but is also associated
with a particular first order phase transition \cite{Babington:2003vm,
  Kirsch:2004km, Mateos:2007vn, Ghoroku:2005tf,Albash:2006ew} which is
not expected to be present in QCD.

We thus see that the string theory gravity dual picture of strongly
coupled gauge theory is beginning to make contact with QCD physics.
Qualitatively the pictures are beginning to match well and in some
cases quantitative predictions are not widely off the mark.  The
possibilities for this technology appear promising. In
this review we will develop each of these subjects pedagogically for
the interested but non-specialist reader.

This review is organized as follows. We begin in section 2 with a
brief description of the AdS/CFT correspondence, including a short
summary of string theory which serves as a reference in subsequent
sections.  In section 3 we describe in detail how flavour degrees
of freedom, i.e.~quarks, may be added to the AdS/CFT
correspondence, keeping the number of flavours $N_f$ much smaller
than the number of colours $N \rightarrow \infty$. In section 4 we
move beyond this limit, the so-called probe limit, and consider
the case that $N_f \sim N$. In section 5 we describe mesons in
further supersymmetric geometries. In section 6 we consider in
detail how chiral symmetry breaking arises in non-supersymmetric
geometries. Section 7 is devoted to the gravity dual description
of field theories at finite temperature. In section 8, the
phenomenological AdS/QCD approach is presented, also referred to
as `bottom-up' approach. We briefly conclude in section 9 with
general comments.

For readers unfamiliar with the subject, we recommend reading the following
sections first (in the order given): 2, 3.1, 3.2, 3.3.1, 6, 7 and 8.

\newpage

\section{Brief introduction to the AdS/CFT correspondence}
\setcounter{equation}{0}\setcounter{figure}{0}\setcounter{table}{0}

\subsection{The basics of string theory} \label{strings}

String theory \cite{Zwiebach:2004tj, Kiritsis:2007zz, Becker:2007zj,
  Green:1987sp, Polchinski:1998rq} plays a major role in the
holographic approach to mesons in strongly coupled gauge theories as
described in this review, so we here provide a very brief overview to
remind the reader and set conventions.

The action of a relativistic string is given by the area of the
world-sheet it sweeps out in time written in Nambu-Goto form as
\begin{equation}
S = T \int d \tau d \sigma \sqrt{ \det P[ G_{ab}] } \, ,  \hspace{1cm} P[
G_{ab}] = G_{\mu \nu} \frac{d X^\mu}{d \sigma^a}\frac{d X^\nu}{d
\sigma^b} \, .
\end{equation}
Here $T \equiv 1/ 2 \pi \alpha'$ is the string tension; $\sigma^a =
(\tau  , \sigma)$ are the time and space coordinates on the worldsheet;
P represents the ``pullback'' of the metric as shown; and $G_{\mu
  \nu}$ is the background metric.

The action can be recast in Polyakov form by introducing a worldsheet
metric $h_{ab}$. The action is then
\begin{equation}
S = - \frac{1}{4 \pi \alpha'} \int d^2 \sigma \sqrt{-h} h^{ab}
\partial_a X^\mu \partial_b X^\nu  G_{\mu \nu} \, ,
\end{equation}
but there is also a constraint
\begin{equation}
T_{ab} = \partial_a X^\mu \partial_b X_\mu - \frac{1}{2} h_{ab} h^{cd}
\partial_c X^\mu \partial_d X_\mu = 0 \,.
\end{equation}
There is sufficient symmetry such that the worldsheet metric can be made
flat, $h^{ab}=\eta^{ab}$, by a conformal transformation (or more precisely by
a Weyl transformation  and reparametrization of
the worldsheet coordinates).

Classically the unexcited string is massless with excitations of
oscillations on the string's surface forming a tower of states with
masses in units of $\sqrt{T}$. The zero point energies of these
oscillations contribute a constant negative shift of this spectrum in
the quantum theory. The only known way to remove the tachyonic modes
is to impose supersymmetry. For this purpose a worldsheet,
two-component real fermion is added to the action. Moreover, in
space-time, the {\it GSO projection} must be imposed to remove states,
leaving a supersymmetric space-time theory.

The worldsheet conformal invariance ($h_{ab} \rightarrow e^\phi
h_{ab}$) is anomalous in the quantum theory, unless the theory
lives in 10 spacetime dimensions.

Oscillations of open strings give rise to massless gauge
multiplets (multiple charges are included via {\it Chan-Paton
factors}, global charges, attached to the ends of the strings,
such that non-abelian gauge symmetries may be realized). Closed
string loops have both left and right moving modes, such that they
naturally generate a massless field that looks like the Lorentz
product of two gauge fields, {\em
  i.e.}\ like a graviton multiplet.

Let us briefly list the spectrum of closed string theory.
It contains the metric,
$G^{MN}$, the scalar dilaton $\Phi$, and a two index
antisymmetric tensor $B^{MN}$. Moreover, the GSO projection acts as
a chiral projection on the spacetime fermions emerging from each
of the left and right moving modes of closed string theory. If the
same chirality is projected in each case, then one obtains type IIA
string theory.  Its  bosonic field content
consists of a gauge field $A_1$ and a three-form  $C_3$.
If the chiral projections are opposite, then type IIB theory
results, with as bosonic field content
a scalar, a two-form $C_2$, and a four-form $C_4$.
Both the type IIA and the type IIB
theories possess $\N=2$ supersymmetry.

Open strings can
also be included into type II string theory, breaking the
supersymmetry to $\N=1$.
Interactions can be introduced by allowing the string worldsheet
to have holes and handles. The dilaton $\Phi$'s action measures
these topology changes so that the quantity $e^\Phi$ plays the role
of the theory's coupling. When open and closed string sectors are
combined the Yang-Mills coupling from the open string sector has
$g^2_{YM} = e^\Phi$.

For the AdS/CFT correspondence applied to $3+1$-dimensional field
theories, ten-dimensional type IIB string theory is of central
importance, and in particular its low-energy limit where strings
become point-like and string theory becomes supergravity.  There
exists no completely satisfactory action for the type IIB
supergravity, since it involves an antisymmetric field $C_4$ with
self-dual field strength $F_5$. However, it is possible to write
an action involving both dualities of $C_4$, and then impose the
self-duality as a supplementary field equation.  In this way one
obtains (see for example
\cite{Howe:1983sra,Schwarz:1983wa,D'Hoker:2002aw})
\begin{eqnarray}
\label{IIBaction}
S_{IIB} &=& \phantom{+} { 1 \over 4 \kappa_B ^2} \int \sqrt G e^{-2\Phi} (2R_G + 8
\p _\mu \Phi \p ^\mu \Phi -  |H_3|^2 )
\\
&& - {1 \over 4 \kappa_B ^2} \int \biggl [ \sqrt G (|F_1|^2 +
|\tilde F_3|^2 + \half  |\tilde F_5|^2) + C_4 \wedge H_3 \wedge F_3
\biggr ] + {\rm fermions} \, ,
\nonumber
\end{eqnarray}
where the field strengths are defined by
\begin{gather}
F_1  = dC \, , \quad  H_3 = dB \, , \quad  F_3  = dC_2 \, , \quad  F_5  =
dC_4\, , \\
\tilde F_3  = F_3 - C H_3 \, , \quad
\tilde F_5  = F_5 -\half A_2 \wedge
H_3 + \half B \wedge F_3 \, , \nonumber
\end{gather}
and we have the additional self-duality condition $* \tilde F_5 = \tilde
F_5$.

\subsubsection{D-branes}

When open strings are included, it turns out to be consistent to
introduce the strings in such a way that their end points are restricted to a
subspace of the full ten dimensions. The resulting hyperplanes, on
which the strings' ends are confined, are called D-branes
\cite{Polchinski:1995mt,Polchinski:1996na,Johnson:2003gi}.
Solitonic solutions of the supergravity actions also exist that
are naturally sourced by these branes. In fact D-branes are the
fundamental electric and magnetic sources of many of the
supergravity antisymmetric forms.

In particular IIA theory allows branes of even dimension that are
electric and magnetic sources for $A_1$ and $C_3$. IIB theory
includes odd dimension branes that are electric and magnetic
sources for the dilaton, two and four index fields.

The action for a D-brane is given by the Dirac-Born-Infeld (DBI)
action which is an extension of the Nambu-Goto form for the
fundamental string - one simply minimizes its world-volume.
There are extra terms originating from the role of the D-branes
as sources for an antisymmetric two-form $F$, including terms of Chern-Simons
type.  $F$ is the gauge field strength tensor describing
gauge fields on the D7 brane probe and $\phi$ the dilaton.
The action, in {\it String frame}, is
\begin{align}
S_{Dp} = & - \mu_p\int \df^{(p+1)}\xi\, ~ e^{-\phi}~
\sqrt{-\det\left(P[G + 2 \pi \alpha' B]_{ab}+2\pi\ap F_{ab}\right)}
\nonumber\\ & \left. \right. \hspace{2cm}+
\frac{(2\pi\alpha')^2}{2} \mu_p\int P[C^{(p+1)}] \wedge F \wedge
F\ ,
\end{align}
where $\mu_p = (2 \pi)^{-p} \alpha'^{- \frac{(p+1)}{2}}$.
Here $B$ is an external antisymmetric two-form which may be present in the
supergravity background.  In principle, the two-form $B$ may also
contribute terms of Chern-Simons form, which are however not relevant for the
examples described in detail in this review.

\subsection{$\N=4$ Super-Yang-Mills theory}

In its original form \cite{Maldacena:1997re},
the AdS/CFT correspondence involves a
highly symmetric quantum field theory in $3+1$ dimensions, $\N=4$ $SU(N)$
supersymmetric Yang-Mills theory.  The field content of
$\N=4$ Super Yang-Mills theory is as follows: There are  a gauge field
$A_\mu$, which is a singlet of the $SU(4)$ global R symmetry group, four Weyl
fermions in the $\bf 4$ of $SU(4)$, and six real scalars in the $\bf 6$ of
$SU(4)$. An important point is that due to the supersymmetry, all these fields
are in the adjoint representation of the gauge group $SU(N)$.

This theory naturally arises on the surface of a D3 brane in type IIB
superstring theory. Open strings generate a massless gauge field in ten
dimensions. When the open string ends are restricted to a 3+1 dimensional
subspace the ten components of the gauge field naturally break
into a 3+1 dimensional gauge field and 6 scalar fields. The
fermionic super-partners naturally separate to complete the 3+1 dimensional
super-multiplets.

The $\N=4$ theory has the property that the beta function of its
unique coupling vanishes to all orders in perturbation theory,
$\beta=0$.  This implies the theory is conformal with conformal
symmetry group $SO(4,2)$ also at the quantum level. Moreover this
theory has a global $SU(4)$ R~symmetry group. The complete
superconformal group is $SU(2,2| 4)$, of which both $SO(4,2)$ and
$SU(4)$ are bosonic subgroups.

\subsection{AdS/CFT correspondence}\label{adsintro}

The AdS/CFT correspondence was first suggested by Maldacena in 1997
\cite{Maldacena:1997re}, using guiding principles from black hole physics.
The string theory origin of AdS/CFT rests in the fact that D3 branes,
{\em i.e.}~$3+1$ dimensional hyperplanes in $9+1$ dimensional space, may be
interpreted from two different points of view.

Firstly, D3 branes are hyperplanes in ten-dimensional space on which
open strings can end. In the low-energy limit where only massless
string degrees of freedom contribute, these open string degrees of
freedom correspond to $\N=4$ Super Yang-Mills theory with gauge group
$U(N)$, where $N$ corresponds to the number of superimposed D3
branes. The gauge group $U(N)$ factorizes into $SU(N) \times
U(1)$. The $U(1)$ factor corresponds to the motion of the center of
mass of the D3 branes. The global symmetries of the theory are the
SO(4,2) superconformal group and the SU(4) R-symmetry (which is
isomorphic to SO(6) )

On the other hand, D3 branes are also solitonic solutions of
ten-dimensional type IIB supergravity, with a metric of the form
\begin{equation}
ds^2 =  \left( 1 + \frac{R^4}{y^4}\right)^{-\frac{1}{2}}  \eta_{ij} dx^i dx^j
\, + \, \left( 1 + \frac{R^4}{y^4}\right)^{\frac{1}{2}} ( dy^2 + y^2
d\Omega_5{}^2) \, .
\label{D3branemetric}
\end{equation}
Here
\begin{equation} \label{importantrelation}
R^4 = 4 \pi g_s N \alpha'{}^2 \,,
\end{equation}
where $\lambda = g_s N = g^2_{YM} N$ is the 't~Hooft coupling, $N$ the
number of D3 branes and $\alpha'$ the inverse string tension
($\alpha'=l_s^2$, $l_s$ string length).  $\eta_{ij}$ is the standard
$3+1$ dimensional Minkowski metric and the $x^i$ are the coordinates
on the stack of D3 branes. $\vec y$ denotes the six spatial
coordinates perpendicular to the brane, $y \equiv \sqrt{y_M y^M}$.
For $y \gg R$ this metric returns to flat $9+1$ dimensional Minkowski
space.  On the other hand, in the {\it near-horizon limit} $y\ll R$,
which is again a low-energy limit, we perform a coordinate
transformation
\begin{equation}
u \equiv \frac{R^2}{y}
\end{equation}
and obtain from (\ref{D3branemetric})
\begin{equation}
ds^2=  R^2 \left( \frac{1}{u^2}  \eta_{ij} dx^i dx^j  + \frac{du^2}{u^2}
\, + \, d\Omega_5{}^2 \right) \, , \label{AdSS5metric}
\end{equation}
which is the metric of $AdS_5 \times S^5$, with $AdS_5$ the five-dimensional
Anti-the Sitter space $ds^2 = \frac{R^2}{u^2} (\eta_{ij} dx^i dx^j + du^2)$.
Here $R$ is the Anti-de Sitter radius. Anti-de Sitter space
has negative constant curvature ${\cal R} = - \frac{d(d-1)}{R^2}$,
and a boundary at $u=0 $.

A further ingredient is that D3 branes carry charge that source a
four-form antisymmetric tensor field $C_4$ in IIB supergravity.
The D3 brane supergravity solution also therefore has a self-dual
five-form field $F_5 = dC_4$, which satisfies
\begin{equation}
\int\limits_{S^5} F_5 = N \, .
\end{equation}

The isometry of the space $AdS_5$ is $SO(4,2)$ (it can be constructed
as a surface embedded in a 4+2 dimensional space). The isometry of the
five sphere is $SO(6)$. This product group matches the maximal bosonic
subgroup of the supergroup $SU(2,2|4)$, which encodes the symmetries
of the $\N=4$ supersymmetric field theory. Note in particular that
$SO(6) \simeq SU(4)$, which is the R~symmetry group of $\N=4$
supersymmetry. Since the global symmetries match it is possible to
consider that these two theories are dual. Note that the gauge
symmetry of the gauge theory is considered a redundant symmetry that
is not manifest in any gauge invariant observable.

The conjecture of Maldacena consists in identifying the two low-energy
theories, {\em i.e.}~$\N=4$ $SU(N)$ Yang-Mills theory, and string
theory on $AdS_5 \times S^5$.  There are three different versions of
this conjecture, depending on the precise form of the limits taken.
The strongest form of the conjecture states that the correspondence
between $\N=4$ $SU(N)$ Yang-Mills theory, and string theory on $AdS_5
\times S^5$, is valid in general. It is not possible to test this
strongest version of the correspondence today since it is not known
how to quantize string theory on curved space backgrounds with
Ramond-Ramond flux. The second form of the correspondence restricts
the duality to the 't~Hooft limit, in which $N \rightarrow \infty$,
while $\lambda= g_{YM}^2 N$ is kept fixed. In this case only planar
diagrams contribute on the field theory side, while the string theory
on $AdS_5 \times S^5$ is restricted to the semiclassical limit in
which the string coupling $g_s \equiv g_{YM}^2$ goes to zero. Finally,
the third form of the correspondence specializes further to the case
in which $\lambda$ is large.   In
this limit, strongly coupled $\N=4$ $SU(N)$ Yang-Mills theory is
mapped to supergravity on $AdS_5 \times S^5$; the inverse string
tension $\alpha'$ goes to zero. In this paper we will be dealing with
this third form of the correspondence.
(\ref{importantrelation}) implies that the AdS radius $R$
remains finite when
 $\lambda$ is large and fixed, $N
\rightarrow \infty$, and $\alpha'$ is small.

\begin{figure}[!ht]
\begin{center}
  \includegraphics[height=6cm]{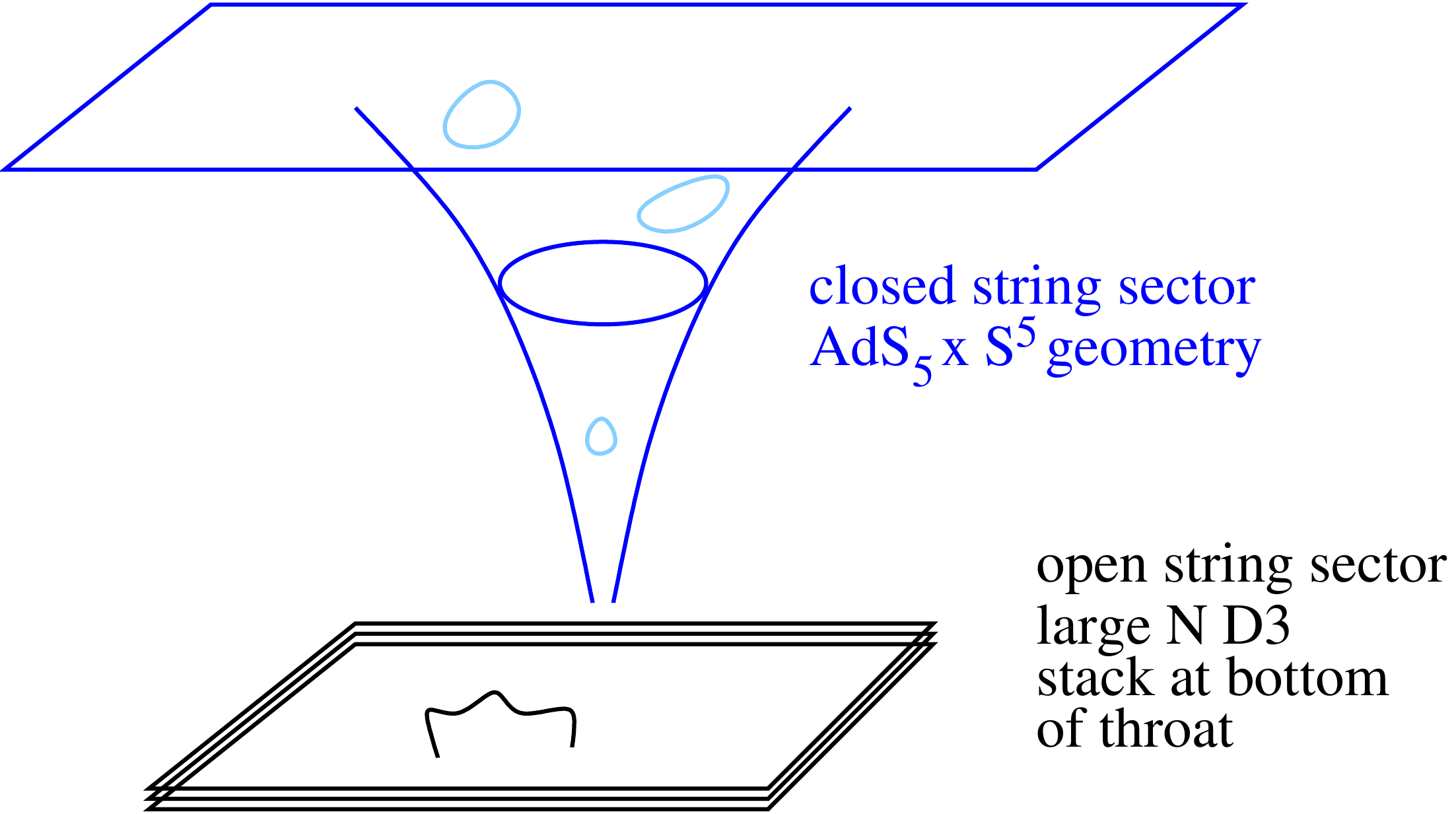}
  \caption{Schematic representation of the AdS/CFT duality.
The D3 branes warp the space into a throat whose near-horizon geometry is
$AdS_5
\times S^5$. Asymptotically, far away from the branes the geometry returns to
flat ten-dimensional space.
Open string degrees of freedom on the D3 branes, which give
rise to $\N=4$ $SU(N)$ Super Yang-Mills
theory, are mapped to closed string excitations in the $AdS_5 \times S^5$
near-horizon geometry.}\label{AdSCFT}
\end{center}
\end{figure}

The gravitational side of the correspondence has an extra non-compact
direction, $y$, relative to the gauge theory and so the correspondence
is described as being holographic \cite{tHooft:1993gx} - the contents
of the 4+1 dimensional theory are encoded by the degrees of freedom in
the 3+1 dimensional gauge theory. To understand what this extra
direction is in the gauge theory it is useful to look at the action of
dilatations (a subgroup of SO(2,4)). The action of a massless scalar
in 3+1 dimensions is invariant under
\begin{equation}
S = \int d^4x (\partial \phi)^2 \, , \hspace{1cm} x
\rightarrow e^\alpha x, \hspace{0.5cm}
\phi \rightarrow e^{- \alpha}\phi \, ,
\end{equation}
with $\alpha$ some arbitrary parameter. The power of the scaling here
tells us that $\phi$ has energy dimension one and $x$ inverse energy
dimensions.  On the gravitational side of the dual this symmetry is a
symmetry of the metric (note supergravity fields do not transform) and
for (\ref{AdSS5metric}) to be invariant we require
\begin{equation}
y \rightarrow e^{- \alpha}y \,.
\end{equation}
We have learnt that the radial direction in AdS scales like a scalar
field under the gauge theory's dilatations and hence is an energy
scale. This leads to the natural interpretation that the holographic
direction is a representation of the renormalization group scale in
the gauge theory.

The AdS/CFT correspondence has been developed further in
\cite{Witten:1998qj,Gubser:1998bc} where a {\it field-operator map}
has been established.  This maps gauge invariant operators of the
$\N=4$ Yang-Mills theory in a particular irreducible representation of
$SU(4)$ to supergravity fields in the same representation.  These
five-dimensional supergravity fields are obtained by Kaluza-Klein
reduction of the original ten-dimensional supergravity fields on the
five-sphere $S^5$. Consider a scalar field in AdS with action
\begin{equation}
S = \int d^4x du \sqrt{-g} \left(g^{ab} \partial_a \phi \partial_b \phi - m^2 \phi^2 \right) \,,
\end{equation}
where $g$ is the determinant of the metric. The solutions of the
equation of motion are of the form
\begin{equation} \label{asymptotic}
\phi(u) \sim u^{4 - \Delta} \phi_0 \, + \, u^\Delta \langle \O \rangle
\end{equation}
with $m^2 = \Delta (\Delta - 4)$.  Since the supergravity field does
not transform under the field theory dilatations and $u$ is an inverse
mass scale, we see that $\phi_0$ and $\langle \O \rangle$ carry
dimension $(4-\Delta)$ and $\Delta$ respectively.  Therefore, as
discussed in \cite{Gubser:1998bc}, the boundary value $\phi_0$ may be
identified with the source of the gauge theory-operator $\cal O$, and
$\langle \O \rangle$ is the vev (vacuum expectation value) of~$\O$.

The AdS/CFT correspondence may then be stated as follows,
\begin{equation}
\langle e^{\int \! d^d x \, \phi_0(\vec{x}) \O(\vec{x})} \rangle_{{\rm CFT}}
= Z_{{\rm Sugra}} \Big|_{\phi(0,\vec{x}) = \phi_0(\vec{x})} \, ,
\end{equation}
{\em i.e.}~the generating functional of particular gauge-invariant operators
in the conformal field theory coincides with the generating functional
for tree diagrams in supergravity, with the boundary values of the
supergravity fields coinciding with the sources.

This suggests that the AdS/CFT correspondence may be tested by
comparing correlation functions of $\N=4$ quantum field theory with
classical correlation functions on $AdS_5$. This is not possible in
general for any correlation function even in the large $N$ limit,
since in the Maldacena limit, the supergravity dual describes $SU(N)$
$\N=4$ Super Yang-Mills at strong coupling.  However, for selected
correlation functions which satisfy non-renormalization theorems such
that they are independent of the coupling, direct comparison is
possible. This applies in particular to the two- and three-point
functions of 1/2 BPS operators \cite{Freedman:1998tz, Lee:1998bxa}.
These operators are annihilated by half of the supersymmetry
generators. Another beautiful test of the AdS/CFT correspondence is
the calculation of the conformal trace anomaly of $\N=4$ theory from
$AdS_5 \times S^5$ supergravity \cite{Henningson:1998gx}.

Let us conclude this introduction to AdS/CFT by noting that up to the
present day, there is no proof of the AdS/CFT correspondence taking
account of its string-theoretical origin. However in the  weakest of
its three forms as discussed above, the huge amount of symmetry
present almost guarantees that the AdS/CFT correspondence should
hold. When proceeding to less symmetrical situations below,
generalized gauge/gravity dualities remain a conjecture though.

\subsection{Holographic RG flows} \label{flows}

A necessary ingredient for obtaining gravity duals of more QCD-like
theories than $\N=4$ Super Yang-Mills theory is to generalize the
correspondence to non-conformal field theories with less or no
supersymmetry, which have a renormalization group flow. In particular,
to obtain theories with a running coupling it is necessary to deform
the five-dimensional AdS space, which has isometry
$SO(4,2)$ \cite{Girardello:1998pd}.
This symmetry corresponds to conformal symmetry in the dual field theory
and thus to a renormalization group fixed point. The simplest way to
do this is to switch on supergravity fields in the bulk which back
react on the metric. The analysis of a scalar field in
(\ref{asymptotic}) is in fact only an asymptotic solution ignoring the
back reaction on the metric. If we switch on a supergravity field the
UV (small $u$) behaviour will be that in (\ref{asymptotic}) - so we
can identify the deforming operator present. Generically in the
interior of the space the geometry will deform from AdS indicating the
loss of conformality.

The simplest example of such a deformation is the multi-centre D3
brane solution \cite{Maldacena:1997re, Kraus:1998hv,Bakas:1999ax}.
These are geometries with a stack of parallel D3
branes present but where the D3 are separated in the six-dimensional
space transverse to their worldvolume
\begin{equation} \label{multi}
ds^2 = H^{-1/2} dx^2 + H^{1/2} (dy^2 + d \Omega_5^2), \hspace{1cm} C_4 = H^{-1} dx^0 \wedge..dx^3
\end{equation}
with
\begin{equation}
H = 1 + \sum_{D3} \left( \frac{1}{|y-y_{D3}|^4 } \right) \,.
\end{equation} The $y_{D3}$ are the positions of the D3 branes.

The function $H$ can be expanded in terms of spherical harmonics
\cite{Kraus:1998hv} on the
five-sphere, labelled by their representation of SO(6),  as follows,
\begin{equation}
H \simeq \frac{R^4}{y^4} \left( ... + \alpha y^4 + 1 +
\frac{\beta}{y^2} Y_{20} + \frac{\gamma}{y^4} Y_{50} + ... \right) \,.
\end{equation}

Each of $\alpha, \beta, ...$ is a deformation of the geometry from AdS
and has a corresponding interpretation as a deformation of the
gauge theory. They correspond to operator vevs in the dual field theory,
which  have been determined using
the symmetries of the set-up in \cite{Skenderis:2006uy,Skenderis:2006di}. 
Here we consider the following example:
$\beta$ must carry field theory energy
dimension of two (to cancel that of $y$) and be in the 20-dimensional
representation of $SO(6)$. There is indeed such an operator in the field
theory, ${\rm Tr}\, \phi^2$. Similarly $\gamma$ matches to ${\rm Tr}\,
\phi^4$. Note these operator vevs are relevant operators and are
unimportant at large $y$ (the UV),  but grow in importance into the IR
(small $y$).

The field theory intepretation of the parameter $\alpha$,
 which corresponds to leaving the near horizon
limit of the geometry, has already been found in
\cite{Intriligator:1999ai,Hashimoto:1999yc,Constable:1999ch}.  Again
from the symmetries we see that it must be an R-charged singlet and of
dimension $-4$ - it corresponds to the coupling of the interaction
term $G {\rm Tr}\, F^4$. This is an irrelevant operator whose
influence is in the UV (at large $y$),  where it grows.

These multi-centre geometries have been explicitly constructed as a
supergravity renormalization group flow in \cite{Freedman:1999gk}.

Other more complicated examples of holographic RG flows 
exist in the literature
\cite{Khavaev:1998fb,Girardello:1999hj,Freedman:1999gp,
Evans:2000ap}.
Generically, the more supersymmetry is retained the more checks of
their agreement with field theory exist. For example, a flow to an
$\N=2$ theory can be found by giving equal mass to four of the six
scalars and two of the four gauginos of the $\N=4$ theory
\cite{Pilch:2000ue,Pilch:2003jg,Brandhuber:2000ct}. This theory
is called the $\N=2^*$ theory and the induced flow of the dilaton can
be matched to the expected running coupling behaviour of the field
theory \cite{Buchel:2000cn,Evans:2000ct}. -- A general field-theoretical
interpretation of holographic RG flows is given in 
\cite{Balasubramanian:1999jd,Erdmenger:2001ja}.

Flows to $\N=1$ \cite{Girardello:1999bd,Pilch:2000fu,Polchinski:2000uf}
and $\N=0$ theories \cite{Distler:1998gb,Gubser:1999pk,Constable:1999ch,
Babington:2002ci,Babington:2002qt}
also exist in the literature. We will
introduce those which are used below in the appropriate sections of the text.

\subsection{Confinement} \label{confine}

The confinement of quarks and gluons within hadrons is a crucial
aspect of QCD and more generically strongly coupled gauge
dynamics. There has been considerable work on how the AdS/CFT
correspondence incorporates confinement which we will briefly
review here.

\subsubsection{Heavy source interaction energy}

The simplest analysis is to look at the interaction energy between two
very heavy static quarks at different separations (field theoretically
this is related to the area or perimeter law of a Wilson loop). In
\cite{Maldacena:1998im,Rey:1998ik} heavy quarks were introduced into
the AdS/CFT correspondence by placing a probe ({\em i.e.}\ non-back
reacting) D3 brane at large radius in AdS (large $y$ in the discussion
above - the field theory UV). Strings stretched from the probe D3 to
the $N$ D3 at the origin are formally very massive gauge bosons
associated with the breaking of the gauge symmetry in the pattern
$SU(N+1) \rightarrow SU(N)$ by a vev for one of the six adjoint
scalars. One can think of these strings equally as heavy sources
though since they are massive objects in the fundamental
representation of SU(N).

\begin{figure}[!ht]
\begin{center}
\includegraphics[height=6.5cm,clip=true,keepaspectratio=true]{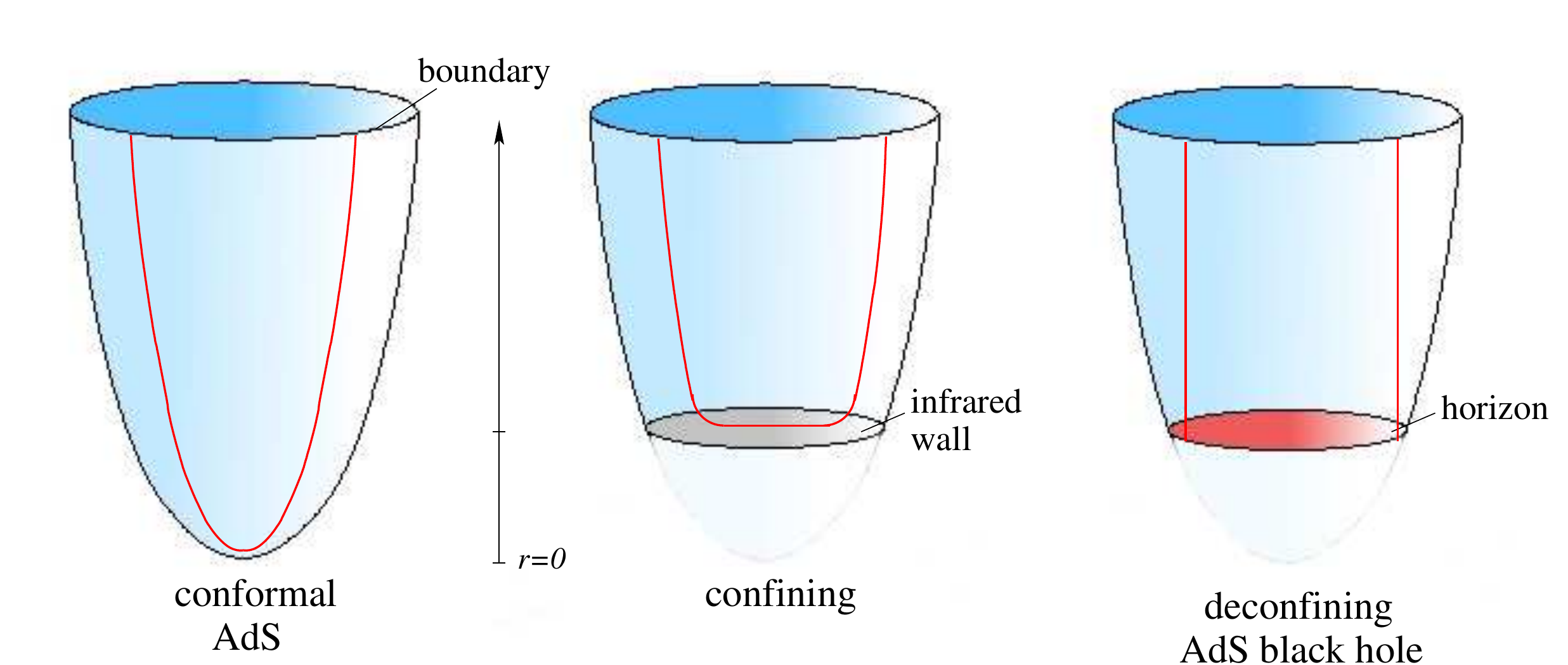}
\caption{\small{Three important configurations of strings
connecting sources on a probe brane. In AdS the strings from two
sources prefer to join than fall independently to $r=0$. They dip
further into the space the more the quarks are separated. In a
confining geometry a blockage forces the string to lie along the
blockage and the energy scales with the separation. Finite
temperature is reproduced by a black hole horizon - once the
strings fall in, the quarks are screened from each other.
}}\label{wilsons}
\end{center}
\end{figure}

When one includes two such strings to represent a quark and an
anti-quark there can be two possible configurations - see figure
\ref{wilsons}. Each string could lie straight in the space between
the probe and the central D3 stack. There would then be no
interaction between the quarks (neglecting the exchange of
supergravity fields). Alternatively it might be energetically
favourable for the strings to join, in which case their energy
would depend on the separation of the quarks on the boundary
probe.

In pure AdS a simple computation with the Nambu-Goto action of the
string \cite{Maldacena:1998im} determines the preferred configuration.
The strings indeed connect and one finds the energy of the
configuration is given by
\begin{equation}
E = - \frac{ 4 \pi^2 (2 g^2_{YM}N)^{1/2}}{\Gamma(1/4)^4 L}
\end{equation}
with $L$ the quark separation. Note the inverse proportionality to $L$
was guaranteed here by the conformal symmetry of the gauge theory. One
finds that as the quarks are separated further, the string connecting
them dips further into the AdS space.

Non-conformal gauge theories such as those induced by a deformation
will be described by some deformed AdS space and the relation between
energy and length can change radically. The simplest toy understanding
of how confinement sets in is as follows. A mass gap in the theory
will be represented by a block in the space stopping the supergravity
fields accessing values of the radius (the renormalization group
scale) below that mass gap energy. Example blockages are divergences
of the supergravity fields or the presence of branes completing the
solution below the gap radius. In these cases the string connecting
two heavy sources will behave for small separations as in AdS but as
the quarks are separated and the strings dip deeper into the interior
of the space they will eventually hit the blockage. At this point they
have little alternative than to lie along the blockage. Now separating
the sources further simply extends the string along the blockage and
the energy must be proportional to the separation $L$. This behaviour
is confinement.

Another useful example is to consider the effects of finite
temperature in this system. If one heats up a gauge theory above the
scale of its mass gap one expects the theory to deconfine. In the dual
gravity description, finite temperature is associated with the
presence of a black hole in the AdS space
\cite{Witten:1998qj,Witten:1998zw}.  A black hole has all the
associated thermodynamic properties (eg temperature and entropy) to be
dual to the equivalent properties of the thermal bath in the field
theory. The position of the black hole's horizon in the radial
direction again cuts off the space at low energies corresponding to
energies below the temperature scale. Consider again the linked string
between two heavy sources as we separate them - now as the string dips
deeper into the space, it will encounter the horizon.  The string must
fall into the black hole and we are left with two disconnected strings
from each source to the horizon. We see that the quarks are screened
from each other since they can now be moved independently.

\subsubsection{A discrete glueball spectrum}

Another clear signal that a theory is confining is if there is a
discrete spectrum of bound states. Below we will discuss in detail
mesons in theories with quarks. Let us briefly review here though
how a discrete glueball spectrum emerges in a gravity dual.

We will look for glueballs associated with the gauge field operator
${\rm Tr}\, F^2$ \cite{Witten:1998qj,Csaki:1998qr}. In the AdS/CFT
correspondence ${\rm Tr}\, F^2$ (conformal dimension $\Delta=4$) is
associated with a massless scalar (the dilaton) with an equation of
motion
\begin{equation}
\partial_r \sqrt{-g} g^{rr} \partial_r \phi + \sqrt{-g} g^{xx}
\partial_x^2 \phi = 0 \,.
\end{equation}
We will look for glueballs as solutions of the form
\begin{equation}
\phi(r,x) = \phi(r) e^{ik.x}, \hspace{1cm} -k^2=M^2 \,.
\end{equation}
In other words we are looking for pure momentum plane wave excitations
of ${\rm Tr}\, F^2$. To find a discrete spectrum we would want the
solutions for $\phi(r)$ only to exist (to be normalizable on the
space) for discrete values of the glueball mass $M$.

Generically in a deformed geometry the metric can be written in the
form
\begin{equation}
ds_5{}^2 = e^{2 A(u) } \eta_{ij} dx^i dx^j + du^2 \, .
\end{equation}
To recover AdS one sets $\exp (A) = R/u$. $u \rightarrow 0$ is the
UV and $u \rightarrow \infty$ the IR. In a deformed geometry $A$
will deviate from the AdS value as one moves into the IR.

Now if we make the transformation $\phi = e^{-3 A/2} \psi$ on the
dilaton's equation of motion it takes the form
\begin{equation}
(- \partial_r^2 + V(r)) \psi = M^2 \psi, \hspace{1cm} V =
\frac{3}{2} A^{''} + \frac{9}{4} (A')^2
\end{equation}
which is a Schr\"odinger equation with energy $M^2$.

In pure AdS the potential is given by $V \sim 1/u^2$ and the spectrum
of the Schr\"odinger equation is continuous - we expect this in a
conformal gauge theory. For a confining geometry though we expect
$A(u)$ to diverge at large $u$ creating a ``hard wall'', at some
$u=u_0$, as discussed in the previous subsection. The potential in the
Schr\"odinger equation is now a well and we expect a discrete energy
spectrum. The glueball spectrum of a theory with a mass gap of this
type in the gravity dual is very likely to generate a discrete
glueball spectrum therefore.

\newpage
\section{AdS/CFT with flavour} \label{flavour}
\setcounter{equation}{0}\setcounter{figure}{0}\setcounter{table}{0}

The original AdS/CFT correspondence only involves fields in the
adjoint representation of the gauge group. To generalize the
correspondence to quark degrees of freedom, which are in the
fundamental representation of the gauge group, additional
ingredients are necessary. The simplest thing is to add a new type
of brane into the configuration in addition to the D3 branes. The
open strings with both ends on the D3 generate the adjoint degrees
of freedom. Strings between the D3 and the new brane have only one
end on the $N$ D3 branes and hence generate matter in the
fundamental representation. Such matter will typically come in
quark super-multiplets because of the underlying supersymmetry of
the string theory. If supersymmetry is broken one expects the
scalar squarks to become massive on the scale of the supersymmetry
breaking whilst the fermionic quarks will be kept massless by
their chiral symmetries.

If the new branes can be separated from the D3 branes in some
direction transverse to both branes, then the minimum length string
between the two branes has none zero energy (length times tension) and
hence the quark is massive ($m_q=L/2 \pi \alpha'$).

Strings with both ends on the flavour brane are in the adjoint of the
U($N_f$) flavour symmetry of the quarks and hence naturally describe
mesonic degrees of freedom. In string theory these states describe
fluctuations of the brane in the background geometry. Small
oscillations of the branes are therefore dual to the gauge theory
mesons.

The need for separating the probe from the D3 brane stack excludes D9 branes
as probes. D3 and D5 brane probes lead to defect field theories discussed
below in section \ref{sec:defects}, if supersymmetry is to be preserved.
This leaves D7 brane probes for adding flavour to a $3+1$-dimensional field
theory.

\subsection{The D3/D7 brane intersection}

\begin{figure}[!ht]
\begin{center}
  \includegraphics[height=6cm]{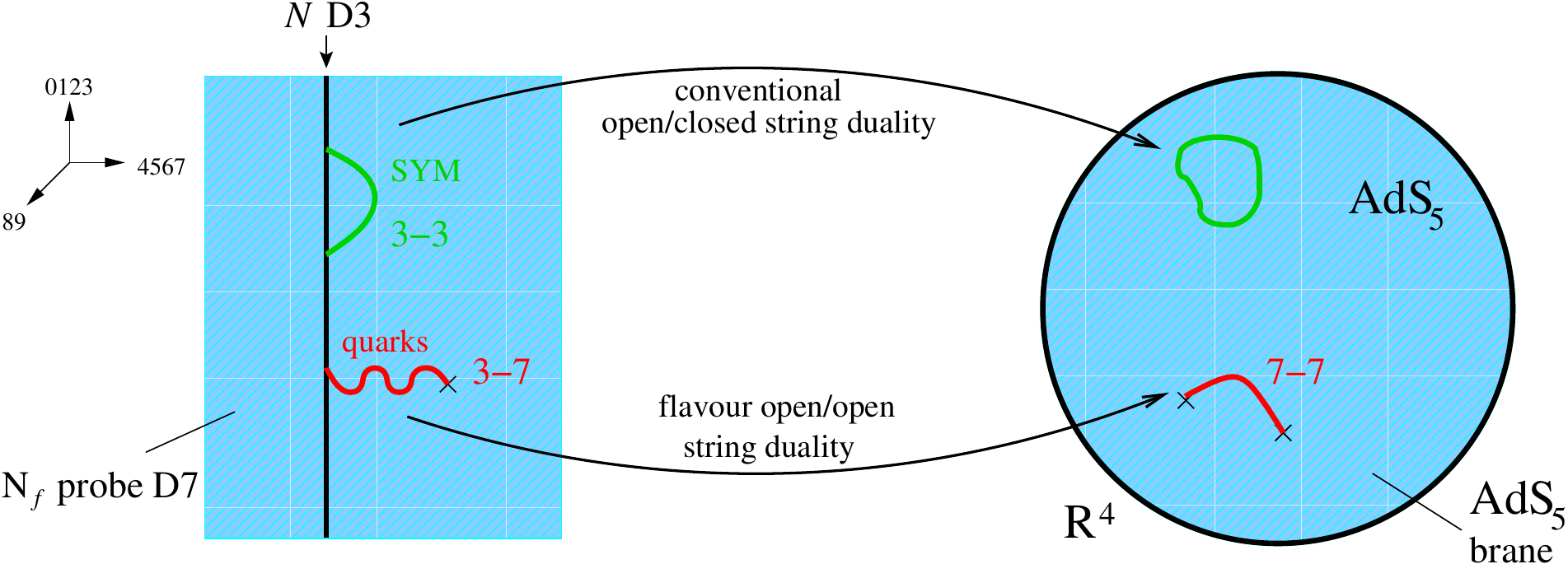}
 \caption{Schematic
  representation of the AdS/CFT duality with added flavour. In
  addition to the original AdS/CFT duality, open string degrees of
  freedom representing quarks are mapped to open strings beginning and
  ending on the D7 probe, which asymptotically near the boundary
 wrap $AdS_5\times S^3$ inside
  $AdS_5\times S^5$. For simplicity, the five-sphere is not drawn in
  this picture.} \label{mapD7}
\end{center}
\end{figure}

The simplest way to obtain quark bilinear operators within
gauge/gravity duality is to add D7 branes \cite{Bertolini:2001qa,Grana:2001xn,
  Karch:2002sh}.  The D7 branes are added in such a way that they
extend in space-time as given in table~\ref{directions}, where $0$ is
the time direction. We thus consider a stack of $N$
coincident D3-branes (along 0123) which is embedded into the world
volume of $N_f$ D7 (probe) branes (along 01234567), as shown (on the
l.h.s.\ of) figure~\ref{mapD7}.
\begin{table}[!ht]
\begin{center}
\begin{tabular}{|r|c|c|c|c|c|c|c|c|c|c|}
\hline
& 0 & 1 & 2 & 3 & 4 & 5 & 6 & 7 & 8 & 9 \\
\hline
D3    & X & X & X & X &   &   &   &   &   &   \\
\hline
D7 & X & X & X   & X   & X & X & X  & X   &   &   \\
\hline
\end{tabular}
\caption{The D3/D7-brane intersection in $9+1$ dimensional flat space.}
\label{directions}
\end{center}
\end{table}

The D3/D7 brane intersection preserves $1/4$ of the total amount of
supersymmetry in type IIB string theory (corresponding to 8
supercharges) and has an $SO(4) \times SO(2)$ isometry in the
directions transverse to the D3-branes. The $SO(4)$ rotates in $x^4,
x^5, x^6, x^7$, while the $SO(2)$ group acts on $x^8,x^9$. Note that
separating the D3-branes from the D7-branes in the 89 direction by a
distance $L$ explicitly breaks the $SO(2)$ group.  These geometrical
symmetries are also realized in the dual field theory.

\subsubsection{Field theory of the D3/D7 brane intersection} \label{sec21}

The field theory corresponding to this brane set-up is a $\N=2$
supersymmetric $U(N)$ gauge theory which, in addition to the degrees
of freedom of $\N=4$ Super Yang-Mills, contains $N_f$ hypermultiplets
in the fundamental representation of the gauge group.

This particular field theory arises as follows. The $\N=4$ super
Yang-Mills multiplet is generated by massless open string modes on the
D3-branes (3-3 strings), whereas the $\N=2$ hypermultiplets descend
from strings stretching between the D3 and the D7 branes (3-7
strings), cf.~figure~\ref{mapD7}. We take a limit in which the 7-7
strings decouple, leaving a purely four-dimensional theory. This
decoupling is achieved by taking the usual large $N$ limit while
keeping the four-dimensional 't~Hooft coupling $\lambda = g^2_{YM} N
= g_s N$ and $N_f$ fixed.  The eight-dimensional 't~Hooft coupling
$\lambda'$ for the $N_f$ D7-branes is $\lambda' =
\lambda (2\pi l_s)^4 N_f/N$ which vanishes in the low-energy $\alpha'
\rightarrow 0$ ({\em i.e.}\ $l_s \rightarrow 0$) limit. The 7-7 strings
therefore do not interact with the other (3-3, 3-7) strings anymore,
and the $U(N_f)$ gauge group on the D7-branes plays the role of a
{\em global} flavour group in the four-dimensional theory.

The Lagrangian of the $\N=2$ world-volume theory can conveniently be
written down in $\N=1$ superspace formalism.  Under $\N=1$
supersymmetry the $\N=4$ vector multiplet decomposes into the vector
multiplet $W_\alpha$ and the three chiral superfields $\Phi_1$,
$\Phi_2$, $\Phi_3$. The $\N=2$ fundamental hypermultiplets can be
written in terms of the $\N=1$ chiral multiplets $Q^r,
\tilde Q_r$ ($r=1,...,N_f$).  The Lagrangian is thus given by
\begin{align}
{\cal L} = {\rm Im} &\left[ \tau \int d^4 \theta
\left( {\rm tr\,} ( \bar \Phi_I e^V \Phi_Ie^ {-V} )+
Q^\dagger_{r} e^V Q^r + \tilde Q^\dagger_{r}  e^{-V} \tilde
Q^r \right) \right. \nonumber\\
&\left. \right. \hspace{1cm}+ \left.\tau \int d^2 \theta ( {\rm tr\,}(W^\alpha W_\alpha) + W ) + c.c.
\right] \,, \label{fieldtheoryaction}
\end{align}
where the superpotential $W$ is
\begin{align}
W={\rm tr\,} (\varepsilon_{IJK} \Phi_I\Phi_J\Phi_K)
+\tilde Q_{r} (m_q+\Phi_3) Q^r  \,, \label{superpotential}
\end{align}
and $\tau$ is the complex gauge coupling.  The beta function of this
theory is $\beta \propto \lambda^2 N_f / N$, which goes to zero for
$N_f$ small, fixed 't~Hooft coupling $\lambda$ and $N \rightarrow
\infty$, such that the theory remains conformal in this limit.

The components of the $\N=1$ superfields and their quantum numbers are
summarized in the table~\ref{tablefields} (see also
\cite{Hong:2003jm}). We will need them for the construction of
operators.  The $SO(2) \simeq U(1)$ isometry corresponds to a $U(1)_R$
R-symmetry in the field theory - note that it is explicitly broken by
a quark mass $m_q \propto L$. The field theory has also a global
$SO(4) \approx SU(2)_\Phi \times SU(2)_{\cal R}$ symmetry which
consists of a $SU(2)_\Phi$ symmetry and a $\N=2$ $SU(2)_{\cal R}$
R-symmetry. The global symmetry $SU(2)_\Phi$ rotates the scalars in
the adjoint hypermultiplet. There is also a baryonic $U(1)_B$ which is
a subgroup of the $U(N_f)$ flavour group. The fundamental superfields
$Q^r$ ($\tilde Q_r$) are charged $+1$ ($-1$) under $U(1)_B$, while the
adjoint fields are inert.

\begin{table}
\begin{center}
\begin{tabular} {|c|c|c|c|c|c|c|c|}
\hline
${\cal N}=2$
&components & spin & $SU(2)_\Phi \times SU(2)_{\cal R}$ & $U(1)_{\cal R}$ &$\Delta$ & $U(N_f)$ & $U(1)_B$\\
\hline
$(\Phi_1, \Phi_2$) & $X^4, X^5, X^6,  X^7$ &$0$& $(\frac{1}{2},\frac{1}{2})$& $0$ & $1$& 1 & 0\\
hyper &$\lambda_1, \lambda_2$ & $\frac 1 2$ & $(\frac{1}{2},0)$& $-1$ & $\frac{3}{2}$& 1 & 0\\
\hline
($\Phi_3$, $W_\alpha$) & $X_V^A=(X^8, X^9)$ & $0$ &$(0,0)$ & $+ 2$ &$1$& 1 & 0 \\
vector&$\lambda_3, \lambda_4$ &$\frac 1 2$& $(0,\frac{1}{2})$& $+1$&$\frac{3}{2}$ & 1
& 0\\
&$v_\mu$& $1$ &$(0,0)$ & $0$& $1$& 1& 0\\
\hline
($Q$, $\tilde Q$) & $q^m=(q, \bar {\tilde q})$&$0$ & $(0,\frac{1}{2})$& $0$&$1$& $N_f$ & +1\\
fund.~hyper &$\psi_i=(\psi, \tilde \psi^\dagger)$ & $\frac{1}{2}$ & $(0,0)$& $\mp 1$& $\frac{3}{2}$ &$N_f$&+1 \\
\hline
\end{tabular}
\end{center}
\caption{Fields of the D3/D7 low-energy effective field theory and
their quantum numbers under the global symmetries. Note that
$U(1)_B \subset U(N_f)$.}
\label{tablefields}
\end{table}

\subsection{The probe brane correspondence} \label{probesupergravity}

The simplest way to analyze the D3/D7 system is to work in the limit
where the D7 branes are treated as probes \cite{Karch:2002sh}.
The term `brane probe' \cite{Johnson:2003gi}
refers to the fact that only a very small number of D7 branes is
added, while the number of D3 branes, $N$, which also determines the
rank of the gauge group $SU(N)$, goes to infinity. In this limit we
neglect the backreaction of the D7 branes on the geometry. Naively it
seems peculiar to treat the large brane as the probe but here one is
working in the $N \rightarrow \infty$ limit so the D3 branes can
dominate. The limit is clearest on the field theory side: the geometry
represents the gauge configuration in which the quarks move. If we
neglect the D7 effects we are simply dropping quark loops from the
gauge background which is simply quenching the gauge theory.  In
section~\ref{secbackreaction} we will discuss the D3/D7 brane
configuration for finite $\frac{N_f}{N}$ including the backreaction
of the flavour branes.

On the supergravity side of the duality, the $\N=4$ degrees of freedom
are described by supergravity on $AdS_5\times S^5$ as before. However
in addition, there are new degrees of freedom corresponding to the D7
brane probe within the ten-dimensional curved space. The low-energy
degrees of freedom of this brane are described by the
Dirac-Born-Infeld action as described below.  These correspond to open
string fluctuations on the D7 probe.  It turns out, as we will show
shortly, that the minimum action configuration for the D7 brane probe
corresponds to a probe configuration which asymptotically near the
boundary wraps an $AdS_5 \times S^3$ subspace of $AdS_5 \times S^5$.

As shown in figure \ref{mapD7}, the new duality conjectured in
\cite{Karch:2002sh} is an open-open string duality, as opposed to the
original AdS/CFT correspondence which is an open-closed string
duality.  The duality states that in addition to the original AdS/CFT
duality, gauge invariant field theory operators involving fundamental
fields are mapped to fluctuations of the D7 brane probe inside 
$AdS_5 \times S^5$. This is also shown in
figure~\ref{mapD7}.

Let us determine the D7 embedding explicitly. The dynamics of the D7
brane probe is described by the combined Dirac-Born-Infeld and
Chern-Simons actions,
\begin{equation}
S_{D7} = -\mu_7\int \df^8\xi\,
\sqrt{-\det\left(P[G]_{ab}+2\pi\ap F_{ab}\right)}
+  \frac{(2\pi\alpha')^2}{2} \mu_7\int P[C^{(4)}] \wedge F \wedge F\, .
\label{daction}
\end{equation}
$\mu_7=[(2\pi)^7g_s\ap^4]^{-1}$ is the D7-brane tension and $P$
denotes the pullback of a bulk field to the world-volume of the brane.
$F_{ab}$ is the world-volume field strength.  The D7-brane action also
contains a fermionic term $S^f_{D7}$ which will be discussed in
section~\ref{secmesino}.

If we write the $AdS_5 \times S^5$ metric in the form
\begin{equation} \label{AdSmetric}
ds^2 \, = \, \frac{r^2}{R^2} \eta_{ij} dx^i dx^j
 \, + \, \frac{R^2}{r^2} (d \rho^2 + \rho^2 d \Omega_3^2 + dw_5^2 + dw_6^2)
 \, ,
\end{equation}
with $\rho^2=w_1^2+...+w_4^2$, $r^2=\rho^2+w_5^2+w_6^2$, then the
action for a static D7 embedding (with $F_{ab}$ zero on its
world-volume) is given up to angular factors from (\ref{daction}) by
\begin{equation}
S_{D7} = -\mu_7\int \df^8\xi\, \rho^3
\sqrt{1 + \dot{w}_5^2 + \dot{w}_6^2} \,,
\end{equation}
where a dot indicates a $\rho$ derivative ({\em e.g.}\ $\dot w_5 \equiv
\partial_\rho w_5$). The ground state configuration of the D7-brane then
corresponds to the solution of the equation of motion
\begin{equation} \label{d7eom} {\frac{d}
{d \rho}} \left[ {\rho^3 \over \sqrt{1 + {\dot w_{5}}^2+ {\dot w_{6}}^2}
}{ d w \over d \rho}\right] = 0 \, ,
\end{equation}
where $w$ denotes either $w_5$ or $w_6$. Clearly the action is
minimized by $w_5,w_6$ being any arbitrary constant. The D7 brane
probe therefore lies flat in the space. The choice of the position in
the $w_5,w_6$ plane corresponds to choosing the quark mass in the
gauge theory action.  That $w_5, w_6$ is constant at all values of
$\rho$ is a statement of the non-renormalization of the mass. The
coordinate $\rho$ is a holographic radial direction of the background
AdS space and therefore corresponds to the renormalization group
scale. The non-renormalization of the mass is an expected
characteristic of supersymmetric gauge theories.

In general, the equations of motion have asymptotic ($\rho\rightarrow \infty$)
solutions of the form
\begin{equation} \label{probeasymptotic}
w = L + \frac{c}{\rho^2} + ...
\end{equation}
$L$ corresponds to the quark mass as discussed. In agreement with the AdS/CFT
result (\ref{asymptotic}), the extra parameter
$c$ must correspond to the vev of an operator with the same symmetries as the
mass and of dimension three (since $\rho$ carries energy dimension).
$c$ is therefore a measure of the quark condensate ($\bar{q}_L q_R$;
more formally it corresponds to $\partial {\cal L} / \partial m$ which
includes scalar squark terms, but we assume that the squarks have zero vev
here). 
Moreover, note that solutions with $c$ non-zero are not regular in AdS space
and these solutions are excluded. This corresponds to a vev for this
operator being forbidden by supersymmetry - it is an F-term of a
chiral superfield.\footnote{For $m_q \neq 0$, consider the term $m_q
  \psi \tilde \psi$ which is the F-term of $m_q Q \tilde Q$.
  Supersymmetry is broken, if $c=\langle \psi \tilde \psi \rangle \neq
  0$. Vice versa, if supersymmetry is preserved, then $c=0$ and the
  embedding profile must be flat.}
-- A detailed discussion 
of relation between the asymptotic behaviour (\ref{asymptotic}) and
(\ref{probeasymptotic}) was given in in \cite{Karch:2005ms} in 
the context of `holographic renormalization' \cite{Skenderis:2002wp}. 

\begin{figure}[t]
\begin{center}
  \includegraphics[height=6cm]{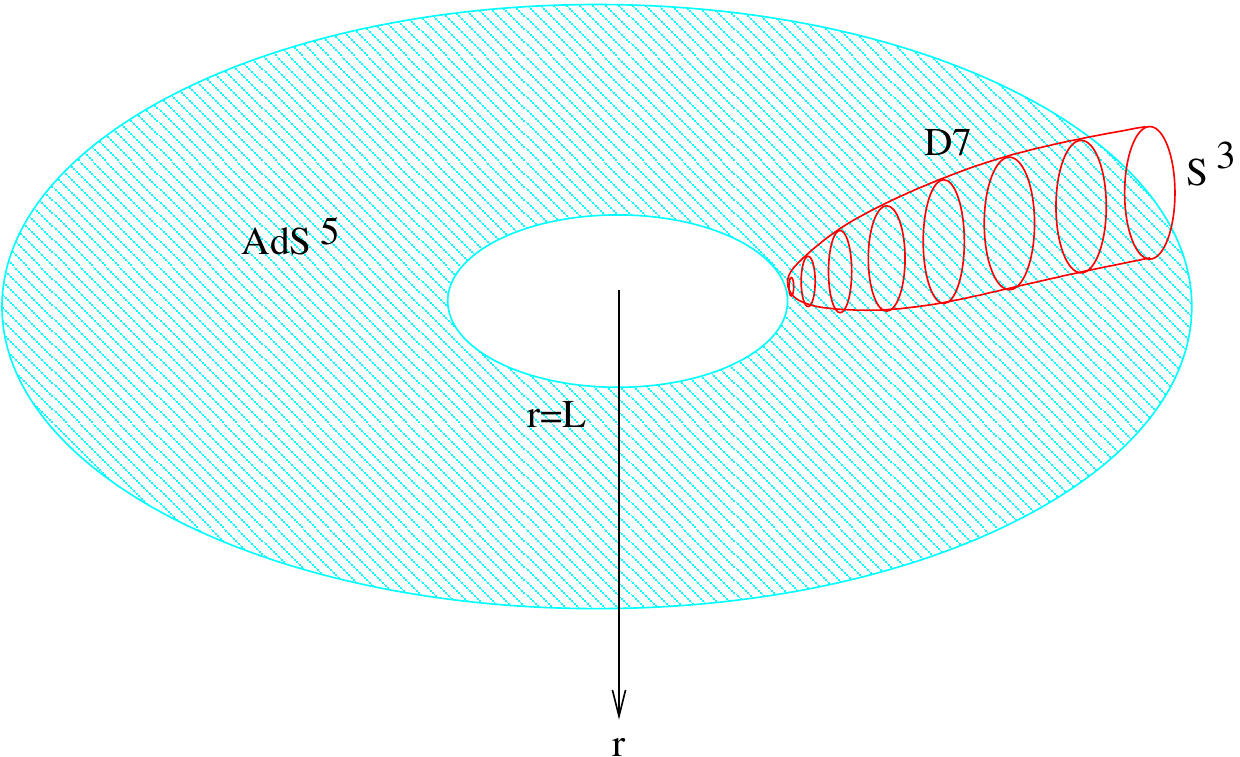}
\caption{Gravity dual in presence
  of a fundamental hypermultiplet with finite mass $m$. The D7 brane
  probe is shown in blue.  In this case, the radius of the $S^3$
  wrapped by the D7 brane probe becomes a function of the AdS radial
  direction $r$. At $r=L$, the radius of the $S^3$ shrinks to zero,
  and the D7 probe does not extend any further into the interior of
  AdS space. (Figure by Zachary Guralnik, from \cite{Babington:2003vm}.)} 
 \label{flow}
\end{center}
\end{figure}
A particularly interesting feature arises if the D7 brane probe is
separated from the stack of D3 branes in either the $w^5$ or $w^6$
directions, where the indices refer to the coordinates given in
(\ref{AdSmetric}). This corresponds to giving a mass to the
fundamental hypermultiplet.  In this case the radius of the $S^3$
becomes a function of the radial coordinate $r$ in ${\rm AdS}_5$. At a
radial distance from the deep interior of the AdS space given by the
hypermultiplet mass, the radius of the $S^3$ shrinks to zero. From a
five-dimensional AdS point of view, the D7 brane probe seems to `end'
at this value of the AdS radial coordinate, as shown in figure~\ref{flow}.

This can be seen as follows.  According to \cite{Kruczenski:2003be},
the induced metric on the D7 brane world-volume is
\begin{equation} \label{inducedmetric}
ds^2 \, = \, \frac{\rho^2+L^2}{R^2} \eta_{ij} dx^i dx^j
 \, + \, \frac{R^2}{\rho^2 + L^2} d\rho^2
 \, + \, \frac{R^2 \rho^2}{\rho^2+L^2 }  d\Omega_3{}^2\, ,
\end{equation}
where $\rho^2 = r^2 -L^2$ and $\Omega_3$ are spherical coordinates in
the 4567-space. For $\rho \rightarrow \infty$,
this is the metric of  $AdS_5 \times S^3$.  When
$\rho=0$ ({\em i.e.}~$r^2=L^2$), the radius of the $S^3$ shrinks to zero.

The scalar mode with dimension $\Delta=3$ ({\em i.e.}~supergravity mass
$M_{\rm sugra}^2=\Delta(\Delta-4)=-3$) maps to the fermion bilinear
$\tilde \psi \psi$ in the dual field theory. This mode corresponds to
an imaginary AdS mass.  However this mass is above the
Breitenlohner-Freedman bound \cite{Breitenlohner:1982bm,
  Breitenlohner:1982jf} for $AdS_5$ ($M^2_{\rm BF} = -4$) and thus
guarantees stability.  For this it is important that the D7 branes do
not carry any net charge from the five-dimensional point of view,
since they wrap a topologically trivial cycle with zero flux.

\begin{figure}[t]
\begin{center}
  \includegraphics[height=4cm]{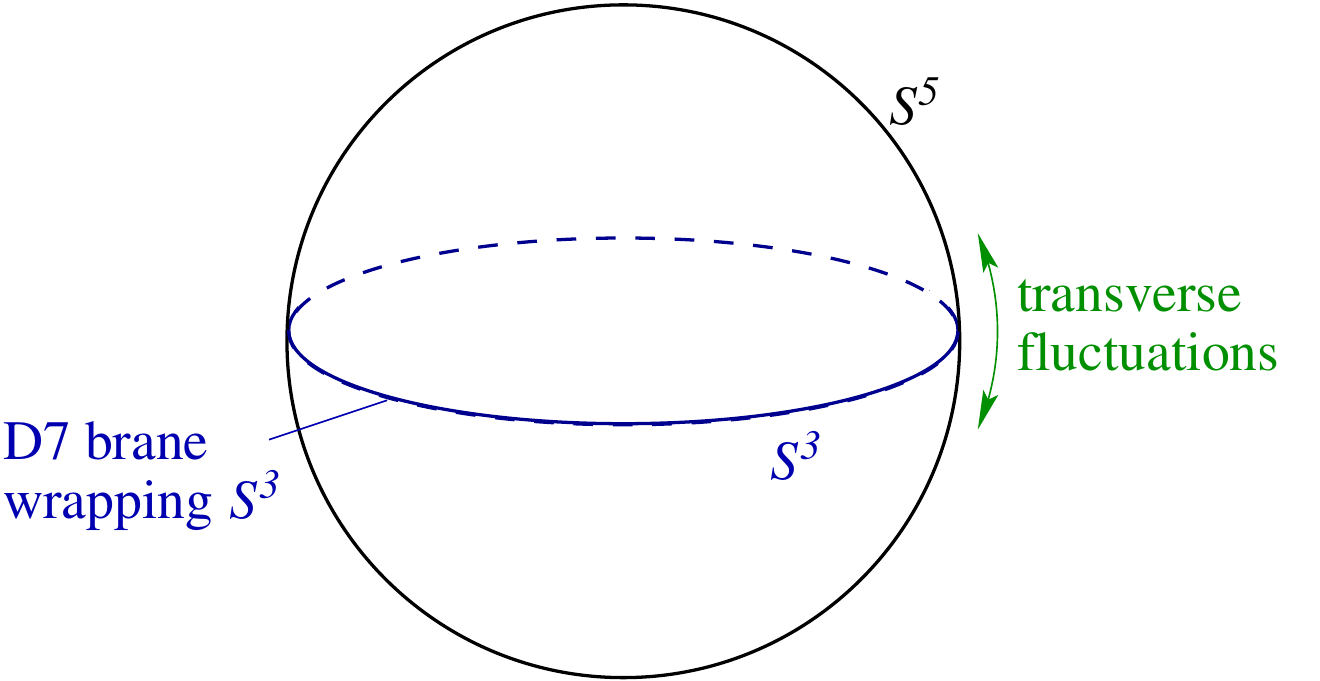}
  \caption{Fluctuations of the $S^3$ wrapped by the D7 probe inside
    $S^5$.  These modes give rise to the meson masses.}
  \label{ball}
\end{center}
\end{figure}

\medskip
\subsection{D7 brane fluctuations and mesons in $\N=2$ theory}
\label{section2.3}

The fluctuations of the D7 brane give rise to the mesons and we can
determine their masses. This is similar to previously studied
supergravity fluctuations which give rise to glueball
masses~\cite{Csaki:1998qr}.

\subsubsection{Scalar field fluctuations (spin $0$)}
\label{scalarfieldfluctuations}

As an example, we discuss the fluctuation modes and meson masses for
the scalar fields, following the discussion of
\cite{Kruczenski:2003be}.

The directions transverse to the D7-brane are chosen to be $w^5$ and
$w^6$, and the embedding is as follows,
\begin{equation}
w_5=0+\delta w_5 \,, \qquad w_6=L+\delta w_6 \,, \label{embed}
\end{equation}
where $\delta w_5$ and $\delta w_6$ are the transverse scalar
fluctuations shown in figure~\ref{ball}.
To calculate the spectra of the worldvolume fields it is sufficient to
work to quadratic order. For the scalars, the relevant part of the
Lagrangian density is
\begin{equation}
\cL \simeq -\mu_7 \sqrt{-\det g_{ab}}\left(1+\frac{1}{2} \frac{R^2}{r^2}
{g^{cd}} \pr_c \Phi \pr_d \Phi \right)\ . \label{lag}
\end{equation}
Here, $\Phi$ is used to denote either (real) fluctuation,
$\delta w_{5, 6}$, and
$g_{ab}$ is the induced metric on the D7 worldvolume as given by
(\ref{inducedmetric}). In spherical coordinates with $r^2= \rho^2 +
L^2$ the equation of motion becomes
\beq
\pr_a\left(\frac{\rho^3\epsilon_3}{\rho^2+L^2}g^{ab}\pr_b \Phi
\right)=0\ .\label{meow}
\eeq
$\epsilon_3$ is the metric on the unit sphere spanned by $(\rho,
\Omega_3)$.

The equation of motion can be expanded as
\beq
\frac{R^4}{(\rho^2+L^2)^2}\pr^\mu\pr_\mu\Phi
+\frac{1}{\rho^3}\pr_\rho(\rho^3\pr_\rho\Phi)
+\frac{1}{\rho^2}\nabla^i \nabla_i \Phi=0\ , \label{mex}
\eeq
where $\nabla_i$ is the covariant derivative on the three-sphere.
Using separation of variables,  an ansatz for the modes may be written as
\beq
\Phi = \phi(\rho) e^{ik \cdot x} \cY^{\ell}(S^3)\ , \label{wavefunk}
\eeq
where $\cY^\ell(S^3)$ are the scalar spherical harmonics on $S^3$,
which  satisfy
\beq
\nabla^i \nabla_i \cY^{\ell} = -\ell(\ell+2) \cY^{\ell}\ .
\label{seer}
\eeq
The meson masses are defined by $M^2 = -k^2$ for the wavevector $k$ introduced
in (\ref{wavefunk}).

Then equation (\ref{mex}) gives rise to an equation for $\phi(\rho)$ that,
with the redefinitions
\beq
\vr = \frac{\rho}{L}\ , \qquad \bar{M}^2 = -\frac{k^2R^4}{L^2}\ ,
\label{redefinition}
\eeq
becomes
\beq
\pr_\vr^2\phi+\frac{3}{\vr}\pr_\vr\phi
+\left(\frac{\bar{M}^2}{(1+\vr^2)^2}
-\frac{\ell(\ell+2)}{\vr^2}\right)\phi=0\ .
\label{meor}
\eeq
This equation may be solved in terms of a hypergeometric function. Imposing
normalizability, the solution is
\beq
\phi(\rho)=\frac{\rho^\ell}{(\rho^2+L^2)^{n+\ell+1}}\
F\left(-(n+\ell+1)\ ,\ -n\ ;\ \ell+2\ ;\ -\rho^2/L^2\right)\
\label{salaam}
\eeq
with
\beq
\bar{M}^2=4(n+\ell+1)(n+\ell+2)\ .
\label{spectre}
\eeq
Using this, and $M^2=-k^2= \bar{M}^2 L^2/R^4$,
the four-dimensional mass spectrum of scalar mesons is given by
\beq
M_s(n,\ell)=\frac{2L}{R^2}\sqrt{(n+\ell+1)(n+\ell+2)}\ .
\label{spectrum}
\eeq
Normalizability of the modes results in a discrete spectrum with
a mass scale set by $L$, the position of the D7-brane.

\subsubsection{Fermionic fluctuations (spin $\frac{1}{2}$)}
\label{secmesino}

The spectrum of fermionic fluctuations of the D7-brane has been
studied in \cite{AKR, Kirsch:2006he}. These fluctuations are dual to
so-called ``mesino'' operators which are the fermionic superpartners
of the mesons.  Typical mesino operators with conformal dimension
$\Delta=\frac{5}{2}$ and $\Delta=\frac{9}{2}$ are ${\cal F} \sim \bar
\psi q$ and ${\cal G} \sim \bar \psi \lambda \psi$, where $\psi$ ($q$)
is a quark (squark) and $\lambda$ an adjoint fermion.  The precise
form of these operators is given in section~\ref{secdictionary}.

The dual fluctuations have spin $\frac{1}{2}$ and are described by the
{\em fermionic} part of the D7-brane action, that is the
supersymmetric completion of the Dirac-Born-Infeld action. This action
is given in an explicit form by \cite{Martucci:2005rb}
\begin{align} \label{fermionicDBI}
S^{f}_{D7}= \frac{\tau_{D7}}{2} \int d^{8}\xi \sqrt{-\det g} {\bar
   \Psi} {\cal P}_- \Gamma^A ( D_A + \frac{1}{8}
   \frac{i}{2 \cdot 5!} F_{NPQRS}
   \Gamma^{NPQRS} \Gamma_A ) \Psi  \,.
\end{align}
Here $\xi^A$ are the world-volume coordinates ($A =0,...,7$) which, in
static gauge, will be identified with the spacetime coordinates $t,
x^1,...,x^7$.  The field $\Psi$ is the ten-dimensional  positive chirality
Majorana-Weyl spinor of type IIB string theory and $\Gamma_A$ is the
pullback of the ten-dimensional
 gamma matrix $\Gamma_{M}$ ($M, N, ...  =0,...,9$),
$\Gamma_A=\Gamma_{M} \partial_A x^{M}$.  The integration goes over the
world-volume of the D7-brane which wraps a
submanifold of \mbox{$AdS_5 \times S^5$} which asymptotes to
$AdS_5 \times S^3$.  The spinor $\Psi =
\Psi (x^M, \theta^m)$ depends on the coordinates $x^M$ of $AdS_5$ and
the three angles $\theta^m=(\theta^1, \theta^2, \theta^3)$ of the
three-sphere $S^3$.  The operator ${\cal P}_-$ is a $\kappa$-symmetry
projector ensuring $\kappa$-symmetry invariance of the action. The
action $S_{D7}=S^b_{D7}+S^f_{D7}$ with $S^b_{D7}$ and $S^f_{D7}$
given by (\ref{daction}) and (\ref{fermionicDBI}) is therefore
invariant under supersymmetries corresponding to any bulk Killing
spinor.

We must now evaluate the five-form $F_{NPQRS}$ as well as the
curved-spacetime covariant derivative $D_M$ on $AdS_5 \times S^5$
\cite{AKR, Kirsch:2006he}.  This will give a Dirac-type equation which
will then be transformed into a second-order differential equation.
The fluctuations are assumed to be of plane-wave type, $\Psi(x, \rho)
= \psi_{\ell,\pm}(\rho) e^{ik_\mu x^\mu} \chi^\pm_\ell $, where
$\psi_{\ell,\pm}$ and $\chi^\pm_\ell $ are spinors on $AdS_5$
and~$S^5$, respectively.\footnote{The $\pm$ signs refer to the
  eigenvalues of the spinor spherical harmonics on $S^3$, $\lambda R =
  \pm(\ell + \frac{3}{2})$.} $M^2=-k^2$ is again interpreted as the
mass of the dual operator.  After a somewhat lengthy calculation,
which we do not present here, one obtains~\cite{AKR}\footnote{For
  overlapping D3/D7 branes ($L=0$) this equation reduces to that found
  in \cite{Kirsch:2006he}.}
\begin{align}
&\left[ \partial_\rho^2 + \frac{3}{\rho}\left(1+\frac{\rho^2}{r^2}\right)
\partial_\rho + \frac{M^2R^4}{r^4} - \frac{3}{4} \frac{\rho^2}{r^4}
+ \frac{1}{r^2} (7-2 m_\ell R+(2m_\ell R-1)\gamma^\rho)
\right.\nonumber\\ &\left.\quad- \frac{1}{\rho^2} \left( (m_\ell R -1)^2
-\textstyle\frac{3}{4} + (m_\ell R -1) \gamma^\rho \right) \right]
\psi_{\ell,\pm}(\rho) = 0  \qquad (\ell \geq 0)\,, \label{fluc}
\end{align}
where $r^2={\rho^2+L^2}$, and the distance $L$ is proportional to the
quark mass $m_q$, $L=2\pi \alpha' m_q$. $m_\ell R$~represents one of the
masses
\begin{align}
m_{\ell,+} R = \textstyle\frac{5}{2} +\ell \,,\qquad
m_{\ell,-} R = -(\textstyle\frac{1}{2} + \ell) \,.
\end{align}

The spin-$\frac{1}{2}$ operators dual to the fluctuations
$\psi_{\ell,\pm}$ will be denoted by ${\cal G}^\ell_\alpha$ and ${\cal
  F}^\ell_\alpha$. The mass-dimension relation for spin-$\frac{1}{2}$
fields, $\vert m \vert = \Delta - 2$, determines the conformal
dimensions of these operators:
\begin{align}
\Delta_{\cal G} =  \textstyle\frac{9}{2} + \ell \,,\qquad
\Delta_{\cal F} =  \textstyle\frac{5}{2} + \ell \,\qquad (\ell \geq 0)\,.
\end{align}
We must also ensure that the operators ${\cal G}^\ell_\alpha$ and
${\cal F}^\ell_\alpha$ have the same $SO(4)$ and $U(1)_R$ quantum
numbers as the fluctuations. For instance, the spinorial spherical
harmonics on $S^3$ transform in the
$(\frac{\ell+1}{2},\frac{\ell}{2})$ and
$(\frac{\ell}{2},\frac{\ell+1}{2})$ of $SO(4)=SU(2)\times SU(2)$,
while the $U(1)_R$ charge is $+1$.  These properties uniquely fix the
structure of ${\cal G}^\ell_\alpha$ and ${\cal F}^\ell_\alpha$.  Their
explicit form is given in section~\ref{secdictionary}.

The fluctuation equation (\ref{fluc}) can now be solved in terms of
hypergeometric functions. For instance, for the fluctuations
$\psi_{\ell,+}$ the solution is given by \cite{AKR}
\begin{align}
\psi_{\ell,+} &=\,u_{\,+}\,\varrho^{\,\ell+1}\,\left(1+\varrho^2
\right)^{\,-{1\over2}\left(2\lambda\,+\,{3\over2}\right)}\,F\,(\,
-\lambda\,,\,\ell-\lambda+2\,,\,\ell+3\,,\,-\varrho^2\,)  \nonumber\\
&\quad+ u_{-}\,\varrho^{\,\ell}\,\left(1+\varrho^2
\right)^{\,-{1\over2}\left(\,2\lambda\,+\,{3\over2}\right)}\,F\,(\,
-\lambda-1\,,\,\ell-\lambda+2\,,\,\ell+2\,,\,-\varrho^2\,)
\,,
\label{psihatmnsol}
\end{align}
where we rescaled $\varrho=\frac{\rho}{L}$, $\bar M^2=
\frac{M^2R^4}{L^2}$, and defined $\lambda$ by $\bar
M^2=4\,\lambda(\lambda+1)$. The spinors satisfy $\gamma^\rho u_\pm =
\pm u_\pm$ and $u_- = \frac{\gamma^\mu k_\mu}{k^2} u_+$.

In order for the solution to be well-behaved at large radii,
$\varrho\to\infty$, the solution is subjected to the quantization
condition
\begin{align}
-n = \ell- \lambda +2 \,.
\end{align}
Solving this for $\bar M^2$, we obtain the fluctuation masses
\begin{align}
\bar M^2_{\cal G}=4(n+\ell+2)(n+\ell+3)\,
\end{align}
which is the spectrum of the operators ${\cal G}^\ell_\alpha$.  The
spectrum of ${\cal F}^\ell_\alpha$ is obtained in a similar way by
solving the equations of motion for $\psi_{\ell,-}$.

\subsubsection{Gauge field fluctuations (spin $1$)}

The fluctuations of the D7 world-volume gauge field $A_M$
($M=0,...,7$) give rise to three further mass spectra denoted by
$M_{I,\pm}, M_{II}$ and $M_{III}$ in \cite{Kruczenski:2003be}. These
spectra are generated by plane-wave fluctuations of the components
$A_i$ (along the $S^3$), $A_\mu$ (along $x^{0,...,3}$) and $A_\rho$
(along the radial direction $\rho$) of the eight-dimensional
world-volume gauge field $A_M=(A_\mu, A_\rho, A_i)$.  Details on the
computation of these spectra can be found in \cite{Kruczenski:2003be}.

\subsubsection{Fluctuation-operator matching}
\label{secdictionary}

So far we discussed the mass spectra of open string fluctuations on
the D7-branes. In order to interpret these spectra as those of
meson-like operators, we must map the fluctuations to the
corresponding meson operators in the dual field theory. In the
following we construct these operators and assign them to the
corresponding open string fluctuations.

As was first found in \cite{Kruczenski:2003be}, the complete set of D7-brane
fluctuations fits into a series of massive gauge supermultiplets of
the $\N=2$ supersymmetry algebra. These multiplets contain
$16(\ell+1)$ states with the masses
\begin{align}
  M^2 = \frac{4L^2}{R^4} (n+ \ell +1)(n+\ell+2) \qquad (n,\ell
  \geqslant 0) \, .
\end{align}
Since the supercharges commute with the generators of the global group
$SU(2)_\Phi$, the $SU(2)_\Phi$ quantum number $\frac{\ell}{2}$ is the
same for all fluctuations in a supermultiplet.

All D7 brane fluctuations and their quantum numbers are listed in
table~\ref{tableop}. The notation of the fluctuations and their mass
spectra is the same as in \cite{Kruczenski:2003be}. The numbers
$(j_\Phi,j_{\cal R})_q$ label a representation of $SO(4) \approx
SU(2)_\Phi \times SU(2)_{\cal R}$, and $q$ is the $U(1)_R$ charge. In
order to count the number of states in a multiplet we must take into
account the degeneracy in the $SU(2)_{\cal R}$ quantum number, {{\em
    i.e.}}\ we count the degrees of freedom of a particular massive
fluctuation and multiply it with $(2j_{\cal R}+1)$. Then, the number
of bosonic components in a multiplet is
\begin{align}
1 (2 (\textstyle\frac{\ell}2+1)+1) + (2+3+1)(2
\textstyle\frac{\ell}2+1) + 1 (2 (\textstyle\frac{\ell}2-1)+1) =
8(\ell+1) \,.
\end{align}
Of course, this agrees with the number of fermionic components,
\begin{align}
4 (2 \textstyle\frac{\ell+1}2+1) +
4 (2 \textstyle\frac{\ell-1}2+1) = 8(\ell+1) \,,
\end{align}
giving altogether $16(\ell+1)$ states.

\begin{table}[ht]
\begin{center}
\begin{tabular}{|c|c|c|c|c|cc|c|c|}
\hline
      &\!\! fluctuation & d.o.f. & $(j_\Phi,j_{\cal R})_q$ & 5d mass & spectrum && op. & $\Delta$\\
\hline  \vspace{-0.4cm}&&&&&&&&\\
mesons & 1 scalar  &1& ($\frac{\ell}{2}$, $\frac{\ell}{2}+ 1$)$_0$ & $m^2=-4$ &  $M_{I,-}(n,\ell+1)$  & $(\ell \geq 0)$ & ${\cal C}^{I\ell}$ & $2$ \\
(bosons)       & 2 scalars &2& ($\frac{\ell}{2}$, $\frac{\ell}{2}$)$_2$  &$m^2=-3$  & $M_s(n,\ell)$ & $(\ell \geq 0)$ & ${\cal M}_s^{A\ell}$  & $3$\\
       & 1 scalar  &1& ($\frac{\ell}{2}$, $\frac{\ell}{2}$)$_0$ & $m^2=-3$  & $M_{III}(n,\ell)$  & $(\ell \geq 1)$ & ${\cal J}^{5\ell}$ & $3$\\
 & 1 vector  &3 & ($\frac{\ell}{2}$, $\frac{\ell}{2}$)$_0$ & $m^2=0$  & $M_{II}(n,\ell)$  & $(\ell \geq 0)$ & ${\cal J}^{\mu\ell}$  & $3$ \\
       & 1 scalar  &1& ($\frac{\ell}{2}$, $\frac{\ell}{2}- 1$)$_0$& $m^2=0$ &  $M_{I,+}(n,\ell-1)$  & $(\ell \geq 2)$ & -- & $4$  \\
\hline \vspace{-0.4cm}&&&&&&&&\\
mesinos  & 1 Dirac  &4 & ($\frac{\ell}{2}$, $\frac{\ell + 1}{2}$)$_1$& $|m|=\frac{1}{2}$& $M_{{\cal F}}(n, \ell)$  & $(\ell \geq 0)$& ${\cal F}^\ell_\alpha$ & $\frac{5}{2}$ \\
(fermions)  & 1 Dirac &4& ($\frac{\ell}{2}$, $\frac{\ell - 1}{2}$)$_1$& $|m|=\frac{5}{2}$& $M_{{\cal G}}(n, \ell-1)$  & $(\ell \geq 1)$ & ${\cal G}^\ell_\alpha$ & $\frac{9}{2}$\\
\hline
\end{tabular}
\end{center}
\caption{Field content of the $\N=2$ supermultiplets in the D3/D7 theory.}\label{tableop}
\end{table}

\medskip We now assign operators to the D7 brane fluctuations
appearing in table~\ref{tableop}. Note that the masses are above the
Breitenlohner-Freedman bound \cite{Breitenlohner:1982bm,
  Breitenlohner:1982jf} and thus admissible, even if their square is
negative.  Open strings are dual to composite operators with
fundamental fields at their ends: scalars $q^m=(q, \bar {\tilde q})^T$
and spinors $\psi_i=(\psi, \tilde \psi^\dagger)^T$. We will refer to
these operators as mesons and their superpartners as {\em mesinos}. We
must ensure that the operators have the same quantum numbers ({\em
  i.e.}~spin, global symmetries, etc.) as the corresponding
fluctuations. Also, the five-dimensional mass of a fluctuation and the
conformal dimension of the dual operator must satisfy a particular
relation depending on the spin, {\em e.g.}\ $m^2=\Delta(\Delta-4)$ for
scalars.

Let us construct gauge invariant operators for the bosonic fluctuations
\cite{Kruczenski:2003be, Hong:2003jm, Kirsch:2006he}. First, there is a
scalar in the \mbox{($\frac{\ell}{2}$, $\frac{\ell}{2}+1$)$_0$} with
5d mass $m^2 = -4 + \ell \geq m^2_{BF}$\footnote{The lowest
fluctuation with has negative mass squared, $m^2=-4$, saturating the
Breitenlohner-Freedman bound, $m^2_{BF}=-d^2/4=-4$ ($d=4$).}  which
corresponds to the $\Delta=\ell+2$ chiral primaries
\begin{align}\label{primaries}
 {\cal C}^{I\ell}= \bar q^m \sigma_{mn}^I X^\ell q^n \,.
\end{align}
Here the Pauli matrices $\sigma_{mn}^I$ ($I=1,2,3$) transform in the
triplet representation of $SU(2)_R$, while $q^m$, $\psi^i$ and
$X^\ell$ have the $SO(4)$ quantum numbers $(0, \frac{1}{2})$, $(0,0)$
and $(\frac \ell 2, \frac \ell 2)$, respectively. $X^\ell$ denotes the
symmetric, traceless operator insertion $X^{\{ i_1} \cdots X^{ i_\ell
\}}$ of $\ell$ adjoint scalars $X^i$ ($i=4,5,6,7$). This operator
insertion generates operators with higher angular momentum $\ell$.

Then, there are 2 scalars in the ($\frac{\ell}{2}$,
$\frac{\ell}{2}$)$_2$ which we are dual to the scalar meson operators
\begin{align} \label{mesonop}
 {\cal M}^{A\ell}_{s} = \bar \psi_i \sigma^A_{ij} X^\ell \psi_j + \bar
 q^m X^A_V X^\ell q^m \,\qquad(i, m=1,2)
\end{align}
which have conformal dimensions $\Delta=\ell+3$. Here $X^A_V$ denotes
the vector $(X^8, X^9)$ and $\sigma^A=(\sigma^1, \sigma^2)$ is a
doublet of Pauli matrices. Both $X^A_V$ and $\sigma^A$ transform in
the ${\mathbf 2}$ of $U(1)_R$. The operators ${\cal M}^{A\ell}_{s}$
thus transform in the $(\frac{\ell}{2}, \frac{\ell}{2})$ of $SO(4)$
and are charged $+2$ under $U(1)_R$.

Next, there is a vector in the ($\frac{\ell}{2}$,
$\frac{\ell}{2}$)$_0$ associated with the $\Delta=\ell+3$ operator
\begin{align}
 {\cal J}^{\mu\ell} &= \bar \psi_i^\alpha \gamma^\mu_{\alpha\beta}
X^\ell \psi_i^\beta
+ i \bar q^m X^\ell D^\mu q^m
- i \bar  D^\mu \bar q^m X^\ell q^m \qquad (\mu=0,1,2,3)
 \,
\end{align}
which we identify as the $U(N_f)$ flavour current.

Finally, there is a (pseudo-)scalar in the ($\frac{\ell}{2}$,
$\frac{\ell}{2}$)$_0$ dual to ${\cal J}^{5\ell-1} = \bar \psi_i^\alpha
\gamma^5_{\alpha\beta} X^{\ell-1} \psi_i^\beta +... $ ($\ell \geq 1$)
and a scalar in the ($\frac{\ell}{2}$, $\frac{\ell}{2}+1$)$_0$
($\ell\geq 2$) which corresponds to a higher descendant of ${\cal
  C}^{I\ell}$. These operators do not appear in the lowest ($\ell=0$)
multiplet.

\medskip

We now turn to the fermionic fluctuations \cite{Kirsch:2006he}. These
fluctuations are dual to so-called {\em mesino} operators, the
superpartners of the meson-like operators studied above. There are two
types of spin-$\frac{1}{2}$ fluctuations with quantum numbers
($\frac{\ell}{2}$, $\frac{\ell + 1}{2}$)$_1$ and ($\frac{\ell}{2}$,
$\frac{\ell - 1}{2}$)$_1$. These correspond to the mesino operators
\begin{align} \label{opF}
{\cal F}^{\ell}_{\alpha} &= \bar q
 X^{\ell} \tilde \psi^\dagger_\alpha  + \tilde \psi_\alpha X^{\ell} q  \,, \\
 {\cal G}^{\ell-1}_{\alpha } &
 = \bar \psi_i \sigma^B_{ij}
 \lambda_{\alpha C} X^{\ell-1} \psi_j
 + \bar q^m X_V^B \lambda_{\alpha C} X^{\ell-1} q^m
 \,,  \qquad(A,B,C=1,2)
\end{align}
which have the conformal dimensions $\Delta=\frac{5}{2}+\ell$ $(\ell
\geq 0)$ and $\Delta=\frac{7}{2}+\ell$ $(\ell \geq 1)$,
respectively. As their bosonic partners, mesinos have fundamental
fields at their ends.  The spinors $\lambda_{\alpha A}$ ($A=1,2$) have
the $SO(4)$ quantum numbers $(\frac{1}{2},0)$ and belong to the
adjoint hypermultiplets $(\Phi_1, \Phi_2)$.

\subsubsection{Interactions}

Form factors for the interactions between the mesons can be computed
from higher order terms in the DBI action. For example if we consider
$N_f$ D7 branes then there are DBI terms of the form

\beq
S \sim \int d^8x \sqrt{-g} g^{ab} g^{\mu \nu} f^{abc} A^a_\mu
A^b_\alpha \partial_\nu A^c_\beta \,,
\eeq
where $f^{abc}$ is a structure constant for the flavour group.
$A_\alpha$ are the scalar fields discussed above and $A_\mu$ describe
the vector mesons.  If we substitute in the solutions for the meson
mass eigenstates we have found above and integrate over the four
directions of the D7 transverse to the D3 we are left with the
effective interaction between two scalars and a vector meson. Equally
one could replace $A_\mu$ by it's non-normalizable solution giving the
coupling of the two scalars to a flavour gauge boson.

These form factors are explicitly computed for the $\N=2$ theory in
\cite{Hong:2003jm} (see also \cite{Hong:2004sa}).  There the form
factors are also Fourier transformed to position space to give an
estimate of the effective size of the mesons.  The typical dimension
is given by the inverse of the meson's mass $\sqrt{g_{YM}^2 N}/ m_q$
\cite{Hong:2003jm,Hong:2004gz}.

\subsubsection{Mesons on the Coulomb branch}

The $\N=4$ gauge theory has a large moduli space on which the six
adjoint scalars have mutually commuting vacuum expectation values.
This corresponds in the gravity dual to separating the D3 branes in
the six transverse directions to their world volume as discussed above
in section~\ref{flows} - the gravity dual is a multi-centre solution
(\ref{multi}). D7 brane probes continue to lie flat in these
geometries since the $H$ factors of the metric cancel from the DBI
action. The adjoint vev should generate a quark mass through the
Yukawa term in the superpotential $\tilde{Q} A Q$.  Mesonic
fluctuations for some sample geometries have been computed in
\cite{Shock:2006fc} and indeed for massless quarks the mesons have
masses proportional to the vacuum expectation value of the scalars.


\subsection{Holographic heavy-light mesons}

\begin{figure}[t]
\begin{center}
  \includegraphics[height=4.5cm,clip=true,keepaspectratio=true]{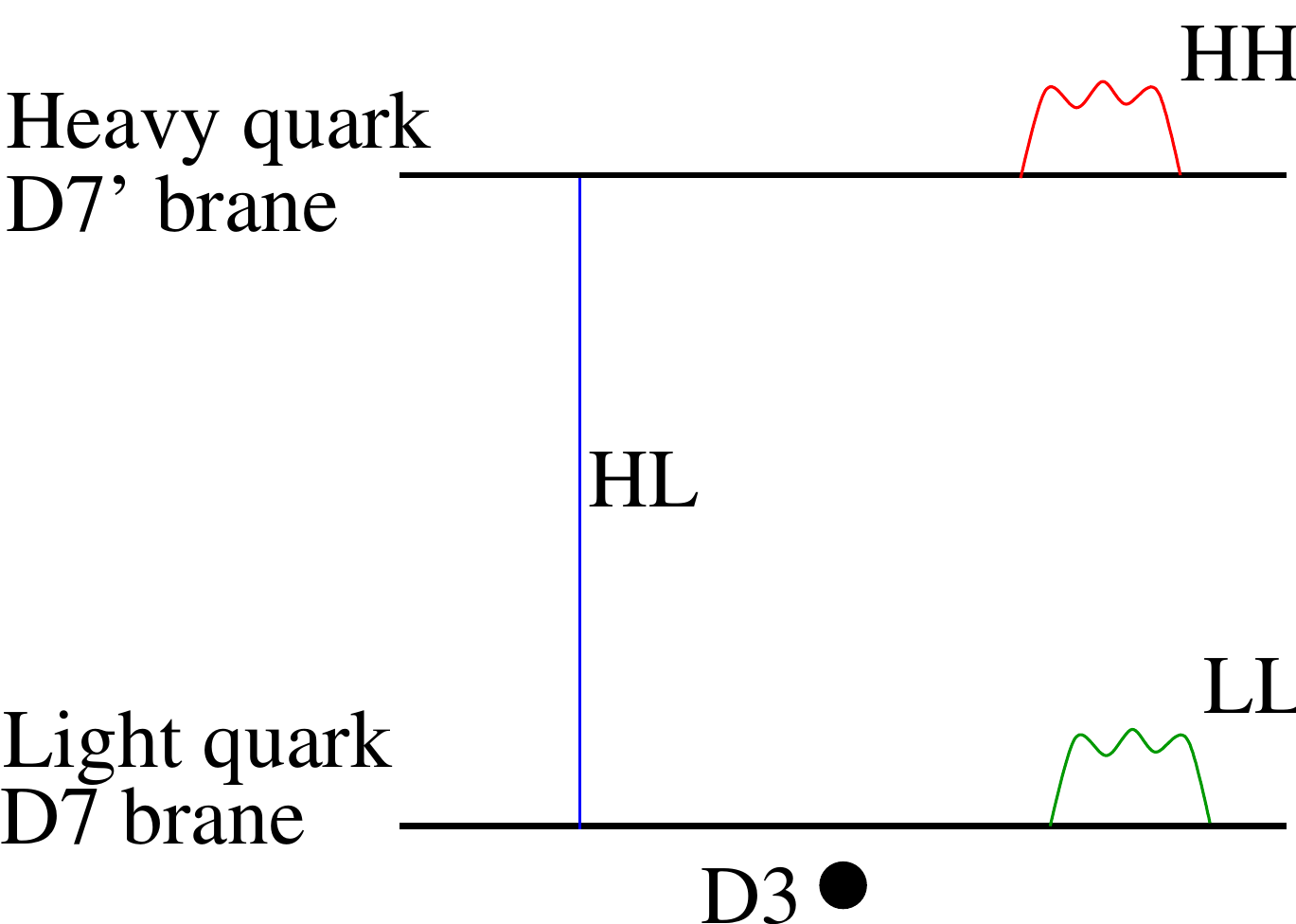}
  \caption{The brane configuration including both a heavy and a light quark.
  The $77$ and $7'7'$ strings are holographically dual
  to light-light and heavy-heavy
  mesons, respectively. Heavy-light mesons are holographically described by
  strings between the two D7 branes -- we work in the semi-classical limit
  where those strings are stretched tight. From \cite{Erdmenger:2006bg}.
 }\label{f:stretch}
\end{center}
\end{figure}

We have seen that meson states made of a quark and its anti-quark are
described by the open string modes on the surface of a D7 brane in a
D3 brane background. We can introduce two quarks with different masses
by including two D7 branes with different separations (in the $w_5$ or
$w_6$ directions) from the D3 branes as shown in
figure~\ref{f:stretch}. The strings stretched between the two D7
branes carry the flavour quantum numbers of each of the two branes and
therefore they have the correct symmetries to holographically describe
the heavy light meson operators.

In AdS the preferred static configuration for these strings is to lie
stretched straight between the D7 branes at constant $\rho$ as if they
were in flat space \cite{Karch:2002xe}.  This can be easily seen from
the Nambu-Goto action of the strings,
\begin{equation}
S_{\rm string} = T \int d^2 \sigma \sqrt{\det G_{MN} {dX^M \over d
\sigma^a} {dX^N \over d \sigma^b} } \,.
\end{equation}
For a static string the determinant is given by the product of the
$G_{tt}$ and $G_{w_5w_5}$ metric components. In AdS this product
cancels to give unity and the action is that in flat space with the
straight string minimizing the action. For the moment we will
concentrate on this configuration and assume we can give the string
some small centre of mass motion without it bending.

As the separation between the two D7 branes is increased the 77$'$
strings grow and naively one expects these states to have a mass given
by the product of their length and tension. This immediately provides
an apparent confusion - the mass of the supergravity state is
holographically related to the dimension of the operator it is
describing in the field theory. Why should the bi-fermionic heavy
light operator's dimension be changing as we increase one quark's
mass? To resolve this confusion it is helpful to look at the Polyakov
form of the action for the string.

We use the gauge-fixed Polyakov string action \beq S_P = -
\frac{T}{2} \int d\sigma d \tau G_{\mu \nu} ( - \dot{X}^\mu
\dot{X}^\nu + X'{} ^{\mu} X'{} ^{\nu}) \, , \end{equation} so we
must also impose the constraint equations \beq G_{\mu \nu}
\dot{X}^\mu X'{}^\nu =0, \hspace{1cm} G_{\mu \nu}(\dot{X}^\mu
\dot{X}^\nu + X'{}^\mu X'{}^\nu)=0  \, . \end{equation} For the
configuration we are considering and for a diagonal metric the
first constraint vanishes ($\dot{w_5}=0$). In flat space the
second equation, after integration over $\sigma$ gives the
familiar $E^2-p^2 = L^2 T^2$ energy momentum relation for the
centre of mass motion with $L$ the length of the string.

In AdS the $x_{//}$ and $\rho$ directions are distinct and we must
be careful. Integrating the metric over $\sigma$ gives

\begin{equation} S_P = - \frac{T L}{2} \int d \tau \left[-
\tilde{G}_{xx} \dot{x}^2 - \tilde{G}_{ww} \dot{w}_i^2 +
\tilde{G}_{ww} \right] \, , \label{poly}
\end{equation} where \beq \tilde{G}_{xx} = {1 \over L} \int_0^L
d \sigma G_{xx} ={1 \over R^2} \left(\rho^2  +\frac{1}{ 3} L^2
\right), \quad \tilde{G}_{ww} = {1 \over L} \int_0^L d
\sigma G_{ww} = \frac{R^2}{\rho L} \arctan (L/\rho) \, .
\end{equation}
These are essentially averages of the metric components along the
stretched string's length.

The constraint, when integrated over $\sigma$, gives
 \begin{equation} \label{Legendre} \tilde{G}^{xx} p_{x}^2 + \tilde{G}^{ww}
p_{w}^2 + T^2 L^2 \tilde{G}_{ww} =0 \, , \end{equation} where
$p_x{}^\alpha \equiv  \delta {\cal L} / \delta \dot x_\alpha$,
$p_w{}^i \equiv  \delta {\cal L} / \delta \dot w_i$. Note that
(\ref{Legendre}) is a simple modification of the usual
$E^2-p^2=m^2$ with the effective mass depending on the $\rho$
position of the string. If we expand for large $\rho$ we obtain
\begin{equation} p_{x}^2 +{\rho^4 \over R^4} p_{w}^2 + L^2 T^2 =0  \, .
\end{equation}

The form of this equation is transparent in terms of the dilatations
in the field theory - $x$ is a length whilst $w$ have energy
dimensions. The factor of $\rho^4$ is clearly necessary. We can now
see that for motion in the $\rho$ direction at large $\rho$ (the UV of
the field theory) the string mass is effectively zero no matter the
length of the string - the ``holographic'' mass determining the
operator dimensions is zero independent of the string length. On the
other hand for motion in the $x$ directions the state has a large mass
if the string is long and this will be reflected by the meson mass
becoming degenerate with the quark mass at large quark mass. Note this
behaviour should be compared with the meson mass made of the heavy
quark and it's anti-quark - that state is lighter with mass suppressed
at large 't~Hooft coupling by $\sqrt{\lambda}$.

We have assumed above that the straight string can be boosted from
rest in the $\rho$ direction. In fact there are not solutions of this
form. We know of no studies of moving strings but presumably the
string bends. We will continue to work here in the
straight string approximation -
this is presumably reasonable for short strings or slow moving
strings.

\subsubsection{Semi-classical action for heavy-light states \label{longhl}}
\label{NickHL}

The classical analysis above of the heavy light strings of course
misses much of the quantum theory - in particular the unexcited string
is the tachyon which is not part of the theory and the lightest state
is the spin one gauge field. What we can learn from (\ref{poly}) above
is that the centre of mass of the string state has the standard action
of a particle in a curved space-time although with metric factors
averaged over its length. We expect an action in 10d of the form
 \begin{equation}
S = {1 \over (2 \pi)^9 \alpha^{'5}}\int d^{10}x \sqrt{-{\rm det}
\tilde{G}_{10}} e^{-\phi}\left({ - 1 \over 4} \tilde{G}^{MN}
\tilde{G}^{KL} F_{MK} F_{LK} + M^2 \tilde{G}_{MN} A^{M}A^{N}\right) \, .
\end{equation}

The ends of these string though are tied to  D7 branes so we must
T-dualize the action twice - the $A^{8,9}$ components of the gauge
field become two scalars, $\phi^a$, with action
\begin{equation}
 S = {1 \over (2 \pi)^9 \alpha^{'5}} \int d^2x \sqrt{-{\rm det} \tilde{G}_{8-9}}
 \int d^8x e^{-\phi} \sqrt{-{\rm det} \tilde{G}_{0-7}} ~
 \left(\tilde{G}^{mn}\tilde{G}_{ww}
 \partial_m \phi^a \partial_n \phi^a + M^2 \phi^{a2} \right) \, .
\end{equation}
The two dimensional integral simply gives an overall factor of $(2 \pi
R)^2$. One must also re-write the dilaton in terms of the dilaton of
the T-dual theory (one equates the string coupling of the two theories
as described in \cite{Myers:1999ps,Johnson:2003gi} - $e^{-\phi_9} =
e^{-\phi_7} \alpha'/R^2$. We have
\begin{equation} S = {1 \over (2 \pi)^7 \alpha^{'4}}
 \int d^8x e^{-\phi} \sqrt{-{\rm det} \tilde{G}_{0-7}} ~ \left(\tilde{G}^{mn}
 \tilde{G}_{ww}
 \partial_m \phi^a \partial_n \phi^a + M^2 \phi^{a2} \right) \, .
\end{equation}

The kinetic term of this action takes the form of the lowest order
expansion of the DBI action for a D7 brane except with metric factors
averaged over the $w_5$ direction. In the limit of very small D7
separation the metric factors simply become those on the D7 branes'
world volumes and these states form part of the non-abelian DBI
action. In addition there is a mass term for the string -- in the
semi-classical limit of a very long string, one has
\begin{equation}
M = L T \tilde{G}_{ww}
\end{equation}
to be consistent with (\ref{Legendre}).

\begin{figure}
\begin{center}
\includegraphics{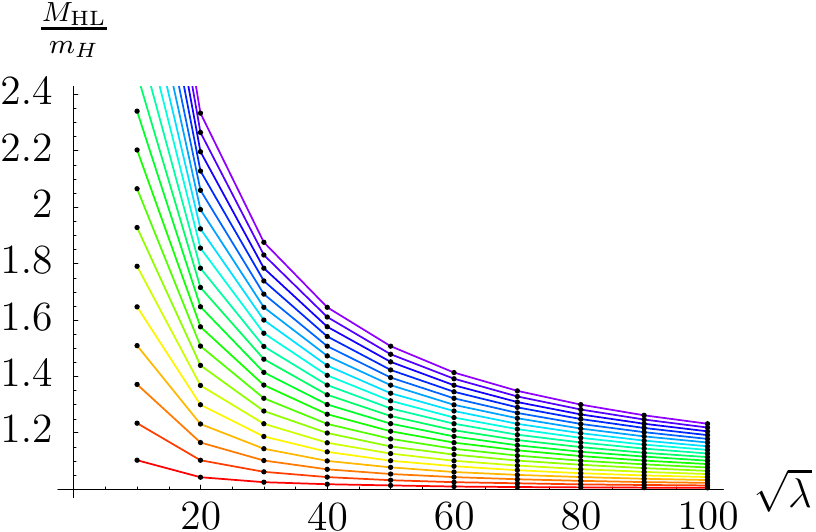}

\caption{The masses $M_{HL}$ of the meson and its excited states for
  the AdS background.  The ratio $M_{HL}/m_H$, with $m_H$ the heavy
  quark mass (the light quark is taken to be massless), is plotted
  against the square root of the 't~Hooft coupling $\lambda$.  We
  observe that in the large $\lambda$ limit, $M_{HL}/m_H$ behaves as
  $1+\text{const}/{\sqrt \lambda} +
  \mathcal{O}(\lambda^{-1})$. From \cite{Erdmenger:2006bg}. \label{f:ads1}}
\end{center}
\end{figure}

The heavy-light mesons and their radially excited partners are then
described by the holographic equation of motion
\begin{equation}
\partial_\rho \sqrt{-{\rm det} \tilde{G}_{0-7}} \partial_\rho \phi +
\sqrt{-{\rm det} \tilde{G}_{0-7}} \tilde{G}_{ww} \tilde{G}^{xx}
\partial_x^2 \phi - \sqrt{-{\rm det} \tilde{G}_{0-7}} \tilde{G}_{ww}^2 LT \phi
=0 \, ,
\end{equation}
with solutions of the form $\phi(x, \rho) = f(r) e^{ik.x}$, $-k^2=M^2$ as
usual. A plot of the solutions from \cite{Erdmenger:2006bg} is shown
in figure~\ref{f:ads1} - the meson masses divided by the heavy quark
mass are plotted as a function of the 't~Hooft coupling. Note that at
large 't~Hooft coupling, the meson mass is just that of the heavy
quark (or long string) as expected.

\subsubsection{Heavy-light mesons from non-abelian DBI action}

A different approach to the holographic description of heavy-light
mesons has been proposed in \cite{Erdmenger:2007vj}. Again in this
case, two D7 brane probes are embedded at different positions into the
ten-dimensional gravity background. Now, however, these two branes are
described by a non-abelian Dirac-Born-Infeld action. In this action,
the world-volume fields are assigned to $U(N_f)$ matrix-valued
functions for $N_f$ D7 branes.  We choose $N_f=2$.  The embedding
configuration of the two D7 branes is determined by the diagonal
components of the scalar fields. The corresponding equation of motion
is solved by the profile functions of two separated branes, one of
which corresponds to the heavy and one to the light quark. The quark
masses are given by the boundary values of the two embedded branes.
The fluctuations of the diagonal elements of the $2\times 2$ flavour
matrices correspond to the light-light and heavy-heavy mesons,
respectively.  On the other hand, the off-diagonal components of the
fluctuations of the fields on the branes are identified with the
heavy-light mesons.\bigskip

\subsubsection*{Embeddings}

The starting point is the non-Abelian Dirac-Born-Infeld action in
curved space proposed by Myers in \cite{Myers:1999ps}. This action
describes the dynamics of $N_f$ D$p$-branes in a background with
metric $G_{mn}$ and is given by
\begin{align}
S_{N_f} = - \tau_p \int d^{p+1} \xi  e^{-\phi} {\,\rm STr} \left(
\sqrt{-\det ( P[G_{rs} + G_{ra} (Q^{-1} -\delta)^{ab} G_{sb}]
+ T^{-1} F_{rs})}
\sqrt{\det Q^a{}_b} \right) \,, \label{two-brane}
\end{align}
where the matrix $Q^a{}_b$ is defined by
\begin{align}
 Q^a{}_b = \delta^a{}_b + i T [X^a,X^c] G_{cb} \, ,\qquad
T^{-1} =2\pi\alpha ',
\end{align}
and $X^a$ are the coordinates transverse to the stack of branes, which
now take values in a $U(N_f)$ algebra.  The symbol ${\rm STr}$ denotes
the symmetrized trace ${\rm STr} (A_1 ...A_n) \equiv \frac{1}{ n!}
{\rm Tr}(A_1...A_n $ + all permutations) and is needed to avoid the
ambiguity of the ordering of the expansion of all fields in the DBI
action.

This non-Abelian DBI action is now  used to find the embedding of
$N_f$ probe D7 branes in different gravity backgrounds. The embedding
profiles correspond to the classical solutions for the scalar fields
in the D7 brane action.  In our case, the scalar fields $X^a$ are
$U(N_f)$ matrix valued functions, which makes it difficult to obtain a
general form of the profile functions. In order to simplify the
problem, we use the diagonal ansatz
\beq \label{diagonal}
 X^a={\rm diag}(w_1^a, \cdots, w_{N_f}^a) \, ,
\eeq
thereby setting all off-diagonal components to zero.  Here each of the
functions $w_i^a$ corresponds to one of the $N_f$ D7 branes.

The quark mass for each flavour is given by the asymptotic value of
$w_i^a$ in the ultraviolet limit. They are the integration constants
and given by hand as parameters of the theory.
The equations of motion for the $w_i^a$ are
obtained from the action
\begin{align}
 S_{N_f} &=
  \tau_7   \int d^{8} \xi \  e^{-\Phi}{\rm STr}  \left(
  \sqrt {-\det ({G}_{rs}  + G_{ab}\partial_r w_i^a \partial_s w_i^b )}
  \right) \nonumber \\
 &= \tau_7   \int d^{8} \xi \  e^{-\Phi}\sum_{i=1}^{N_f}
  \sqrt {-\det ({G}_{rs}+ G_{ab}\partial_r w_i^a \partial_s w_i^b )}\,
  \label{two-brane-3}
\end{align}
which is Eq.~(\ref{two-brane}) for the embedding (\ref{diagonal})
and $p=7$. The essential point is here that for the diagonal ansatz
(\ref{diagonal}), we obtain $N_f$ decoupled equations of motion for
the $w_i^a$, such that the embeddings of each of the probe branes is
independent of the other. In other words, for diagonal embeddings the
non-Abelian DBI action reduces to the sum of $N_f$ abelian DBI
actions.\bigskip

\noindent{\bf Fluctuations} \bigskip

We now consider the scalar and vector meson spectra obtained by
considering the fluctuations about the background given.  At this
stage, we restrict to the case of $N_f=2$ flavours or two D7 branes
such that the scalar and vector fields in the non-Abelian DBI action
are represented by $2\times 2$-matrices. For the classical embedding,
we choose the diagonal configuration given by
\beq
 \bar{X}^8=0\,,\qquad
 \bar{X}^9=\left(\begin{array}{cc}
            w_1&0\\
            0&w_2 \end{array}\right)\, .
\eeq
In terms of the Pauli matrices
\beq
 \tau^0={1\over 2}
        \left(\begin{array}{rr}
            1&0\\
            0&1 \end{array}\right)\, ,\,\,
\tau^1={1\over 2}
       \left(\begin{array}{rr}
            0&1\\
            1&0 \end{array}\right)\, ,\,\,
\tau^2={1\over 2}
       \left(\begin{array}{rr}
            0&-i\\
            i&0 \end{array}\right)\, ,\,\,
\tau^3={1\over 2}
       \left(\begin{array}{rr}
            1&0\\
            0&-1 \end{array}\right)\,,
\eeq
$\bar X^9$ can be rewritten as
\beq
 \bar{X}^9=w\tau_0+v\tau_3\,, \quad w_1=(w+v)/2\,, \quad w_2=(w-v)/2\,,
 \label{profile}
\eeq
where $v=w_1-w_2$. The asymptotic boundary values of $w_1$ and $w_2$
correspond to the heavy and light quark masses, respectively.  When
$v=0$, the two branes are at the same place, $w_1=w_2=w$,
corresponding to a $U(2)$ flavour symmetry. For $v \neq 0$ this
flavour symmetry is explicitly broken.

The scalar and gauge field fluctuations are taken to be of the form $(a=8, 9)$
\beq
X^9=\bar{X}^9+\phi^9\,, \quad X^8=\phi^8\,, \label{expansions}
\eeq
\beq \phi^a=\phi^a_0\tau^0+\phi^a_i\tau^i\, , \quad
A^r=A^r_0\tau^0+A^r_i\tau^i\, , \label{flucs}
\eeq
and can be written as
\beq
\phi^a = \left(\begin{array}{cc}
    \phi^a_+ & \phi^a_{12}\\
    \phi^a_{21}& \phi^a_- \end{array}\right) \,,
\eeq
similarly $A^r$.  The diagonal elements $\phi^a_\pm=\phi^a_0 \pm
\phi^a_3$ describe fluctuations of each brane and are dual to the
heavy-heavy and light-light mesons. On the other hand, the
off-diagonal elements $\phi^a_{12}=\phi^a_1-i\phi^a_2$ and
$\phi^a_{21} = \phi^a_1+i\phi^a_2$ correspond to fluctuations of
strings stretched between the two branes and are dual to the
heavy-light mesons. The mass of this last type of fluctuations will
depend on $v$. A similar structure emerges also for gauge field
fluctuations $A_r$.

These meson mass spectra are obtained by solving the linearized
equation of motions for the field fluctuations. For the $AdS_5 \times S^5$
background, the heavy-light meson masses are obtained from
\beq
\left(\partial_{\rho}^2+{3\over \rho}\partial_{\rho} -{l(l+2)\over
    \rho^2}+{M^2-v^2\over 2}\left( \left({R^2\over
        \rho^2+w_1^2}\right)^2+ \left({R^2\over
        \rho^2+w_2^2}\right)^2\right) \right) \phi=0 \,.
\label{H-L-phi-2}
\eeq
For $w_1=w_2=w$, we get $v=0$ and the equation reduces to the one
given by Kruczenski {\em et al} \cite{Kruczenski:2003be} which can be
solved analytically, as described in section~\ref{meor} above.

A central point is that the $\lambda$ dependence of the heavy-light meson
mass obtained from the non-abelian DBI action as described here
coincides with the one obtained using the Polyakov action approach
discussed in section~\ref{longhl} above.  A finite contribution to the
mass remains in the limit of $\lambda\to \infty$. This contribution
corresponds to the minimum energy of a classical string connecting two
separated D7 branes, and thus is equivalent to the mass obtained from
the Polyakov action.

In general, (\ref{H-L-phi-2}) must be solved numerically. However,
for a heavy-light meson with a very heavy quark, $w_2 \gg
w_1$, the term in (\ref{H-L-phi-2}) involving $w_2$
is much smaller than the one involving $w_1$ and may be neglected.
In this case, the heavy-light meson mass is found to be
\beq
 {M}^2_{HL} = {16 w_1^2\over R^4}+{v^2\over (2\pi\alpha')^2} \label{HL-mass}
= 16 \pi {m_L^2 \over \lambda }
+{\textstyle }(m_H - m_L)^2  \,,
\eeq
where we reintroduced the string tension $T=1/(2\pi\alpha')$ (which
was set to one above) and defined the quark masses $m_{L,H}=
w_{1,2}/(2\pi \alpha')$ as the distances $w_{1,2}$ in units of $T$.
(\ref{HL-mass}) implies that the mass of HL mesons has two different
contributions. The first term proportional to ${m_{L}^2 \over
  \sqrt{\lambda}}$ has the same dependence on the 't~Hooft coupling as
in the single flavour case \cite{Kruczenski:2003be}. The second term
is dominant at large 't~Hooft coupling ($\lambda \rightarrow \infty$),
where the mass of the HL mesons is approximated by the second term,
\beq \label{MHL} M_{HL} \approx {v\over 2\pi\alpha'} = m_H- m_L \,.
\eeq In this strong-coupling regime, the heavy-light meson mass
depends solely on the difference of the two quark masses.  This is
consistent with the result obtained in \cite{Erdmenger:2006bg},
discussed in section~\ref{NickHL} above.

\begin{figure}[htbp]
\begin{center}
\voffset=15cm
 \includegraphics[scale=0.73]{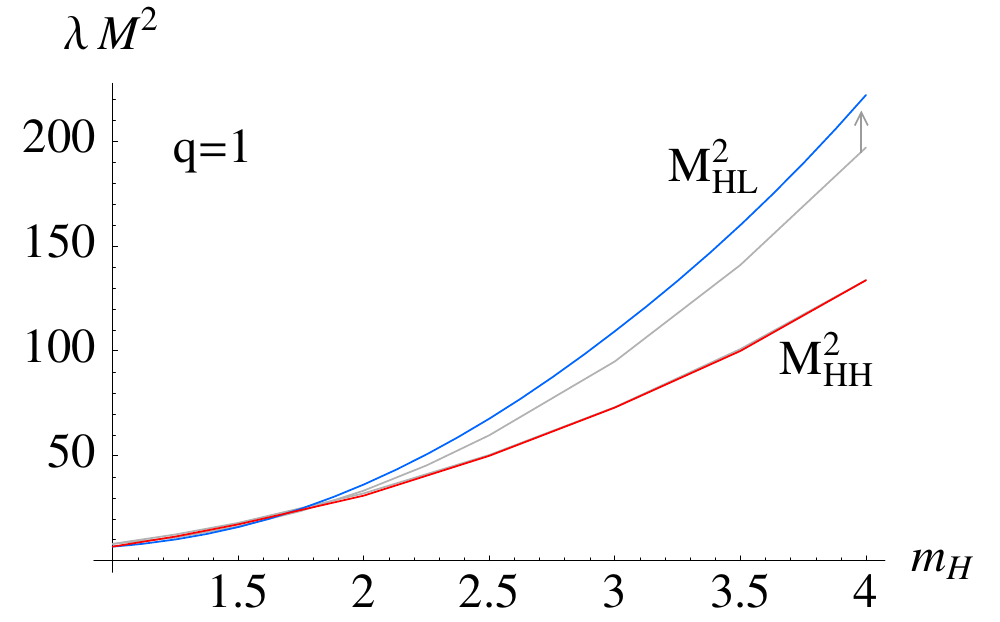}
 \includegraphics[scale=0.73]{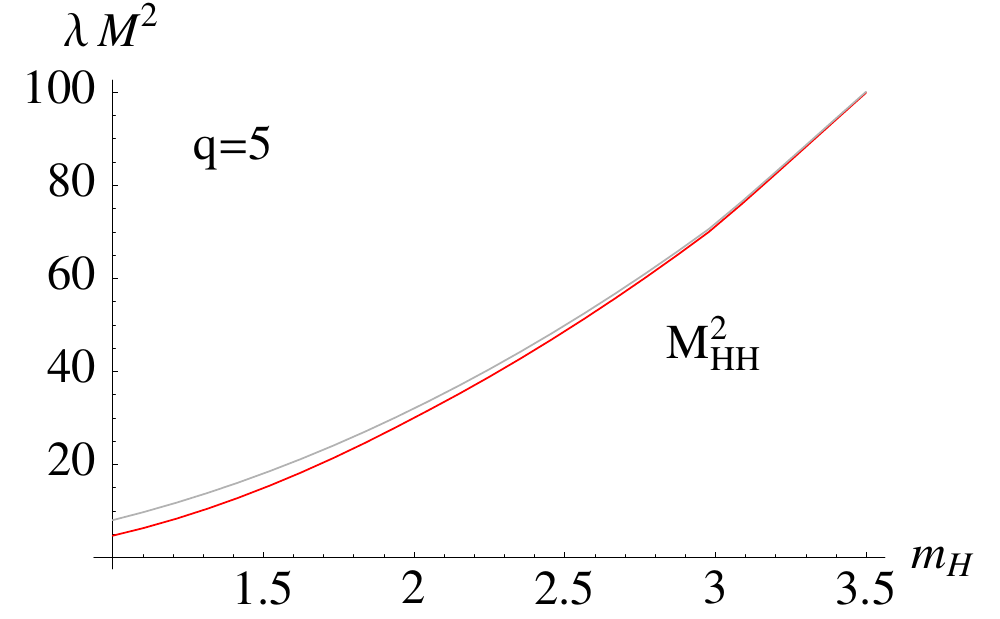}
\caption{Meson masses for non-zero $q$. The red and blue
  curves show $M_{HL}$, $M_{HH}$ for $\lambda=3^4$, $q=1$ (left) and
  $q=5$ (right). The grey curves
  show the corresponding meson masses for $q=0$. The presence of $q$
  increases the HL meson masses. The lambda dependence remains unchanged.
From \cite{Erdmenger:2007vj}. 
\label{m-w2}}
\end{center}
\end{figure}

This $\lambda$ dependence persists if instead of the $AdS_5 \times
S^5$ background, we consider a deformed gravity background as
introduced in section~\ref{flows} above.  As an example, we consider
the ${\rm D3} + {\rm D(-1)}$ gravity background of~\cite{Liu:1999fc}.
This is an example of a dilaton flow background, in which the dilaton
has a non-trivial profile.  The field theory dual to this background
is a confining $\N = 1$ supersymmetric theory, in which a condensate
$q \equiv \pi^2 \langle F^2 \rangle$ is switched on.  The background
in string frame is given by a non-trivial dilaton $\Phi$ and axion
$\chi$ \cite{Liu:1999fc}, \beq ds^2_{10}= e^{\Phi/2} \left(
  \frac{r^2}{R^2}A^2(r)\eta_{\mu\nu}dx^\mu dx^\nu + \frac{R^2}{r^2}
  dr^2+R^2 d\Omega_5^2 \right) \ \, ,
\label{non-SUSY-sol}
\eeq
where
\beq
A=1, \quad e^\Phi=
1+\frac{q}{r^4} \ , \quad \chi=-e^{-\Phi}+\chi_0 \ .
\label{dilaton}
\eeq
In this case, there is only a $\N=1$ supersymmetry remaining for the
background with D7 brane probes embedded. Therefore, vector and scalar
mesons are no longer degenerate, as shown in figure~\ref{m-w2}. For
large heavy quark masses, supersymmetry and thus the meson mass
degeneracy are restored.

It is instructive to compare the $\lambda$ dependence of the meson
spectra with the $\lambda$ dependence of the tension. For a classical
string stretched between the two D7 brane probes, the string tension
is independent of $\lambda$, in agreement with the heavy-light meson
mass result found both in the Polyakov and in the non-abelian DBI
approach.  For heavy-light mesons, this tension contributes to the
meson mass even if the distance $L$ between the quark and anti-quark
in the four-dimensional boundary space is zero, in which case it
contributes $E=m_H - m_L$ to the Wilson line energy.  For the
heavy-heavy and light-light mesons, the string tension scales as
$m_q^2/\sqrt{\lambda}$ for small $L$ \cite{Kruczenski:2003be,
  Ghoroku:2004sp}. At large $L$, when the dual gauge theory is in the
quark confinement phase, there is a long-range linear potential for
all the mesons considered.
\begin{figure}[t]
\begin{center}
\voffset=15cm
 \includegraphics[width=8cm]{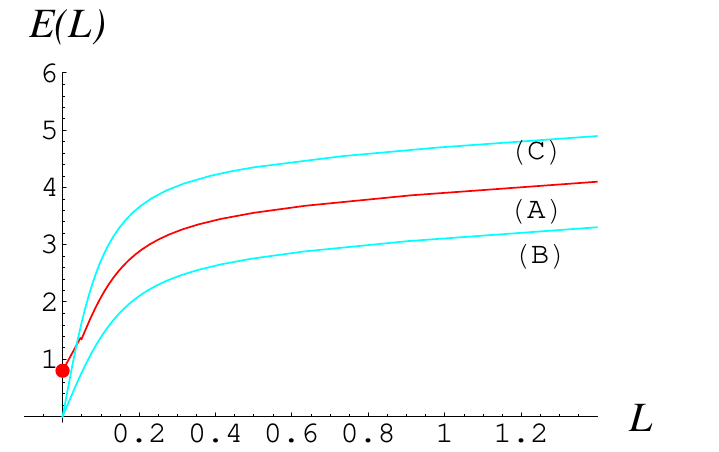}
 \includegraphics[width=7cm]{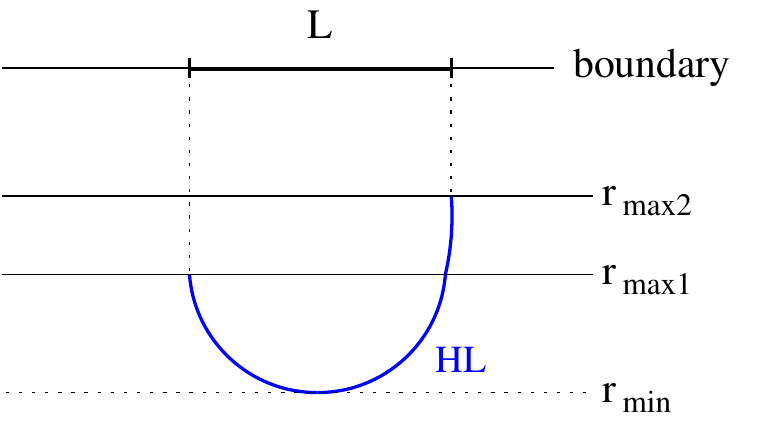}
\caption{a) Numerical plots of the energy $E(L)$ for  HL mesons (A),
 LL mesons (B) and HH mesons (C).  The circle at the endpoint of
  the curve (A) shows a finite string energy $E$ at length $L=0$.
  Here we set $q=5$ and $R=1$, and the brane
  positions are taken at $r_{max1}=10$ and $r_{max2}=15$,
  respectively. b) Schematic plot of the Wilson loop. Figure from 
\cite{Erdmenger:2007vj}. }
\label{Wilson-L}
\end{center}
\end{figure}


\subsection{Mesons with large spin ($J \gg 1$)}\label{largespin}

So far we discussed mesons with spin $0$ and $1$ (and mesinos with
spin $\frac{1}{2}$). The calculation of the spectrum of mesons
with higher four-dimensional spin $J$ would require the
quantization of open strings on the D7-branes, which is difficult.
However, meson operators with large spin have small anomalous
dimensions and quantum corrections are
negligible~\cite{Gubser:2002tv}. Large spin mesons therefore have
a dual description in terms of a classical rotating string. In the
following we show how Regge trajectories in the ${\cal
  N}=2$ theory of the D3/D7 system can be computed by means of a
semi-classical string computation.

\begin{figure}
\vspace{0.5cm}
\begin{center}
\hspace{1.2cm}\includegraphics[scale=0.6]{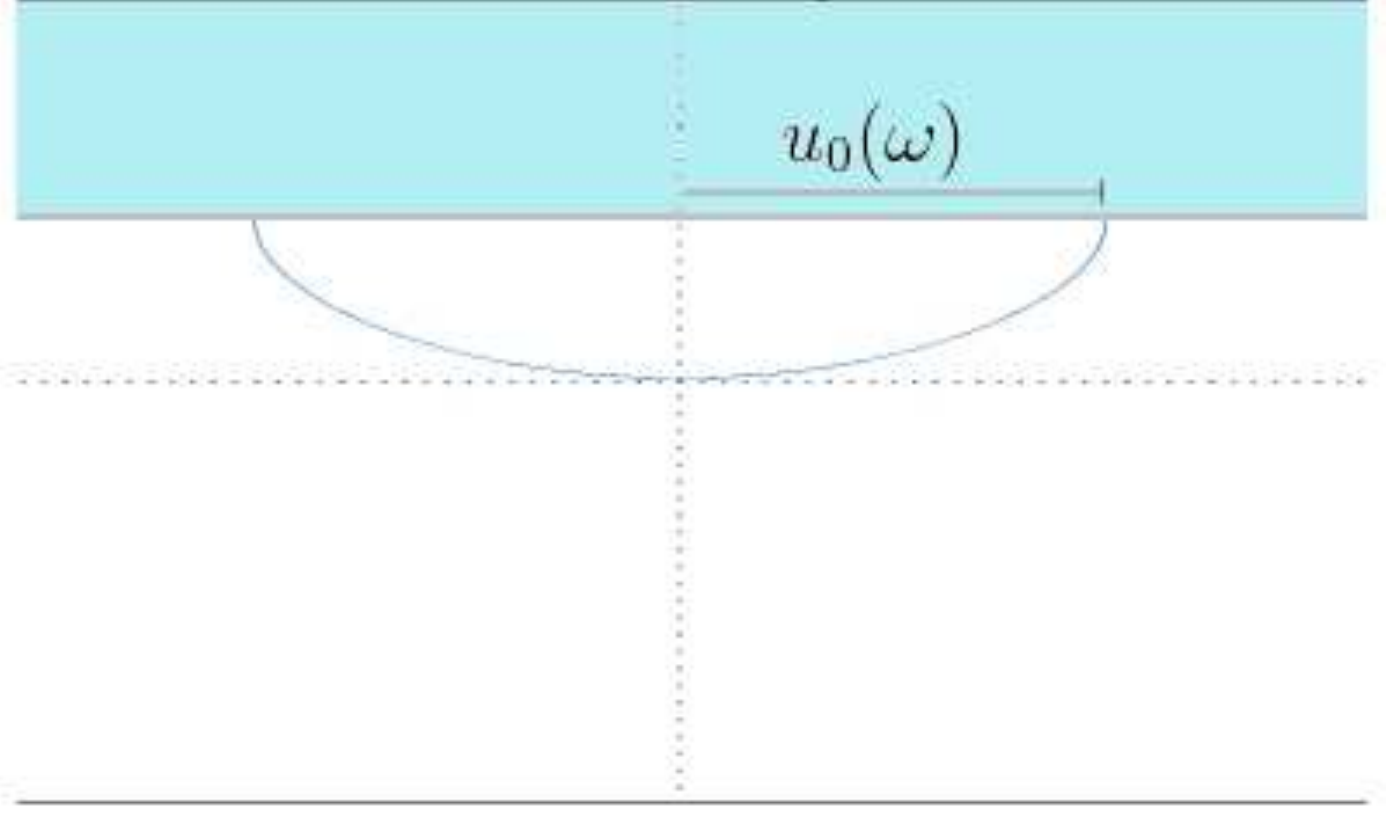}
\end{center}
\caption{Example of a string profile $ z(u)$.} \label{figbc}

\vspace{-7.cm} \hspace{8cm} boundary \vspace{6.5cm}

\vspace{-0.1cm}
\vspace{-1.7cm} \hspace{3.6cm} $ z=\infty$  \vspace{1.2cm}

\vspace{-4.4cm} \hspace{3.8cm} $z_{0}$  \vspace{3.9cm}

\vspace{-5.3cm} \hspace{3.8cm} $ z_R$ \vspace{4.8cm}

\vspace{-6.6cm} \hspace{3.6cm} $ z=0$
\vspace{6.6cm}
\end{figure}

Following \cite{Gubser:2002tv, Kruczenski:2003be}, we consider a
classical open string which rotates in an $AdS_5 \times S^5$
background and ends on a probe D7-brane. This string is dual to a
meson with large spin $J$ in the $\N=2$ theory located on the D3/D7
intersection. We start from the classical Nambu-Goto action in the
form
\begin{align}
S = -T_s \int d\tau d\sigma \sqrt{ (\dot X \cdot X')^2 - \dot X^2
X'^2} \, ,
\end{align}
where dots and primes denote differentiation with respect to $\tau$
and $\sigma$, respectively. The scalar product is taken using the
$AdS_5 \times S^5$ metric. We parameterize the $AdS_5$ metric as
\begin{align}
ds^2 = \frac{R^2}{z^2} \left(-dt^2 + du^2 +u^2 d\varphi^2 +dx_3^2 +
dz^2
\right)\,,
\end{align}
where $u$ and $\varphi$ are the coordinates of the plane of rotation
$x^1-x^2$.  The string has length $2u_0$ and stretches from $-u_0$ to
$+u_0$ along the $u$ direction. The end points of the string are
attached to a probe D7-brane located a distance $z_R$ in the radial
direction.  An example of a spinning string is shown in
figure~\ref{figbc}.

An appropriate ansatz for a string rotating with constant angular
velocity $\omega$ is
\begin{align} t=\tau \,, \quad \varphi =\omega\tau \,, \quad
u=u(\sigma) \,, \quad z=z(\sigma) \,.
\end{align}
With this ansatz the Nambu-Goto Lagrangian takes the form
$(T_s=1)$
\begin{align}
{\cal L} = - \frac{R^2}{z^2}
\sqrt{(1-\omega^2 u^2)(u'{}^2 +z'{}^2 )} \,.
\end{align}
It is convenient to use the rescaled coordinates
\begin{align}
\tilde u = \omega u \,,\quad \tilde z={\omega}z \,.
\end{align}
In these coordinates, the energy and the angular momentum of
the spinning string are given by
\begin{align}
E&=\int d\sigma \left(\omega \frac{\partial {\cal L}}{\partial \omega}
- {\cal L} \right) =  \int d\sigma \frac{\omega}{\cal E}
\frac{R^2}{\tilde z^2} \sqrt{ \tilde u'{}^2
  + \tilde z'^2 } \,,
\label{energy}\\
J&=\int d\sigma \frac{\partial {\cal L}}{\partial \omega} = \int
d\sigma \frac{\tilde u^2}{\cal E} \frac{R^2}{\tilde z^2} \sqrt{\tilde
u'^2 + \tilde z'^2 } \,, \label{spin}
\end{align}
where we defined ${\cal E} = \sqrt{1- \tilde u^2}$.

In the gauge $\tilde u=\sigma$, we find the following equation of
motion for $\tilde z( \tilde u)$:
\begin{align}
  \frac{\tilde z''}{1+\tilde z'^2} - \frac{\tilde u}{\cal E} \tilde z'
+\frac{2}{\tilde z} = 0 \,. \label{eomstring2a}
\end{align}
The solutions of this equation provide the embedding profiles
$\tilde z( \tilde u)$ of the spinning string.

Eq.~(\ref{eomstring2a}) is a nonlinear differential equation of second
order which requires two boundary conditions. These can be obtained
from the usual open string boundary terms
\begin{align}
\left.\frac{\partial {\cal L}}{\partial \tilde u'} \delta \tilde
u \right\vert_{\sigma=0,\pi}
=
\left.\frac{\partial {\cal L}}{\partial \tilde z'} \delta \tilde
z \right\vert_{\sigma=0,\pi}=0\,
\end{align}
for strings ending on a probe D7-brane at $\tilde z=\tilde z_R=const$.
Such strings have a Neumann boundary condition in the $\tilde u$
direction and a Dirichlet boundary condition in the $\tilde z$
direction, {\it i.e.}\ $\delta \tilde u\vert_{\sigma=0,\pi}$ is
arbitrary, whereas $\delta \tilde z \vert_{\sigma=0,\pi}=0$. The
latter condition holds, if we set $\tilde z(\pm \tilde u_0)=\tilde z_R
= const$. The remaining condition $\partial {\cal L}/{\partial \tilde
  u'} \vert_{\sigma=0, \pi}=0$ is satisfied, if \mbox{$\tilde
  u'\vert_{\sigma=0,\pi}=0$}.  Using the gauge {$\tilde z=\sigma$}, we
see that this corresponds to ${\partial \tilde z}/{\partial \tilde
  u}\vert_{\tilde u=\pm \tilde u_0} \rightarrow \infty$ which means
that the string ends orthogonally on the D7 brane at $\tilde z_R$.  In
actual computations of the string profile, the orthogonality condition
is inexpedient. We therefore use the fact that the solutions are
symmetric around $\tilde u=0$, where they have their only maximum, and
impose the boundary condition $\tilde z'(0)=0$.

\begin{figure}
\vspace{0.5cm}
\begin{center}
\hspace{1.2cm}\includegraphics[scale=0.8]{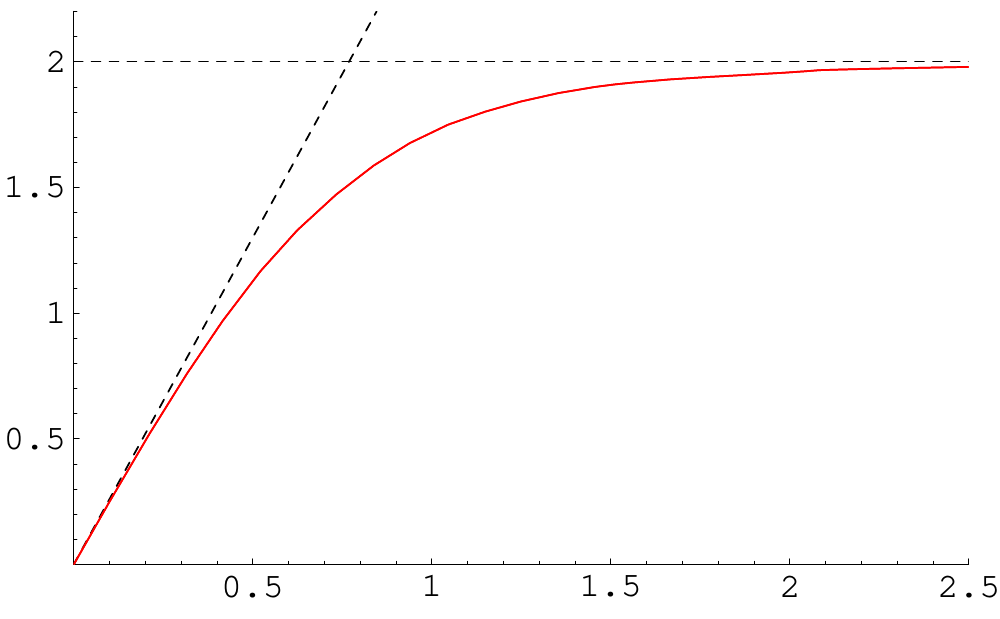}
\end{center}
\caption{Numerical Chew-Frautschi plot. The horizontal line (dashed)
corresponds to the rest mass of the quark-antiquark pair.}
\label{regge}
\vspace{-5cm} \hspace{3.3cm} $E/m_q$ \vspace{4.5cm}

\vspace{-1.8cm} \hspace{8cm} $\sqrt{J}/\lambda^{\frac{1}{4}}$  \vspace{1.9cm}
\end{figure}

\medskip
The Regge trajectories $E(J)$ can now be obtained as follows.
We first solve the equation of motion (\ref{eomstring2a}) for the
string profile $\tilde z(\tilde u)$ by integrating (\ref{eomstring2a})
from $-\tilde u_0$ to $+\tilde u_0$. In the shooting technique, we set
$\tilde z(0)=\tilde z_0 =const$, $\tilde z'(0)=0$ such that $\tilde
z(\pm \tilde R_0)=\tilde z_R$. This yields the string length
$u_0=\tilde u_0/\omega$ as the location at which $\tilde z'(\tilde
u_0) \rightarrow \infty$.  A typical profile is shown in
figure~\ref{figbc}. Then, substituting the profiles $\tilde z(\tilde u)
\equiv 0$ into Eqns.~(\ref{energy}) and (\ref{spin}), we determine the
energy $E(\omega)$ and the angular momentum $J(\omega)$ of the
spinning string for various values of the angular velocity $\omega$.
The Regge trajectory then corresponds to a curve in the $\sqrt{J}-E$
plane parameterized by~$\omega$, as shown in figure~\ref{regge}.

The $\N=2$ theory we consider is not confining and we therefore expect
some deviations from the linear Regge behaviour of QCD.  We observe
that for small spin $J$ the meson mass approximately follows a linear
Regge trajectory, whereas for large $J$ the energy $E(J)$ asymptotes to
the rest mass energy. This can be understood from the behaviour of the
string length as a function of the spin.  At small spin values the
length of the string is much smaller than the scale of the space, and
the string is effectively rotating in flat space leading to a linear
Regge behaviour. At large spin the string is larger than the size of
the space. Here the string rotates very slowly and the energy is that
of particles moving in a Coulomb potential \cite{Kruczenski:2003be}.
The binding energy of the quark-antiquark pair thus vanishes at large
spin values and $E(J)$ asymptotes to the rest mass $E = 2m$.

Analysis of high spin mesons with constituent quarks with different
masses can be found in \cite{Paredes:2004is}. In these cases with more
than one D7 brane at different positions high spin mesons can decay.
If the string in figure~\ref{figbc} dips sufficiently far into the
interior of the space that it meets a second D7 brane then the string
can split into two segments between the two different branes. The rate
for this process has been computed in
\cite{Cotrone:2005fr,Bigazzi:2006jt}.


\subsection{The squark sector from instantons on the D7 probe}

Since the original configuration of \cite{Karch:2002sh} is
supersymmetric, in addition to the fundamental fermion bilinear there
is also a squark (scalar) present in the D3/D7 system. For two
coincident D7 branes, the vev of this squark bilinear has been shown
to be dual to the radius of an $SU(2)$ instanton on the D7 brane probe
\cite{Guralnik:2004ve,Guralnik:2004wq,Guralnik:2005jg}. The vector
meson spectrum for this background has been calculated in
\cite{Erdmenger:2005bj}. For the part of the Higgs branch dual to a
single instanton, the spectrum is computed as function of the
instanton size.  It turns out that the zero size and infinite size
limits are equivalent modulo a singular gauge transformation: In the
dual large $N$ gauge theory, this is an equivalence between the
spectrum of the $SU(N)$ theory and the $SU(N-1)$ theory obtained
by taking the Higgs vev to infinity.  The spectral flow between these
limits leads to a non-trivial re-arrangement of the mass eigenstates
and global charges. In particular, the flow takes vector mesons in the
$(0,0)$ representation of the global $SU(2)_L \times SU(2)_R$
symmetry, which is unbroken at the origin of moduli space, to vector
mesons in the representation $(1,1)$.

For the field theory given by (\ref{fieldtheoryaction}) with
(\ref{superpotential}), on the Higgs branch the vector multiplet
moduli $\phi_3$ vanish while $q^i$ and $\tilde q_i$ have non-zero
expectation values. Here the lower-case letters denote the scalar
components of the corresponding superfields. There are also mixed
Coulomb-Higgs vacua, for which both $q^i,\tilde q_i$ and $\phi_3$ have
non-zero expectation values.

For non-zero $m$ and vanishing $\phi_3$, the fundamental
hypermultiplets are massive and there is no Higgs branch.  However
there is a mixed Coulomb-Higgs branch when $\phi_3$ has an expectation
value such that some components of the hypermultiplets are massless.
An example of a point on a mixed Coulomb-Higgs branch is given by a
diagonal $\phi_3$ for which all but the last $k$ entries on the
diagonal are vanishing, with the last $k$ entries equal to $-m$.  In
this case, the F-flatness equations $\tilde q_i (\phi_3+m) =
(\phi_3+m) q^i = 0$ permit fundamental hypermultiplet expectation
values in which only the last $k$ entries of $q^i$ and $\tilde q_i$
are non-zero.

On the supergravity side, the effective action describing D7-branes in
a curved background is given by (\ref{daction}).  Since we need to
consider at least two flavours (two D7's) in order to have a Higgs
branch, we have to consider the non-Abelian version of
(\ref{daction}).

At leading order in $\alpha'$, field strengths which are self dual
with respect to the flat four dimensional metric $ds^2 = \sum_{m=1}^4
dy^mdy^m$ solve the corresponding equations of motion, due to a
conspiracy between the Chern-Simons and DBI terms. Here the $y^m$
denote the $4,5,6,7$-directions wrapped by the D7 brane probe.
Inserting the explicit AdS background values for the metric and
Ramond-Ramond four-form into the action (\ref{daction}), with
non-trivial field strengths only in the $4,5,6,7$-directions labelled
by $y^m$, gives
\begin{align}\label{theactn}
\begin{split}
  S &= \frac{\mu_7 (2\pi\alpha')^2}{4} \int\, d^4x\,d^4y\, H(r)^{-1} \,
       \left(-\frac{1}{2}\epsilon_{mnrs}F_{mn}F_{rs} +
       F_{mn}F_{mn}\right) = \\
    &= \frac{\mu_7 (2\pi\alpha')^2}{2} \int \, d^4x\,d^4y\,  H(r)^{-1} F_-^2 \, ,
\end{split}
\end{align}
to leading order in $\alpha'$, where $r^2 = y^m y^m + (2\pi\alpha'
m)^2$ and $F^-_{mn} = \frac{1}{2}(F_{mn}-\frac{1}{2}
\epsilon_{mnrs}F_{rs})$. Field strengths $F^-_{mn} = 0$, which are
self-dual with respect to the flat metric $dy^mdy^m$, manifestly solve
the equations of motion.  These solutions correspond to points on the
Higgs branch of the dual $\N=2$ theory.  Strictly speaking, this is a
point on the mixed Coulomb-Higgs branch if $m \ne 0$. In order to
neglect the back-reaction due to dissolved D3-branes, we are
considering a portion of the moduli space for which the instanton
number $k$ is fixed in the large $N$ limit.

In \cite{Arean:2007nh}, it was found that the instanton is also a
solution of the action to all orders in $\alpha'$.

The AdS/CFT dictionary for the Higgs branch is obtained by considering
the symmetries in both field theory and supergravity as usual. On both
sides, for $m \neq 0$ there is a $SO(2,4) \times SU(2)_L \times
SU(2)_R \times U(1)_R \times SU(2)_f$ symmetry, where $SU(2)_f$ stands
for the flavour symmetry present if two coincident D7 branes are
considered.  We focus on that part of the Higgs branch, or mixed
Coulomb-Higgs branch, which is dual to a single instanton centered at
the origin $y^m=0$. The instanton, in ``singular gauge,'' is given by
\begin{align}\label{thinst}
A_\mu = 0,\qquad A_m = \frac{2\Lambda^2 \bar\sigma_{nm}y_n}{y^2(y^2 +
\Lambda^2)} \,,
\end{align}
where $\Lambda$ is the instanton size, and
\begin{align}
  \bar\sigma_{mn} &\equiv \frac{1}{4}(\bar\sigma_m \sigma_n
  - \bar\sigma_n \sigma_m) \, , &
  \sigma_m &\equiv (i\vec\tau, 1_{2\times 2})\, , \nonumber \\
  \sigma_{mn}  & \equiv \frac{1}{4}(\sigma_m \bar\sigma_n
  - \sigma_n \bar\sigma_m)\, , &
  \bar\sigma_m & \equiv \sigma_m^{\dagger} = (-i\vec\tau, 1_{2\times2})\,.
\end{align}
with $\vec\tau$ being the three Pauli-matrices. We choose singular
gauge, as opposed to the regular gauge in which $A_n = 2
\sigma_{mn}y^m/(y^2 + \Lambda^2)$, because of the improved asymptotic
behaviour at large $y$.  The instanton \eqref{thinst} breaks the
symmetries to
\begin{align}G = SO(1,3) \times SU(2)_L \times \diag (SU(2)_R
\times SU(2)_f)\,,
\end{align}
and corresponds to a point on the Higgs branch
\begin{gather}
q_{i \alpha }  =  \, v \, \varepsilon_{i \alpha
} \, ,  \qquad v = \frac{\Lambda}{2\pi \alpha'} \, ,
\end{gather}
where $q_{i \alpha  }$ are scalar components of the fundamental
hypermultiplets, labelled by a $SU(2)_f$ index $i=1,2$, and a
$SU(2)_R$ index $\alpha =1,2$.  All the broken symmetries are
restored in the ultraviolet (large $r$), where the theory becomes
conformal.

\begin{figure}
\begin{center}
  \includegraphics[width=0.65\linewidth]{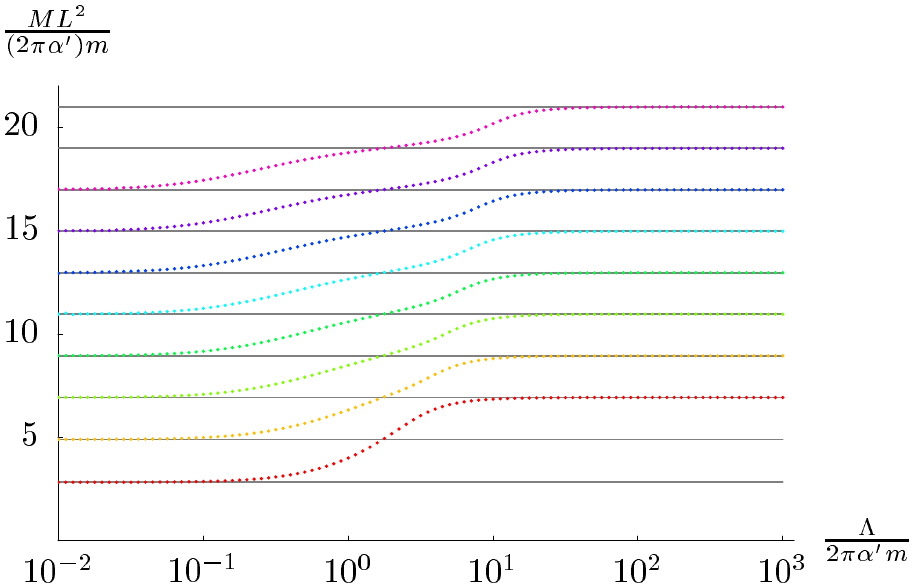}
\caption[Meson Masses]{Meson masses as function of the Higgs vev, from 
\cite{Erdmenger:2005bj}.
 Each dotted line represents a regular solution of the equation of
 motion, corresponding to a vector multiplet of mesons.  The
 vertical axis
 is $\sqrt{\lambda}M/m$ where $M$ is the meson mass, $\lambda$
 the 't~Hooft coupling and $m$ the quark mass. The horizontal axis
 is $v/m$ where $v= \Lambda/2\pi \alpha' $ is the Higgs VEV.
 In the limits of zero and infinite instanton size (Higgs vev), one
        recovers the spectrum (gray horizontal lines)
        obtained analytically in the absence of an instanton background by
        \protect\cite{Kruczenski:2003be}.
        } \label{fig:instanton}
\end{center}
\end{figure}
The simplest non-Abelian ansatz for fluctuations ${\cal
A}_\mu$ about the instanton background is given by
\begin{gather} \label{ansatz}
  {\cal A}_\mu{}^{(a)} = \xi_\mu(k) f(y) e^{ik_\mu x_\mu} \tau^a \, ,
  \qquad y^2 \equiv y^m y^m\,  ,
\end{gather}
which is a singlet under $SU(2)_L$ and a triplet under
$\diag(SU(2)_R\times SU(2)_f)$. $\tau^a$ are the three Pauli matrices.
With this ansatz, the vector meson masses are obtained in direct
analogy to the method presented in section
\ref{scalarfieldfluctuations} above.  The result for the vector meson
masses is shown in figure~\ref{fig:instanton}, where the meson masses
in the presence of the instanton are shown as dotted lines, while the
full horizontal lines correspond to the meson spectrum without an
instanton found in \cite{Kruczenski:2003be}, discussed above in
\ref{scalarfieldfluctuations}. We see that the spectrum is shifted by
two levels when moving from zero to infinite instanton size. This may
be understood as follows.  In singular gauge, the infinite size
instanton is given by
\begin{align}
A_n = 2 \frac{\bar\sigma_{mn}y^m}{y^2} \, .
\end{align}
By virtue of the singular gauge transformation
\begin{align}\label{gtrans}
U = \sigma^m y^m/|y|\,,
\end{align}
$A_n$ may be set to zero, which allows for direct comparison with the
meson spectrum of \cite{Kruczenski:2003be}. However performing the
same singular gauge transformation on the fluctuations (\ref{ansatz}),
we obtain fluctuations in a higher spherical harmonic $\ell=2$ on
$S^3$.  The fact that $\ell=2$ is in exact agreement with the level
shift observed in figure~\ref{fig:instanton}.

The instanton ansatz has also been used for studying the Higgs
potential in gravity duals with less supersymmetry
\cite{Apreda:2005yz, Apreda:2005hj}.  In particular, it has been used
to show with gauge/gravity dual methods that an isospin chemical
potential leads to instabilities in supersymmetric theories, in
agreement with field theory results \cite{Harnik:2003ke}.

\subsection{Summary}

We have introduced quarks into the AdS/CFT Correspondence, in the
quenched limit $N_f \ll N$, by including probe D7 branes. By `quenched' we
mean that there are no quark loops contributing to the gauge propagators.
The gauge theory resulting from adding D7 probes
has $\N=2$ supersymmetry.  The D3 and D7 branes
can be separated in two directions and the separation gives a mass to
the quarks. Fluctuations of the D7 brane in these two directions are
dual to scalar and pseudo-scalar mesons. A gauge field on the D7
world-volume is dual to the vector mesons of the gauge theory.

We have been able to compute a number of meson masses in this strongly
coupled supersymmetric model. If the quark mass is zero, the theory is
conformal and there are no bound states. When the quark mass, $m_q=L/2
\pi \alpha'$, is none zero the masses of mesons made from a single
quark flavour are generically given by
\begin{equation}
M \sim \frac{2 L}{R^2} n \sim \frac{2 m_q\,n}{\sqrt{g_{YM}^2 N}} \,,
\end{equation}
where we have used (\ref{importantrelation}) and $n$ is the radial
excitation number of the meson.

These mesons are very tightly bound - their mass is suppressed
relative to the mass of the quarks they are made of by the 't~Hooft
coupling which is formally infinite.  Note also that these states do
not show Regge behaviour ($M \sim \sqrt{n}$).  We will discuss the
relation of these results to QCD later in section~\ref{secAdSQCD} on
AdS/QCD.

We have also looked at highly spinning strings and strings that are
dual to heavy light mesons. In each case these strings are extended
and a semi-classical approximation can be used. These mesons have
masses of order the quark mass of their contents ({\it i.e.}\ not
suppressed by the 't~Hooft coupling). This separation in masses of
different states is rather unlike QCD.

\newpage

\section{Beyond the probe approximation (backreaction)}
\setcounter{equation}{0}\setcounter{figure}{0}\setcounter{table}{0}

The computations of quark and meson behaviour reviewed so far has been
restricted to the probe brane approximation or, equivalently in the
gauge theory, the quenched approximation. A significant limitation of
the probe approximation is that the number of flavours must be much
smaller than the number of colours, $N_f \ll N$.  Similarly to the
quenched approximation in (lattice) gauge theory, the probe
approximation ignores the effects of the creation and annihilation of
virtual quark-anti-quark pairs on the gauge degrees of freedom.  An
obvious consequence of quenching is that potentially interesting quark
contributions to the theory's $\beta$-function are lost.  An
unquenched computation, in which $N$ and $N_f$ are of the same order,
requires that we go beyond the probe approximation. Virtual quark
loops can be taken into account by including the backreaction of the
flavour branes. In this section we discuss the simplest supersymmetric
example of a supergravity solution which involves the backreaction of
the flavour brane on the supergravity geometry. These computations are
much harder than the probe computations and progress in more QCD-like
theories is so far limited.

\subsection{Fully-localized D3/D7 brane intersection} \label{secbackreaction}

As an example we consider the fully-localized D3/D7 intersection in
flat space which has been constructed in a series of papers
\cite{Aharony:1998xz, Grana:2001xn, Burrington:2004id, Kirsch:2005uy}.
Before discussing the corresponding supergravity solution, we will
review the D3/D7 world-volume field theory at finite $N_f/N$, where,
as in section~\ref{flavour}, $N$ and $N_f$ are the number of D3 and
D7-branes, respectively.

\subsubsection{The $\N=2$ field theory at finite $N_f/N$} \label{secbetafu}

Many aspects of the $\N=2$ field theory located on the D3/D7 brane
intersection have already been discussed in section~\ref{sec21}. The
main difference from the quenched theory in the probe limit is
that the theory has a positive one-loop beta function proportional
to $\beta^\lambda_{\N=2} \sim \lambda^2 \frac{N_f}{N}$
\cite{Hong:2003jm, Kirsch:2005uy}, {\em i.e.}\ it is not conformal.
Since the theory is $\N=2$ supersymmetric, this is the exact
(all-order) perturbative beta function - possible non-perturbative
instanton contributions are ignored here. Note that the beta
function vanishes in the conformal (probe) limit $\frac{N_f}{N}
\rightarrow 0$ and the gauge coupling is constant, in agreement
with the discussion in section~\ref{flavour}. For finite values of
the quotient~$\frac{N_f}{N}$, the perturbative gauge coupling
$\alpha=g_{YM}^2/4\pi$ is given by
\begin{align} \label{ymcoupling}
\alpha(Q^2) = \frac{1}{{\beta_0} \log
 \frac{\Lambda^2_L}{Q^2}} \qquad{\textmd{with}} \qquad \Lambda^2_L =
 \mu^2 e^{{1}/({\alpha(\mu^2)\beta_0 })}\,,
\end{align}
where $Q^2$ is the energy scale, $\mu^2$ a reference scale and
$\beta_0=N_f/4\pi$.  The gauge coupling has logarithmic behaviour and
runs into an ultraviolet Landau pole at the scale $\Lambda_L$.

Another interesting feature of the $\N=2$ theory, not present in
the probe limit, is the {\em chiral (or axial) anomaly}.  In the
chiral limit $m_q \rightarrow 0$, the classical $\N=2$ theory
features a chiral $U(1)_{\cal R}$ symmetry corresponding to
$SO(2)$ rotations in $X^{8,9}$. At the quantum level, this
symmetry is explicitly broken by the chiral anomaly which is
proportional to
\begin{align}
\textstyle \frac{N_f}{N} {\rm Tr\,} F \wedge F \,.
\end{align}
The anomaly-free (unbroken) subgroup of $U(1)_{\cal R}$ is therefore
the discrete group $\ZZ_{2Nf}$. This symmetry rotates the fundamental
spinors as $\psi \rightarrow e^{{-i\pi}/{N_f}} \psi$, $\tilde \psi
\rightarrow e^{{-i\pi}/{N_f}} \tilde \psi$, while the scalar
$X=X^8+iX^9$ of the (adjoint) gauge multiplet transforms as $X
\rightarrow e^{{i2\pi}/{N_f}} X$.

In the next section we will discuss the dual supergravity description
of the perturbative field theory ignoring instanton effects. The
breaking of $U(1)_{\cal R}$ to $\ZZ_{2Nf}$ will not be visible in this
solution. We will however come back to the chiral anomaly and its
realization in the D3/D7 system in section~\ref{sec413}.

\subsubsection{The D3/D7 supergravity solution}

We will see that the running gauge coupling $\alpha(Q^2)$ and the
non-trivial theta angle $\theta_{YM}$ of the $\N=2$ theory can be
recovered from the fully-localized D3/D7 intersection.\footnote{In a
  fully-localized brane solution the branes are located at a fix
  location in the transverse direction.  One should compare this to a
  solution in which branes are ``smeared'' over some direction.} For
simplicity we will work in the case of massless quarks - the D7 branes
are at the origin in the $w_5,w_6$ plane. The D3/D7 supergravity
solution is given by
\begin{align} \label{d3d7solution}
ds_{10}^2 &= h^{-1/2} \, \eta_{\mu\nu} dx^{\mu} dx^\nu + h^{1/2}
\big(d\rho^2+\rho^2d\Omega^2_3 + e^{-\phi} (d\w^2+\w^2 d\theta^2)
\big) \,,
\end{align}
where the (near-core) warp factor $h=h(\rho,\w)$ is\footnote{For
simplicity, we restrict to the near-core region of the D3/D7
intersection (small values of $w$).  The near-core approximation
corresponds to the IR region of the dual field theory. The full warp
factor is known in terms of a uniformly converging series expansion
\cite{Kirsch:2005uy}.}
\begin{align}
h(\rho,\w) = 1+\frac{R^4}{(\rho^2+ e^{-\phi} \w^2)^2} \ ,
\end{align}
with $R^4=4\pi g_s N \alpha'^2$. The axion $\chi$ and dilaton
$\phi$ are given by
\begin{align}
 \chi(\theta)&=\frac{N_f}{2\pi} \theta \,,\qquad e^{-\phi(\w)}=
 \beta_0 \log \frac{\w_\Lambda^2}{\w^2} \,, \label{D7dilaton}
\end{align}
where we choose the integration constant $\w_\Lambda$ to be
\begin{align}
 \w^2_\Lambda&=\w_0^2\, e^{1/(g_s \beta_0)} \,,\qquad
 \beta_0=\frac{N_f}{4\pi}\,.
\end{align}
Here $x^\mu$ parameterizes the directions along the D3-branes (0123),
while $\rho$ is the radial direction in the four-plane transverse to
the D3-branes, but along the D7-branes (4567) and $w$, $\theta$ are
the radial and angular direction in the two-plane transverse to the
D7-branes. The background also contains the RR four-form potential of
the D3-brane solution. It has been shown in \cite{Grana:2001xn,
  Burrington:2004id} that this background preserves eight
supercharges. For $N_f=0$, the background reduces to that of a stack
of $N$ D3-branes (note here that $e^{\phi}=g_s$ and rescale $\w^2
\rightarrow \w^2 g_s$).  For $N=0$, the background is that of $N_f$
D7-branes.

\begin{figure}[t]
\begin{center}
  \includegraphics[height=5cm]{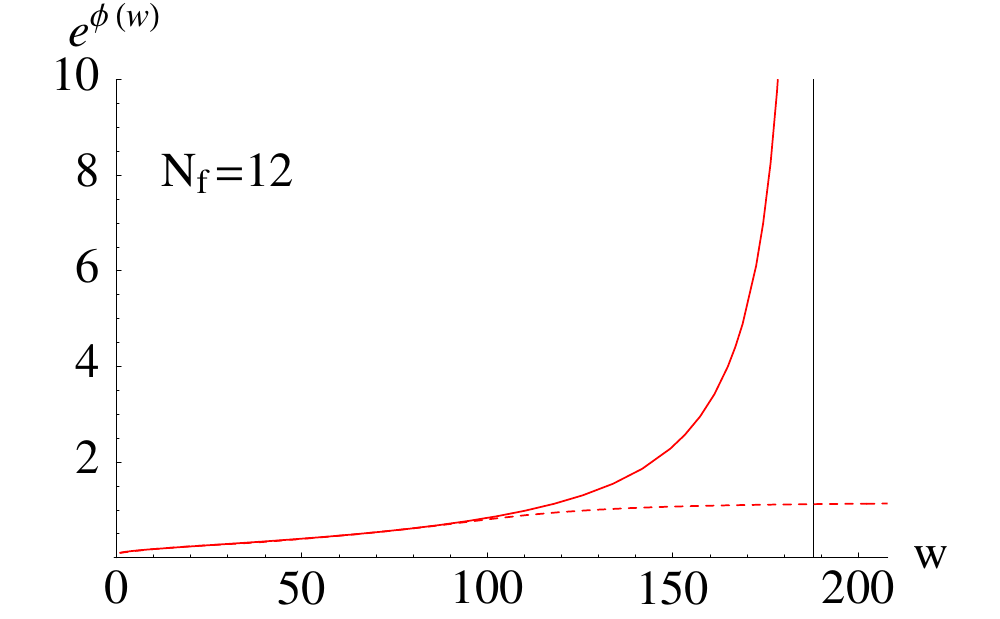}
  \includegraphics[height=5cm]{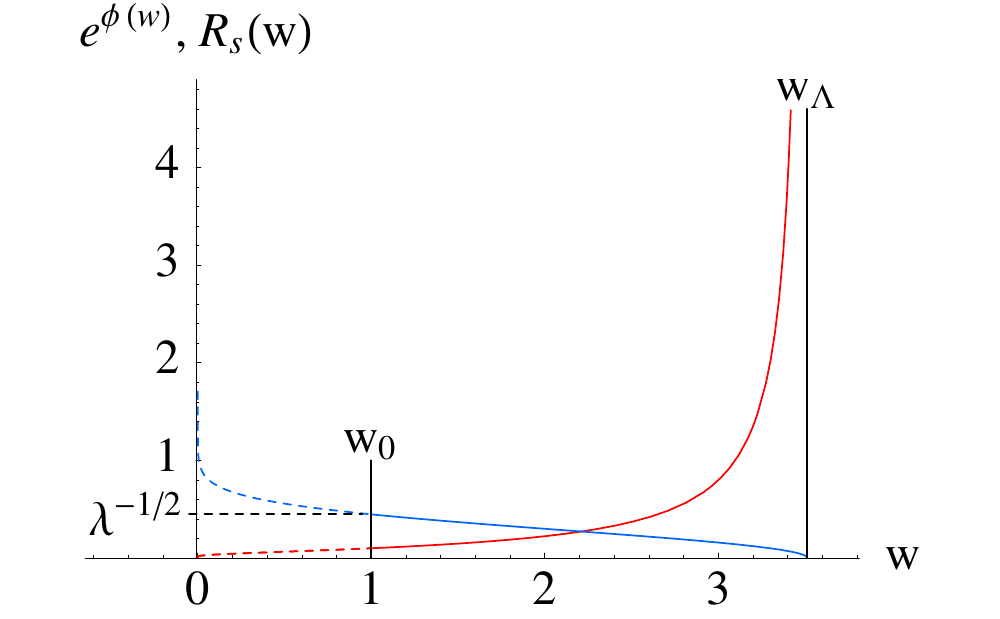} \caption{a) The
  logarithmic (red curve) and the full regular (dashed curve) dilaton
  profile (for $N_f=12$, $g_s=0.1$, $\theta=\pi/N_f$). The regular
  profile approaches $e^\phi=2/\sqrt{3}$ ({\em i.e.}\ $\tau=j^{-1}(0)$
  \cite{Greene:1989ya}) at $w\rightarrow \infty$. b)~Dilaton (red
  curve) and curvature ${\cal R}_s(w)$ (blue curve) (for $N_f=N=50$,
  $g_s=0.1$).}  \label{figbackreaction}
\end{center}
\end{figure}

Let us consider the case $N=0$ in more detail. Then, the warp factor
becomes one, $h=1$, and we recover the D7-brane solution in the
so-called {\em weak-coupling} approximation. Here the complete
D7-brane dilaton profile \cite{Greene:1989ya} (shown as a dashed curve
in figure~\ref{figbackreaction}a) is approximated by the logarithmic
profile~(\ref{D7dilaton}) (solid curve in figure~\ref{figbackreaction}a)
at small radii $w$.  Extrapolation of the logarithmic behaviour to
larger values of $w$ shows an apparent dilaton divergence at the
scale~$\w_\Lambda$.

This perfectly reflects the perturbative aspects of the dual $\N=2$
field theory, as can be seen as follows. Comparing (\ref{D7dilaton})
with (\ref{ymcoupling}), shows an intriguing similarity between the
weak-coupling dilaton profile and the perturbative gauge coupling
\cite{Burrington:2004id, Kirsch:2005uy}. In fact, let us assume that
the $\w$ direction in the background corresponds to the energy scale
$Q$ in the $\N=2$ theory, $Q=\w/(2\pi\alpha')$, and set
$\mu=\w_0/(2\pi\alpha')$. If we then identify
\begin{align}
\alpha(Q^2) = g^2_{YM}(Q^2)/4\pi = e^{\phi(w)} \,,\qquad
\alpha(\mu^2)=g^2_{YM}(\mu^2)/4\pi=g_s \,, \label{alphaym}
\end{align}
we observe that there is a direct correspondence between the running
of the gauge coupling~(\ref{ymcoupling}) and the logarithmic dilaton
profile (\ref{D7dilaton}). Moreover, the string coupling
$g_s=e^{\phi(w_0)}$ is fixed at $w_0$, not at infinity. This
corresponds to fixing the gauge coupling $g^2_{YM}(\mu^2)$ at some
reference scale~$\mu$. The above identification implies in particular
that the UV Landau pole at $\Lambda_L$ $(g_{YM}\to \infty)$ is mapped
to the dilaton divergence at $\w_\Lambda$, $\Lambda_L=\w_\Lambda/(2\pi
\alpha'{}^2)$. Of course, here we map one pathology to another: The
perturbative field theory becomes strongly coupled at the Landau pole
$\Lambda_L$, while the supergravity solution breaks down at some
distance $\w_\Lambda$. In principle, both sides must be cured at these
scales. This issue will be addressed in the next section.

We also find that the Yang-Mills theta angle $\theta_{YM}$ is
reflected by a nontrivial axion profile $\chi$ in the supergravity
background, $\chi =\frac{\theta_{YM}}{2\pi}=\frac{\theta N_f}{2\pi}$.

\medskip In order to obtain a supergravity theory dual to the $\N=2$
field theory, we consider the background (\ref{d3d7solution}) at large
't~Hooft coupling $\lambda$ and fixed $N_f/N$. In this limit ($\lambda
\gg 1$), we may drop the ``1'' in the warp factor $h(\rho,\w)$ and the
D3-branes are replaced by their near-horizon geometry. Note that the
D7-branes do not disappear; open strings ending on the D7-branes are
kept in this limit as signaled by the curvature singularity at the
location $w=0$ of the D7-branes.  (This singularity might be resolved
within classical string theory by $\alpha'$~corrections.)

The background is a good supergravity solution in the regime of small
effective string coupling and small curvature, $e^\phi \ll 1$ and
$\alpha' {\cal R}_s \ll 1$. The first requirement is satisfied for
radii $\w \ll \w_\Lambda$ corresponding to energies much below the
Landau pole. The curvature measured in string units is $\alpha' {\cal
  R}_s \sim \sqrt{e^{-\phi}/N}$ \cite{Grana:2001xn} and diverges for
$w \rightarrow 0$. However, at the (infrared) cut-off $w_0 (\propto
\mu)$ the curvature becomes
\begin{align}
 \alpha' {\cal R}_s \sim \sqrt{\frac{1}{4\pi}\frac{N_f}{N} \log
 \frac{w_0^2}{w^2} + \frac{1}{g_sN}} \stackrel{w=w_0}{=}
 \frac{1}{\sqrt{\lambda}}   \,,
\end{align}
which is small in the large 't~Hooft coupling limit.  The background
is thus a valid supergravity solution in the regime $\w_0 \leq w \ll
w_\Lambda$ (corresponding to energies $\mu \leq Q \ll \Lambda_L$ in
the field theory). figure~\ref{figbackreaction}b shows a plot of the
scalar curvature and the dilaton in this regime.

\medskip
There are some subtleties to this construction which have
been addressed at length in \cite{Kirsch:2005uy}. First, the dilaton
diverges at $w_\Lambda$ and one might worry about the absence of a
true boundary at $\w \rightarrow \infty$.  Note however that massless
open string states (related to the field theory) precisely map into
massless closed string states (generating supergravity)
\cite{Kirsch:2005uy}, {\em i.e.}\ there is no mixing with massive
states. The gauge/gravity duality therefore works even without a true
boundary.  Second, since the D7-branes are codimension-two branes,
there are uncancelled tadpoles in the string background.  Tadpole
divergences usually correspond to gauge anomalies and indicate an
inconsistency in the theory.  However, as it was found first in
\cite{Leigh:1998hj}, logarithmic tadpoles do not correspond to gauge
anomalies, but reflect the fact that the dual gauge theory is not
conformally invariant. In fact, such tadpoles provide the correct
one-loop running of the gauge coupling. Third, the full D7-brane
geometry contains an asymptotic deficit angle of $2\pi N_f/12$ which
restricts the number of flavours to $N_f \leq 12$ ($N_f=24$)
\cite{Greene:1989ya}. However, the background (\ref{d3d7solution}) is
a valid supergravity solution for any $N_f$, which corresponds to the
fact that there is no restriction on $N_f$ in the perturbative field
theory. So, as long as we stay on the supergravity level and we do not
want to extend (\ref{d3d7solution}) to a full string theory solution,
we may use it for any~$N_f$.

In summary, one finds that at large 't~Hooft coupling $\lambda$
and fixed $N_f/N$ the D3/D7 solution (\ref{d3d7solution})
perfectly reflects the perturbative aspects of the dual $\N=2$
field theory at low energies, such as the ultra-violet Landau-pole
and the non-trivial theta-angle.

\subsubsection{Non-perturbative completion and $U(1)_{\cal R}$ chiral anomaly}
\label{sec413}

The supergravity background discussed in the previous section
describes only the perturbative regime of the $\N=2$ theory. In the
strong-coupling regime the gauge coupling is corrected by instanton
contributions which may cure the ultra-violet Landau pole of the
perturbative field theory. We now investigate to what extent a D3/D7
intersection including the full axion-dilaton of the D7-branes could
represent an ultra-violet completion of the perturbative field theory,
at least for $N_f \leq 12$ ($N_f=24$).  We will refer to this
intersection as the {\em complete} D3/D7 intersection as opposed to
its logarithmic approximation discussed above.

\begin{figure}[t]
\begin{center}
  \includegraphics[width=8.5cm]{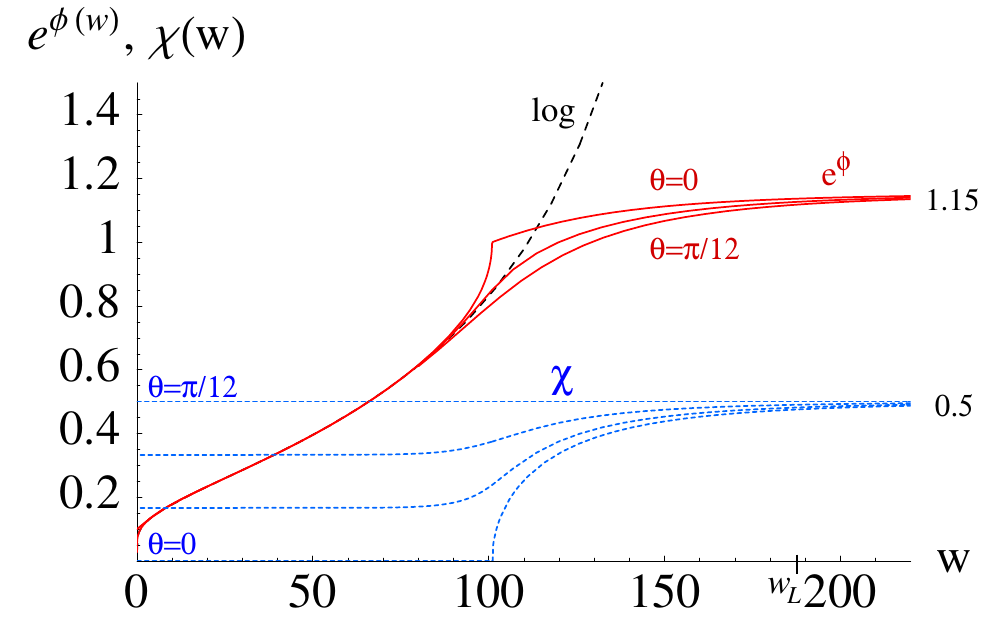}
  \caption{Axion (blue dotted) and dilaton (red solid) for different
    values of $\theta$ ($N_f=12$).}
\label{dilatonfig}
\end{center}
\end{figure}

\begin{figure}[ht]
\begin{center}
  \includegraphics[width=11cm]{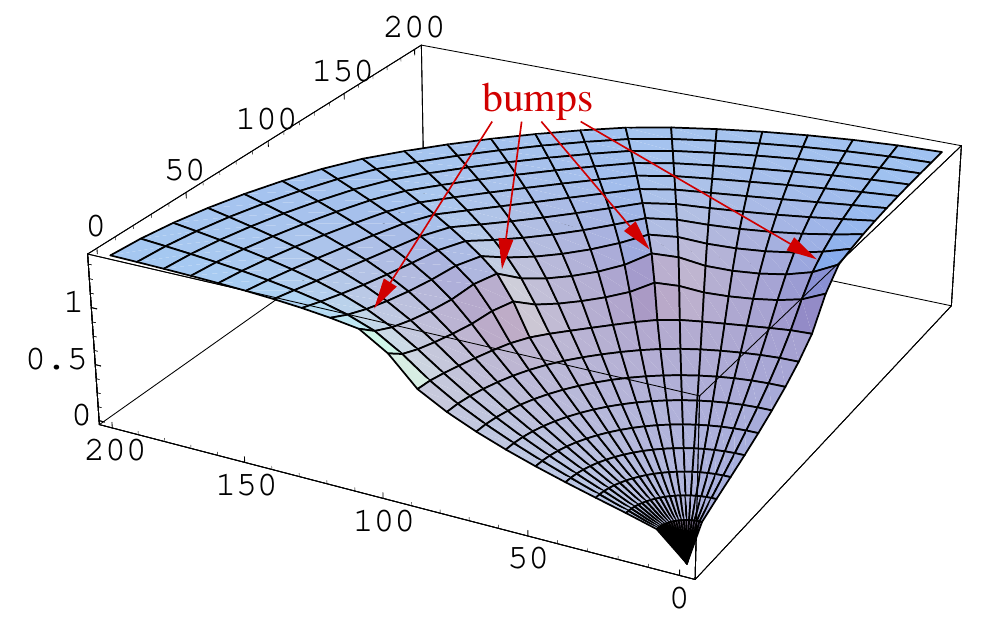} \caption{``Bowl with bumps'':
  D7 brane profile $\exp({\phi(w,\theta)})$ (for $N_f=12$
  flavours). Only the first quadrant is shown.}\label{bumps}
\end{center}
\end{figure}

The full axion-dilaton of the D7-branes is a solution to
$j(\tau)=(w_\Lambda/w)e^{-i\theta}$, where $\tau=\chi+ie^{-\phi}$ and
$j(\tau)$ is the modular $j$-function. Some solutions are plotted in
figure~\ref{dilatonfig} which shows the profiles $e^{\phi(w)}$ and
$\chi(w)$ for different values of the angular direction $\theta$. At
small radii $w$ the dilaton has logarithmic behaviour and is
independent of $\theta$. In the strong-coupling region at the scale
$w_\Lambda$ the full D7 brane dilaton deviates from its logarithmic
approximation. The profile also becomes dependent on $\theta$ leading
to $N_f$ equally distributed ``bumps'' at the angles $2\pi k/N_f$
($k=1,...,N_f$), as shown in figure~\ref{bumps}. Asymptotically, the
dilaton approaches the constant value $1/\sin(2\pi/3) \approx 1.15$.

It is interesting to observe that the profile is invariant under
rotations of $2\pi/N_f$. This is a direct consequence of the chiral
anomaly which breaks the chiral $U(1)_{\cal R}$ symmetry down to
$\ZZ_{2N_f}$.  (Recall that the complex transverse direction
parameterized by $X=w e^{i\theta}$ rotates under $\ZZ_{2N_f}$ as $X
\rightarrow e^{{i2\pi}/{N_f}} X$.)  The chiral anomaly is a
non-perturbative effect and can therefore not be seen in the
logarithmic approximation.  The full axion-dilaton profile of the
D7-branes however nicely demonstrates this anomaly. This is an
important consistency check of the complete D3/D7 intersection as a UV
completion of the perturbative $\N=2$ theory. -- Note that so far we
only discussed the axion-dilaton, but not the metric of the complete
D3/D7 system. It is believed that in this case the warp factor
$h(\rho,\w)$ appearing in the metric ansatz (\ref{d3d7solution}) can
only be computed numerically, which we will not do here.

\subsubsection{Meson computation}

A full computation of the mesonic spectrum of a back-reacted solution
is hard since one would need to look at fluctuations of the full
geometry. Instead we will return to probe methods. We will embed a
probe D7 brane into the back reacted geometry and study the scalar
meson spectrum of the probe. This will at least provide some insight
into the effects of flavour on the meson spectrum.

Firstly we must check that it is possible to embed a probe D7 brane in
the back-reacted D3/D7 geometry (\ref{d3d7solution}). The D7 action
takes the usual Dirac-Born-Infeld
form (in Einstein frame),  with a coupling to the
field $C_8$ which is dual to the dilaton. We find
\begin{equation}
S \sim \int d \rho e^\phi \rho^3 \left( \sqrt{1 + e^{-\phi}
(\partial_\rho w)^2} - 1 \right) \,.
\end{equation}
Note that the factors of $h$ cancel. The cancellation between the
leading DBI term and the Wess-Zumino term is the usual cancellation
that occurs for a D7 probe in the geometry of a stack of many D7
branes. This action is clearly minimized when $\partial_\rho w = 0$
when the action is zero (one can also explicitly solve the equation of
motion). The D7 probe therefore lies flat in the geometry just as in
the usual probe computation - the value of $w$ the probe is placed at
determines the quark mass.

If the quarks are all massless then the $\N=2$ gauge theory is
conformal. We will therefore consider the case of $N_f$ back-reacted
(unquenched) quark flavours and a single quenched massive flavour. The
scalar meson masses are given by fluctuations of the probe in the $w$
directions about its position $d$. The Lagrangian for such linearized
fluctuations is given by
\begin{equation}
{\cal L} = \frac{1}{2} \rho^3 e^{\phi(d)}\left( \frac{R^4}{(\rho^2
+ e^{-\phi(d)} d^2)^2 } ( \partial_x \phi)^2 + (\partial_\rho
\phi)^2 \right) \,.
\end{equation}
Up to an overall constant this is just the Lagrangian from the probe
computation (\ref{mex}) but with
\begin{equation}
  d ~\rightarrow~ {d}\left({\beta_0 \ln \frac{w_\Lambda^2}{d^2}}
  \right)^{\frac{1}{2}} \,.
\end{equation}
The meson spectrum is therefore, replacing $d = 2 \pi \alpha' m_q$
with $m_q$ the quark mass

\begin{equation}
M^2 = \frac{8 \pi}{g_s N} ~ \left({\beta_0 \ln
\frac{\Lambda^2}{m_q^2}}\right)~ m_q^2~ (n+1)(n+2) \,.
\end{equation}
By Eq.~(\ref{alphaym}) the meson mass $M$ is proportional to
\begin{align}
\frac{g_{YM}(\mu^2)}{g_{YM}(m_q^2)} < 1 \, .
\end{align}
The effect of unquenching the quarks is just to replace the gauge
coupling $g^2_{YM}=4\pi g_s$ with the appropriate renormalized value
at the scale of the quark mass.

\newpage

\section{More supersymmetric mesons}
\setcounter{equation}{0}\setcounter{figure}{0}\setcounter{table}{0}

There has been considerable work on including quark fields into
gravity duals of gauge theories with less (but none zero)
supersymmetry. Typically these geometries are more complicated than
AdS so even probe computations are hard work. We will be brief in our
review of this work having spent considerable time on the simplest
$\N=2$ theory and wishing to proceed to models with dynamical chiral
symmetry breaking in the spirit of QCD.  This section is intended as a
guide to references for those who wish to pursue them.

\subsection{Klebanov's duals}

Klebanov, with a variety of collaborators, has studied models in which
D3 branes and fractional D3 are placed on a conifold singularity
\cite{Klebanov:1998hh,Klebanov:1999rd,Klebanov:2000nc,
  Klebanov:1999tb,Klebanov:2000hb}.  A variety of $\N=1$ gauge
theories with a product gauge group structure of the form $SU(N)
\times SU(M)$ can be realized. The adjoint fields of the naive
$SU(N+M)$ group on the D3 branes divide into adjoints of the
sub-groups plus bi-fundamental fields. These theories display a
chain of Seiberg dualities before developing a mass gap in the IR
(corresponding to a deformation of the conifold).

D7 brane probes were first introduced into this theory in a
supersymmetry preserving fashion adding extra massive quark
supermultiplets in \cite{Sakai:2003wu}.  The meson spectrum was
computed and displays a mass gap for the vector, scalar and
pseudo-scalar mesons (see also
\cite{Kuperstein:2004hy,Levi:2005hh}).  A more complete set of
probe embeddings were found in \cite{Arean:2004mm}.

A perturbative analysis of the backreaction due to the introduction of
D7 branes in \cite{Ouyang:2003df} found evidence of Seiberg duality
\cite{Intriligator:1995au} in these theories.

Impressively, fully back-reacted solutions for D7 branes in the
$SU(N)\times SU(N)$ theory were constructed in \cite{Benini:2006hh}.
Many of the symmetry properties of the theory were reproduced and the
running gauge coupling correctly matches gauge theory expectations.
These methods were extended to the general $SU(N) \times SU(M)$ theory
in \cite{Benini:2007gx}.  Again symmetry properties and the running
gauge coupling of the field theory were correctly reproduced on the
gravity side. Seiberg's duality is also manifest in the solutions.
These techniques were also used to find embeddings in the geometry
$AdS_5\times L^{abc}$ in \cite{Canoura:2006es}.

\subsection{B fields in the background:
Polchinski-Strassler dual} \label{secps}

Interesting features arise if an antisymmetric two-form (which enters
the supergravity theory as described in section~\ref{strings}) is
turned on. The two from enters the probe DBI action as in
Eq.~(\ref{daction}). A prime example is the Polchinski-Strassler
background \cite{Polchinski:2000uf}, in which a $B$ field, together
with a non-trivial $C_2$, is switched on in the six directions
perpendicular to the boundary of $AdS_5$. The supersymmetry
representation of $B$ is chosen such that the $B$ field is dual to
mass terms for the adjoint chiral multiplets in the dual $\N=4$ gauge
theory. In the supergravity picture, the D3 branes are polarized into
D5 branes by virtue of the Myers effect \cite{Myers:1999ps}. The
supergravity solution is only known as a perturbative expansion as one
moves into the interior of the space, towards the IR of the field
theory.

The embedding of D7 branes into this background for the so-called $\N
=2^*$ and $\N=1^*$ theories, with a massive hypermultiplet and an
equal mass for the three chiral multiplets, respectively, has been
considered in \cite{Apreda:2006bu} and \cite{Sieg:2007by}.  This
requires the explicit construction of the deformed gravity background
to second order in the masses.  For the $\N =2^*$ case, it was shown
in \cite{Apreda:2006bu} that the meson mass obtained from the D7 probe
brane fluctuations receives a contribution from the adjoint scalar
quark mass, such that there is a mass gap.  Recently
\cite{Penati:2007vj}, the embedding of a D7 brane probe into the
Lunin-Maldacena background \cite{Lunin:2005jy} was considered, which
is dual to a $\N=1$ supersymmetric marginal deformation of $\N=4$
super Yang-Mills theory. Here, a Zeeman-like spitting of the mass
spectrum is observed.

\subsection{Maldacena-Nu$\rm \tilde{n}$ez dual}

The Maldacena-Nu$\rm \tilde{n}$ez background \cite{Maldacena:2000yy} is dual to an
$\N=1$ theory on the world-volume of a D5 brane wrapped on a 2-sphere
and therefore describes a relative of $\N=1$ Yang Mills theory with
additional Kaluza Klein modes. The dual encodes the condensation of
gauginos and the $N$ discrete vacua of the theory.

Probe D5 branes have been used to introduce matter fields into this
theory with $\N=1$ supersymmetry preserved in \cite{Wang:2003yc}
\cite{Nunez:2003cf}. The scalar and vector meson masses were computed
numerically in \cite{Nunez:2003cf} and are both compatible with the
formula

\begin{equation}
M_{n,l} = \sqrt{m^2(r_*,\lambda) n^2 + l^2}, \hspace{1cm} m(r_*,\lambda)
= \frac{\pi}{2 \Lambda} + r_*^2 \left(
\frac{0.23}{\Lambda} + \frac{0.53}{\Lambda^3}
\right) \,,
\end{equation}
where $n$ is the radial excitation number, $l$ the R-charge, $r_*$ is
a measure of the quark mass and $\Lambda$ is the strong coupling scale
of the underlying Yang Mills theory.

A solution for the backreacted version of this theory has been found
in \cite{Casero:2006pt}. The geometry encodes many of the properties
of the theory including confinement and a running coupling.

Quarks and mesons are also investigated in an alternative $\N=2$
wrapped D5 brane theory in \cite{Paredes:2006wb}.

\subsection{Defect theories} \label{sec:defects}

Probe techniques similar to those we have described for mesons have
been used in gravity duals to include matter fields in the fundamental
representation in gauge theories
on defects, i.e.~on subspaces in $2+1$ or $1+1$ dimensions.
The first such examples, in the $\N=4$ theory, were
explored in
\cite{Karch:2000gx,DeWolfe:2001pq,Erdmenger:2002ex,Skenderis:2002vf,Constable:2002xt}.
Closed form expressions for the
masses of lower dimension supersymmetric mesons in D3-D5 and D3-D3
systems were found in \cite{Arean:2006pk, Arean:2006vg}. Other
examples include \cite{Yamaguchi:2003ay, Basu:2006eb, Canoura:2005fc,
  Myers:2006qr}.  A back-reacted D2-D6 system, in which both the adjoint and
the fundamental degrees of freedom live in $2+1$ dimensions, can be found in
\cite{Cherkis:2002ir},  and further analysis is in
\cite{Erdmenger:2004dk}.

\subsection{Non-commutativity}

Quarks have been introduced into non-commutative gauge theories using
probe techniques in \cite{Arean:2005ar}.

\newpage
\section{Chiral symmetry breaking } \label{sectioncsb}
\setcounter{equation}{0}\setcounter{figure}{0}\setcounter{table}{0}

In the sections above we have introduced quarks into the basic AdS/CFT
Correspondence. These supersymmetric theories display bound mesonic
spectra but unlike in QCD become conformal theories in the limit where
the quark masses vanish. In QCD there is a dynamical mass generation
mechanism (chiral symmetry breaking) that ensures the bound states
remain massive as the quark masses fall to zero. In addition there are
a special set of bound states, the pions, that are anomalously light
because they are the (pseudo-)Goldstone bosons of the symmetry
breaking. A gravity dual must capture these crucial pieces of dynamics
if it is to describe QCD successfully - we describe a number of string
constructions that achieve these goals in this section. In each case
there is a very appealing geometric realization of the symmetry
breaking providing a pleasing intuitive picture.

\subsection{Chiral symmetry breaking in field theory}

We begin with a brief summary of chiral symmetry breaking in gauge theory, in
order to compare with the gravity description below.
Consider the Lagrangian of massless QCD,
\begin{gather} \label{m0qcd}
{\cal L}_{QCD} |_{m=0} \, = \, - \frac{1}{4} F^a_{\mu \nu} F^{a \mu \nu}
\, + \, \bar \psi_L \slash \!\!\!\! {D}   \psi_L \, + \, \bar \psi_R \slash
\!\!\!\! D \psi_R \, .
\end{gather}
$\psi_L$ and $\psi_R$ are the chiral projections of the Dirac spinor $\psi$.
In the massless case, the left-handed and right-handed fields have separate
invariances under flavour symmetry. For the case of three flavours
$u,d,s$ we have
\begin{gather}
\psi_L \rightarrow \exp (-i \theta_L \cdot \lambda) \psi_L \, , \qquad
\psi_R \rightarrow \exp (-i \theta_R \cdot \lambda) \psi_R \, ,
\end{gather}
where $\lambda^a$ , $a=1, \dots 8$ are the SU(3) Gell-Mann matrices.
These transformations may also be expressed as vector and axial-vector
transformations,
\begin{gather}
\psi \rightarrow \exp (-i \theta_V \cdot \lambda) \psi \, , \qquad
\psi \rightarrow \exp (-i \theta_A \cdot \lambda \gamma_5) \psi \, ,
\end{gather}
with $\theta_V= (\theta_L+\theta_R)/2$, $\theta_A = (\theta_L -
\theta_R)/2$.  The Lagrangian (\ref{m0qcd}) is thus invariant under
$SU(3)_L\times SU(3)_R$ or $SU(3)_V\times SU(3)_A$.

One might have expected a $U(3)_L\times U(3)_R$ global symmetry. It
turns out that in QCD $U(1)_A$ is anomalous \cite{Witten:1979vv,
tHooft:1986nc}, and thus
not present in the quantum theory - gauge configurations with
non-trivial winding number make $\partial_\mu J^\mu_{U(1)_A}
\neq 0$ through the `triangle' quark loop graph. The only exception is
when $N_f \ll N$ when the triangle graph becomes suppressed in a $1/N$
expansion. The $U(N_f)_A$ symmetry is thus present at large $N$. The
vector U(1) is baryon number and is a spectator to the symmetry breaking.

This chiral symmetry may be broken {\it explicitly} if a mass term is
present in the Lagrangian,
\begin{gather}
{\cal L}_{m} = - m \bar \psi \psi \, .
\end{gather}

There is another {\it spontaneous} breaking of chiral symmetry in QCD
though - the strong dynamics triggers the formation of a vev for the
operator
\begin{gather}
\langle \bar \psi \psi \rangle = \langle \bar \psi_L \psi_R \rangle + h.c. \neq 0 \, .
\end{gather}

In both symmetry breaking cases, the flavour symmetry is broken down
to a single vector $SU(3)_V$ factor,
\begin{gather}
SU(3) \times SU(3) \rightarrow SU(3)_V \, .
\end{gather}
Goldstone's theorem though tells us that for a spontaneously broken
symmetry 8 massless Goldstone bosons are expected, one for each broken
generator. In QCD these are quark bound states, the $\pi^\pm, \pi^0,
K^\pm, K^0, \bar K^0$ and the $\eta$. In the large $N$ limit where the
$U(1)_A$ symmetry is restored the $\eta'$ joins these particles as a
Goldstone boson.

A low energy effective action for the Goldstone modes, which are
lighter than all other QCD bound states, may be written (see for
example \cite{Georgi:1985kw} usual formulation is to write the
Goldstone fields, $\pi^a$, as part of a field
\begin{equation}
U = e^{i 2 \pi^a \lambda^a/ f_\pi} \,,
\end{equation}
where $f_\pi$ is the pion decay constant. $U$ transforms under the
underlying chiral symmetries as $L^\dagger U R$ and its vev (the $3\times 3$
unit matrix) breaks this symmetry to the diagonal. The effective
Lagrangian can be constructed as a derivative expansion with leading
term
\begin{equation}
{\cal L} = f_\pi^2 {\,\rm Tr\,} \partial^\mu U^\dagger \partial_\mu U + ...
\end{equation}

If a small explicit breaking by a quark mass term is present, the
Goldstone bosons acquire mass to become pseudo-Goldstone bosons. Since
the $3\times3$ mass matrix transforms under the (now spurious) chiral
symmetries as $L^\dagger M_q R$ we can add a term to the low energy
action
\begin{equation}
\Delta {\cal L} = \nu^3 {\,\rm Tr\,} M_q^\dagger U \,,
\end{equation}
where $\nu^3$ is some dimension 3 coefficient that measures the
size of the quark condensate and must be fitted phenomenologically.
This term generates  a mass for the Goldstones with $M^2_{\pi} \sim M_q$.

We will see below how this symmetry breaking is realized in
gravity duals. In the first examples, we will make use of the
large $N$ limit of the AdS/CFT Correspondence and realize the
breaking of a simple $U(1)_A$ symmetry, under which $\psi_L$ and
$\psi_R$ transform as
\begin{gather}
\psi_L \rightarrow e^{i \alpha} \psi_L \, , \qquad
\psi_R \rightarrow e^{-i
\alpha} \psi_R    \, .
\end{gather}
The associated Goldstone boson has the quantum numbers of the $\eta'$
particle although it's behaviour is more akin to the pions. We will
also describe a model that can realise the full non-abelian chiral
symmetry breaking pattern as seen in QCD.

\subsection{D7 probes in non-supersymmetric  backgrounds} \label{D7nonsusy}

To see chiral symmetry breaking in the pattern of QCD in the AdS/CFT
Correspondence it is necessary to break supersymmetry completely. The
operator $\bar \psi \psi$ is the F-term of a composite chiral
superfield $\tilde{Q}Q$ - its vev would break supersymmetry and so it
would not be expected to be non-zero in a supersymmetric theory's
ground state.

So far we have encountered two different generalizations of the
AdS/CFT correspondence: On the one hand the deformation of the $AdS_5
\times S^5$ space, described in section~\ref{flows}, leads to
holographic RG flows which in particular cases flow to confining
gauge theories. On the other hand, we have discussed the addition of
flavour arising from the addition of D7 brane probes in section
\ref{flavour}.  The idea is now to combine these two generalizations
of the AdS/CFT correspondence and to add D7 brane probes to deformed
gravity backgrounds. As we discuss below, and first shown in
\cite{Babington:2003vm}, this leads to a dual gravity description of
chiral symmetry breaking and Goldstone bosons.

\subsubsection{Constable-Myers background} \label{sectionConstableMyers}

A prototype example for a confining gravity background in which
supersymmetry is completely broken is the metric constructed by
Constable and Myers in \cite{Constable:1999ch}. This background is an
example of a {\it dilaton flow} (see also \cite{Gubser:1999pk}), in
which the dilaton -- which is constant for the supergravity background
dual to $\N=4$ Super Yang-Mills -- has a non-trivial profile, {\em
  i.e.}~depends on the radial coordinate in deformed AdS space. At the
supergravity level one simply searches for a solution of the IIB
equations of motion with the dilaton switched on.  More physical but
also more complicated examples could be considered but this geometry
provides an easy starting point.  We will interpret the flow in terms
of the field theory shortly\footnote{ Other very similar examples of
  chiral symmetry breaking by embedding D7 brane probes into different
  dilaton-flow geometries have been found for instance in
  \cite{Ghoroku:2004sp, Brevik:2005fs} and \cite{JohannesPhDthesis}.}.

We choose a convenient coordinate system for the gravity background of
Constable and Myers \cite{Constable:1999ch} such that in {Einstein
  frame}, the geometry is given by
\begin{align} ds^2 & =  H^{-1/2} \left( { w^4 + b^4 \over w^4-b^4}
\right)^{\delta/4} \, \sum\limits_{j=0}^{3} dx_{j}^2
+ H^{1/2} \left( {w^4 + b^4 \over w^4-
b^4}\right)^{(2-\delta)/4} {w^4 - b^4 \over w^4 } \sum_{i=1}^6
dw_i^2 \,,
\end{align}
where $b$ is the scale of the geometry that determines the size of the
deformation ($\delta = R^4/(2 b^4)$ with $R$ the AdS radius) and
\vspace{-0.2cm}
\begin{align} H =  \left(  { w^4
+ b^4 \over w^4 - b^4}\right)^{\delta} - 1 \, ,  \;\;
w^2 = \sum\limits_{i=1}^{6} w_i{}^2
\, .
\end{align}
In this coordinate system, the dilaton and four-form are,
with $\Delta^2 + \delta^2 =10$,
\beq e^{2 \phi} = e^{2 \phi_0} \left( { w^4 + b^4 \over
w^4 - b^4} \right)^{\Delta}, \quad C_{(4)} = - {1 \over 4}
H^{-1} dt \wedge dx \wedge dy \wedge dz \,.
\eeq
This geometry returns to $AdS_5 \times S^5$ in the UV as may be seen
by explicitly expanding at large~$w$.

The field theory dual is therefore the $\N=4$ Super Yang-Mills theory
in the far UV. In the IR it is deformed by the parameter $b$ which
sets the conformal symmetry breaking scale - it will determine the
scale equivalent to $\Lambda_{QCD}$ in the gauge theory
\begin{equation}
\Lambda_b = \frac{b}{2 \pi \alpha'} \, .
\end{equation}

The SO(6) symmetry of the geometry is unbroken so the equivalent
deformation in the gauge theory must not break the R-symmetry. We also
see that $b$ enters with the radial direction in AdS $w$ and $b^4$
must therefore correspond to an operator of dimension four. There is a
natural dimension four R-chargeless operator in the field theory which
is ${\rm Tr}\, F^2$.  This is a geometry therefore describing the
$\N=4$ gauge theory with a source forcing it off its supersymmetric
vacuum. Note that ${\rm Tr}\, F^2$ is the F-term of a composite
operator of the product of two chiral superfields fields $W$ and hence
a vev for the operator clearly breaks supersymmetry.

Note that the running of the dilaton in the gauge theory corresponds
to a running coupling. Indeed the dilaton and geometry blow up at the
scale $\Lambda_b$ consistent with the interpretation of that scale
with $\Lambda_{QCD}$.  On the gravity side singularities ought to be
identified with a source - in this non-supersymmetric case this
identification is unclear but one might imagine that the D3 branes of
the geometry have moved out from the origin to $b$ and complete the
geometry. We will escape resolving this issue below because the D7
branes we will embed will not penetrate as far in as $b$.

This field theory of course has extra adjoint degrees of freedom as
compared to QCD. However it has been shown to be confining by
calculating the Wilson loop, which has an area law.  We will therefore
take it as a model for a confining $SU(N)$ theory at large $N$.

The next step is to add quarks \cite{Babington:2003vm,Evans:2004ia}.
We will use an embedded probe D7 brane as discussed in section
\ref{flavour}. The D7-brane will be embedded, in the static gauge,
with world-volume coordinates identified with $x_{0,1,2,3}$ and
$w_{1,2,3,4}$.  Transverse fluctuations will be parameterized by $w_5$
and $w_6$ - it is convenient to define a coordinate $\rho$ such that
$\sum_{i=1}^4 dw_i^2 = d\rho^2 + \rho^2 d\Omega_3^2$ and the radial
coordinate is given by $w^2=\rho^2 + w_5{}^2 + w_6{}^2$.

In the field theory we have introduced $\N=2$ quark hypermultiplets
with the usual superpotential coupling to the $\N=4$ fields is
$\tilde{Q} \Phi Q$. Note there is a U(1)$_R$ symmetry under which $Q$
and $\tilde{Q}$ both have charge $-1$ and $\Phi$ has charge $+2$. This
symmetry is analogous to U(1)$_A$ in that a vev for the fermionic
quark bilinear $\bar \psi \psi$ will break the symmetry. Geometrically
this symmetry corresponds to rotations in the $w_5-w_6$ plane.  With
the supersymmetry breaking induced by $b^4$ in the geometry, we expect
the scalar quarks in the hypermultiplets to become massive (that the
moduli space is lifted and the scalar vevs pinned at zero was checked
in \cite{Apreda:2005hj}).

The Constable-Myers background is convenient for embedding a D7 brane
probe since it preserves $SO(6)$ symmetry. The embedding functions
determining the minimum energy configuration of the D7 probe are
functions of $\rho$ only, {\em i.e.}~essentially of the energy scale. As
alluded to above the D7 brane probes giving rise to chiral symmetry
breaking are embedded in a perfectly regular way avoiding the naked
singularity in the IR at $b$.

The Dirac-Born-Infeld action of the D7-brane probe in
the Constable-Myers background takes the form
\begin{align}   \label{CMD7} S_{D7} \, = \, - T_7 R^4
\int d^8 \xi~ \epsilon_3 ~  e^{ \phi} { \cal G}(\rho,w_5,w_6)
 \Big( 1 + g^{ab} g_{55} \partial_a w_5
\partial_b w_5  + g^{ab} g_{66} \partial_a w_6
\partial_b w_6 \Big)^{1/2},
\end{align}
where
\begin{align} & {\cal G}(\rho,w_5,w_6)
 = \rho^3 {( (\rho^2 + w_5^2
+ w_6^2)^2 + b^4) ( (\rho^2 + w_5^2 + w_6^2)^2 - b^4) \over
(\rho^2 + w_5^2 + w_6^2)^4} \,. \nonumber
\end{align}
Here we have rescaled $w$ and $b$ in units of $R$ as in
\cite{Evans:2005ti} so that factors of $R$ only occur as an overall
factor on the embedding Lagrangian. Note that the factors of $\alpha'$
cancel between $R^4$ and $T_7$ leaving the free energy proportional to
$1/g_s$ - in the usual 't~Hooft limit ($N \rightarrow \infty$ with
$g_s N$ fixed) the free energy grows as $N$ as one would expect.

From these equations we derive the corresponding equation of
motion. We look for classical solutions of the form $w_6 =
w_6(\rho),\, w_5 =0$. The equation of motion reads
\beq \label{eommc}{ d \over d \rho} \left[ {e^{\phi}  { \cal
G}(\rho,w_6) \over \sqrt{ 1 + (\partial_\rho w_6)^2}}
(\partial_\rho w_6)\right] - \sqrt{ 1 + (\partial_\rho w_6)^2} { d
\over d w_6} \left[ e^{\phi} { \cal G}(\rho,w_6) \right] = 0\,.
\eeq The last term in the above is a potential-like term that
is evaluated to be

\beq { d \over d w_6} \left[ e^{\phi} { \cal G}(\rho,w_6)
\right] = { 4 b^4 \rho^3 w_6 \over (\rho^2 + w_6^2)^5} \left( {
(\rho^2 + w_6^2)^2 + b^4 \over  (\rho^2 + w_6^2)^2 - b^4}
\right)^{\Delta/2} (2 b^4  - \Delta (\rho^2 + w_6^2)^2)\,. \eeq\\
Numerically we find solutions with the asymptotic behaviour $w_6 \sim
m + c/\rho^2$. The identification of these constants as field theory
operators requires a coordinate transformation because the scalar
kinetic term is not of the usual canonical AdS form. Transforming to
coordinates \cite{Karch:2002sh} in which the kinetic term has
canonical form, we see that $m$ has dimension 1 and $c$ has dimension
3. These coefficients are then identified with the quark mass $m_q$
and condensate $\langle \bar{\psi} \psi \rangle$ respectively, in
agreement with the usual AdS/CFT dictionary obtained from the
asymptotic boundary behaviour (\ref{asymptotic}).

\begin{figure}[!h]
\begin{center}
\includegraphics[height=7cm,clip=true,keepaspectratio=true]{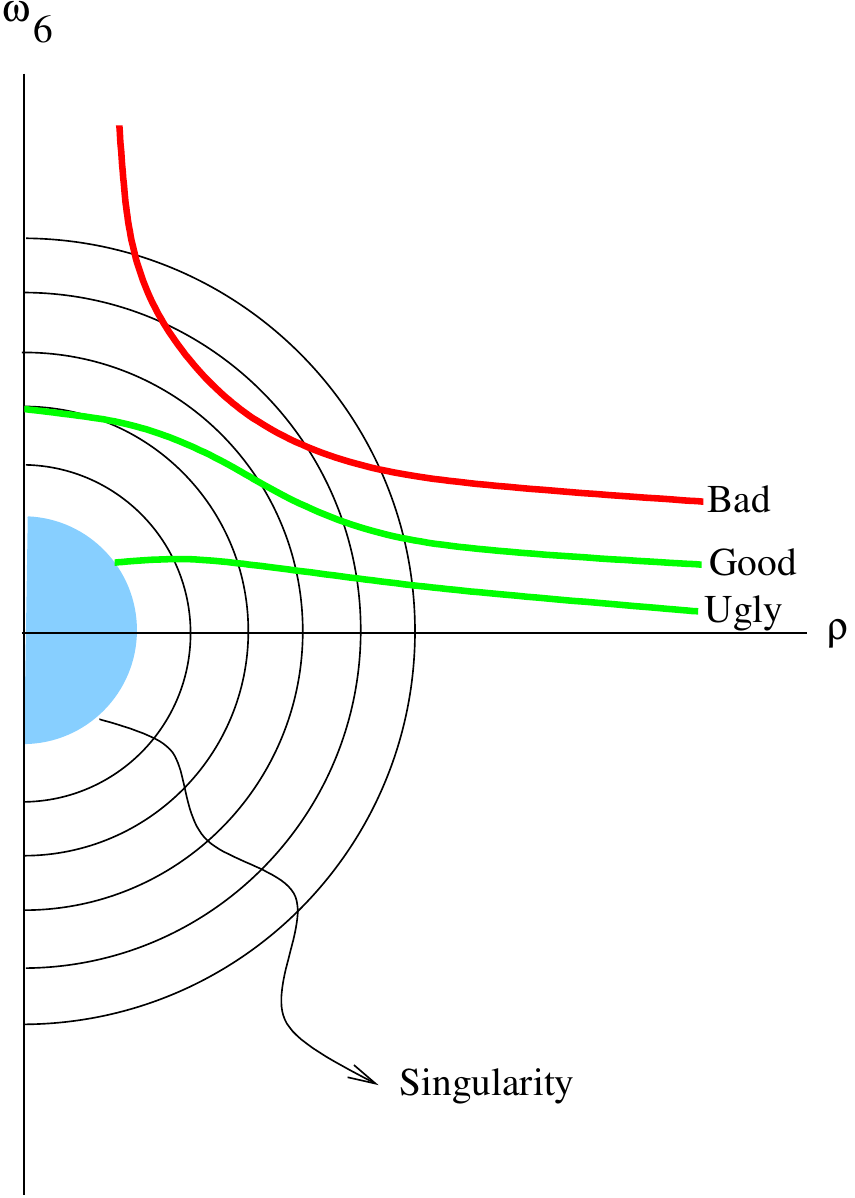}
\caption{\small{Different possibilities for solutions of the
D7-brane equations of motion.  The semicircles are lines of
constant $r$, which should be interpreted as a scale in the dual
Yang-Mills theory. The ``Bad'' curve cannot be interpreted as an
RG flow.  The other curves have an RG flow interpretation, however
the infrared (small r) region of the ``Ugly'' curve is outside the
range of validity of supergravity. Figure from 
\cite{Babington:2003vm}.}}\label{singularregion}
\end{center}
\end{figure}

Due to the singularity in the background, we have to impose a
regularity constraint on the brane embedding, which amounts to a
boundary condition for the equation of motion determining the
embedding. This is illustrated in figure~\ref{singularregion}.  Brane
embeddings reaching the singularity are excluded since they enter a
region of strong curvature where the supergravity approximation is no
longer valid. In addition, embeddings which intersect the circles of
constant energy twice cannot be interpreted as a RG flows and thus are
unphysical.  A boundary condition which selects the physical
embeddings is to require the first derivative of the embedding
functions to vanish at $\rho=0$. In the picture of the Karch/Katz RG
flow discussed in section~\ref{probesupergravity}, this amounts to
requiring the (now deformed) $S^3$ to shrink to zero at this point.

\begin{figure}[!h]
\begin{center}
\includegraphics[height=6cm,clip=true,keepaspectratio=true]{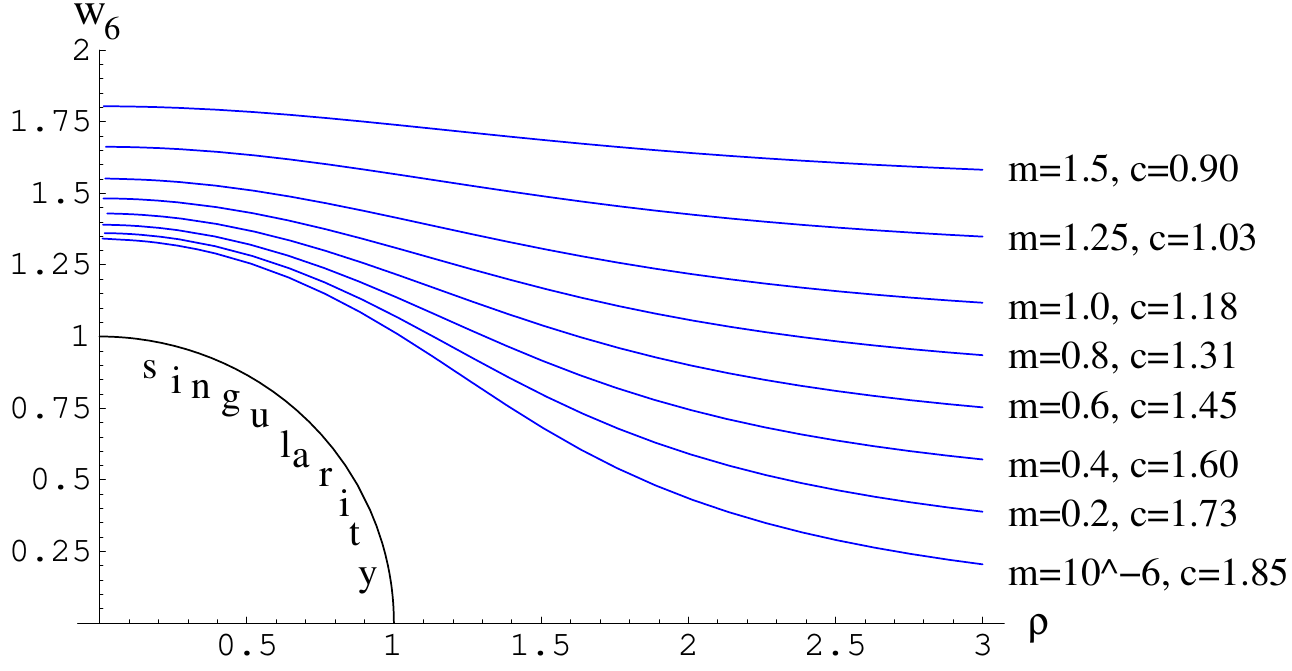}
\caption{Regular solutions in the Constable-Myers background.
From \cite{Babington:2003vm}.}
\label{CMembeddings}
\end{center}
\end{figure}

\begin{figure}[!h]
\begin{center}
\includegraphics[height=5.5cm,clip=true,keepaspectratio=true]{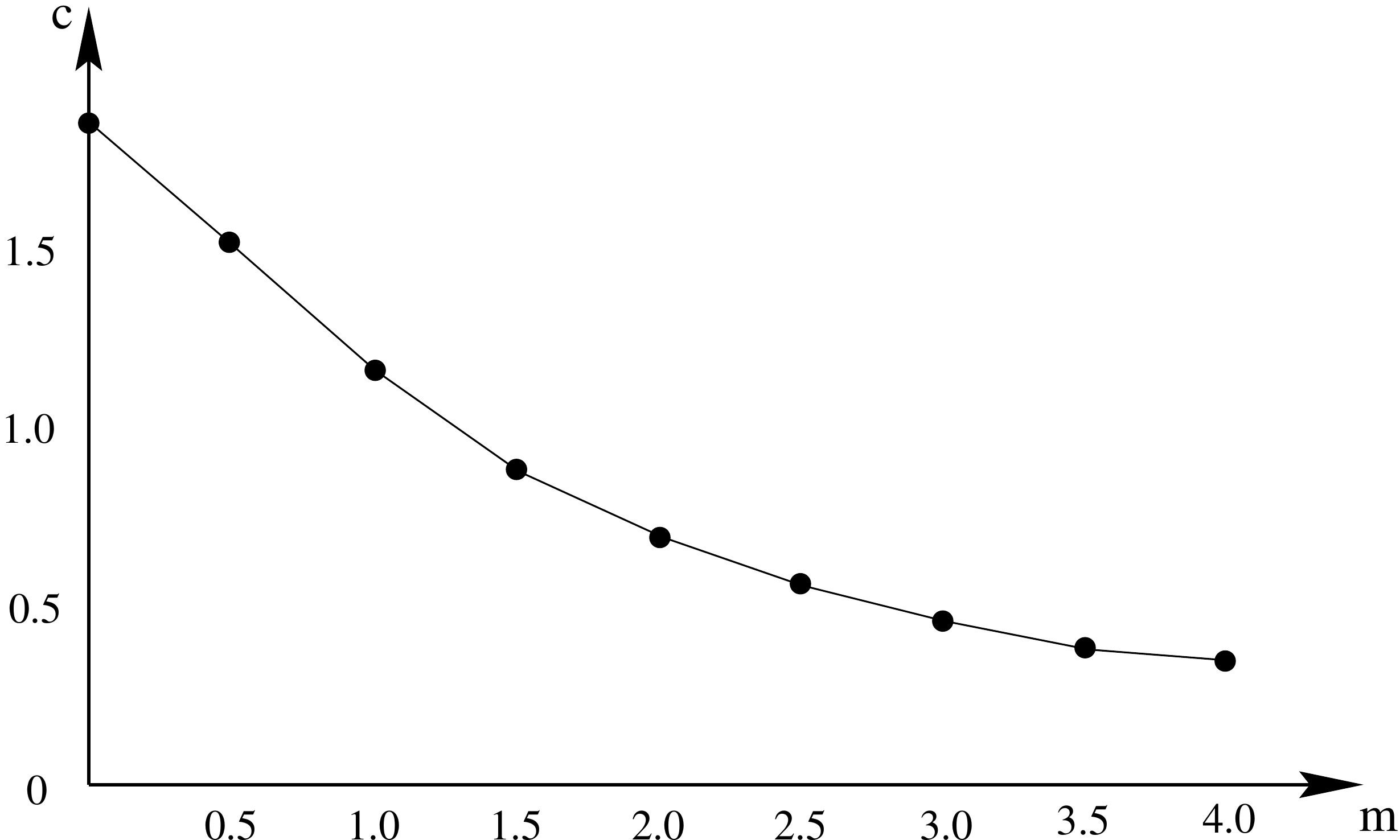}
\caption{A plot of the condensate parameter $c$ vs quark mass $m$ for
the regular solutions of the equation of motion in the Constable-Myers
background. $c$ and $m$ are given in units set by the length scale $b$.
From \cite{Babington:2003vm}.}
\label{quiv8}
\end{center}
\end{figure}

We now calculate the embedding functions for the D7 brane probe by
solving the equations of motion obtained from the DBI action
(\ref{CMD7}).  The numerical result is displayed in figure
\ref{CMembeddings}. For each of these embeddings we fix two boundary
conditions, as required for solving a second order differential
equation: For regularity we require the first derivative of the
embedding to vanish at $\rho=0$. Secondly, the absolute value of the
embedding function $w$ at the boundary $\rho \rightarrow \infty$ fixes
the value of the quark mass in units of the scale $b$.  The condensate
$c \equiv \langle \bar \psi \psi \rangle$ in units of $b$ may then be
read off from the asymptotic behaviour of the embedding at $\rho
\rightarrow \infty$, where the embedding behaves as
\begin{equation}
w \sim m + \frac{c}{\rho^2} \, .
\end{equation}
We see an interesting screening effect in figure~\ref{CMembeddings}:
The regular solutions appear to be repelled by the singularity, rather
than just being straight lines as in the supersymmetric case.  This
can be related to spontaneous chiral symmetry breaking by a quark
condensate: In fact, as is seen from figures \ref{CMembeddings} and
\ref{quiv8}, there is a regular embedding with non-zero condensate
even for $m \rightarrow 0$. This corresponds exactly to spontaneous
chiral symmetry breaking by a quark condensate!  Moreover, at large
$m$ we have $c\sim 1/m$, as expected from field theory.  - Notice also
the finite distance on the $w$-axis between the singularity and the
embedding with $m \rightarrow 0$.

Remember that the D7 and D3 branes can in fact be separated in the
full $w_5-w_6$ plane and therefore a D7 brane lying on the axis of
that plane asymptotically (so the bare mass of the quark is zero and
there is a good U(1)$_A$ symmetry in the UV) is deflected out onto any
point on a circle in the plane. That circle represents the vacuum
manifold of the breaking of the U(1) symmetry - we see a completely
geometric realization of the symmetry breaking.

\subsubsection{Goldstone boson} \label{sec:Goldstone}
\begin{figure}[!h]
\begin{center}
\includegraphics[height=6cm,clip=true,keepaspectratio=true]{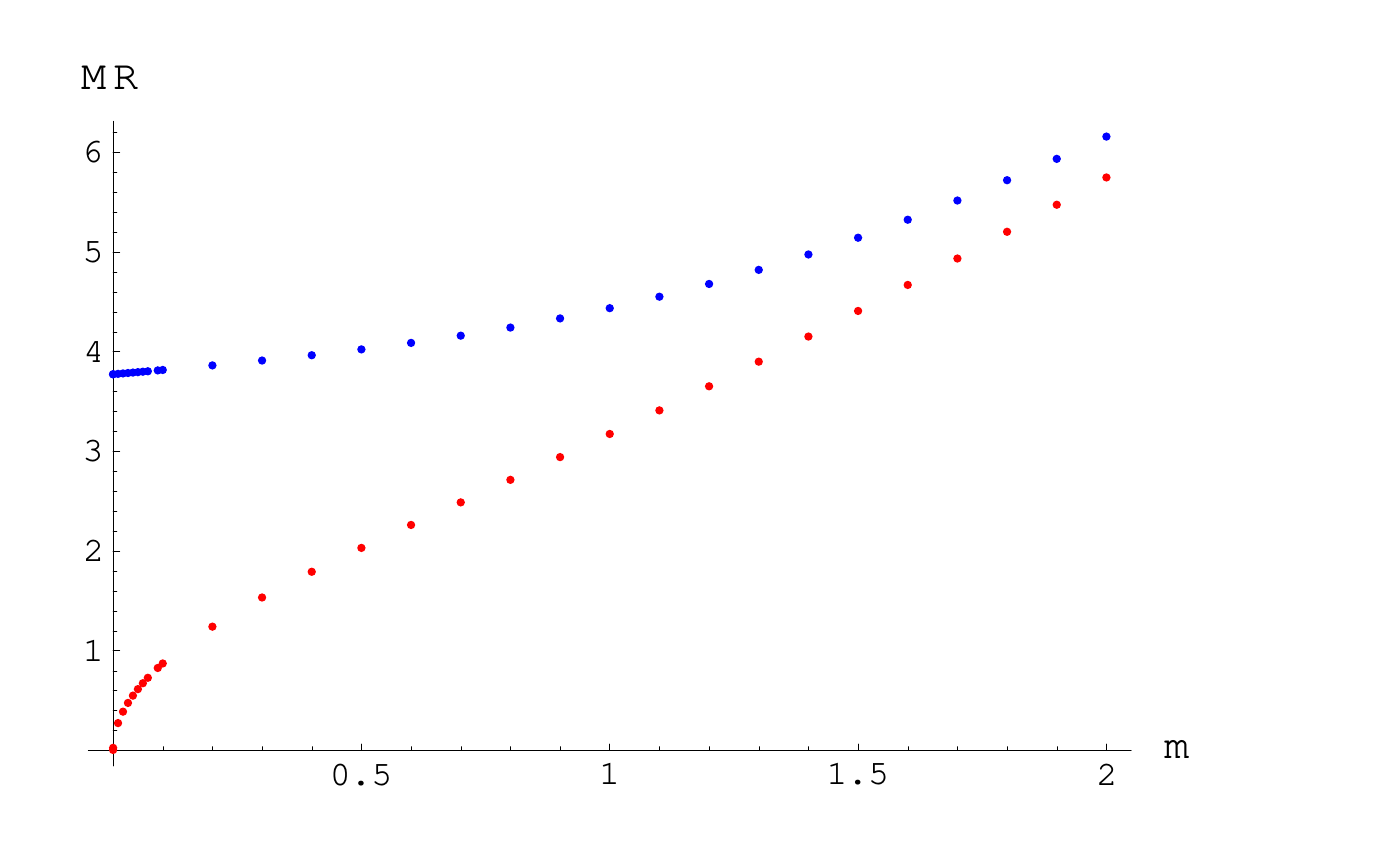}
\caption{
Masses of the lowest-lying meson masses for fluctuations about the D7
brane embedding in radial and angular direction, as a function of the
quark mass.  The angular fluctuation mode gives rise to a
(pseudo-)Goldstone mode. Since the spontaneously broken symmetry is
$U(1)_A$, the Goldstone boson may be interpreted as the $\eta'$, which
is a Goldstone boson of this symmetry in $SU(N)$ gauge theory for $N
\rightarrow \infty$. Figure provided by J.~Gro\ss e.  }
\label{Goldstone}
\end{center}
\end{figure}

Since there is spontaneous symmetry breaking for $m \negthinspace
\rightarrow \negthinspace 0$, we expect a Goldstone boson
ana\-lo\-gous to the $\eta'$, in the meson spectrum. Clearly
fluctuations in the angular direction in the $w_5-w_6$ plane ({\em
  i.e.}\ along the vacuum manifold) will generate these massless
states.  Solving the supergravity equation of motion for D7 probe
brane fluctuations in the two directions transverse to probe, $(
\delta w_5 = f(r)\sin( k \cdot x) \,$, $\delta w_6 = h(r)\sin( k \cdot
x) $) around the D7~brane probe embedding shown in figure~\ref{quiv8},
the meson masses are given by $M^2 = - k^2$.  There are indeed two
distinct mesons (see figure~\ref{Goldstone}): One is massive for every
$m$, and \mbox{corresponds to} fluctuations in the radial transverse
direction, the other, corresponding to the $U(1)$ symmetric
fluctuation, is massless for $m=0$ and is thus a Goldstone boson.  It
may be identified with the $\eta'$, which becomes a $U(1)_A$ Goldstone
boson for $N \rightarrow \infty$.  At finite $N$, pure stringy
corrections will give the $\eta'$ a non-zero mass in the gravity
picture, similarly to instantons in the field theory dual
\cite{Barbon:2004dq,Armoni:2004dc}.

Another important property of the model of \cite{Babington:2003vm} is
the small quark mass behaviour of the meson mass, proportional to the
square root of $m$, thus satisfying the Gell-Mann--Oakes--Renner
relation \cite{GellMann:1968rz} of chiral QCD.  Also the linear
asymptotics for large $m$ correctly reproduce the field theory
results.  In \cite{Evans:2004ia} the R-scaling of the Goldstone's mass
for small quark mass was determined as
\begin{equation}
\frac{M^2_\pi}{\Lambda_b^2} =
2.75 \sqrt{\frac{\pi}{g_s N}}\frac{m_q}{\Lambda_b} \,.
\end{equation}

\subsubsection{Vector mesons} \label{sec:CMvectors}

The vector mesons in the model are, as in the basic D7 brane
embeddings, described by the gauge fields in the DBI action describing
the D7 branes. Again solutions of the form $A^\mu = g(\rho) \sin(k \cdot x)
\epsilon^\mu$ provide the masses of the $\rho$ and its radially
excited states \cite{Evans:2004ia}.  The $n=0$, unexcited, state has
mass
\begin{equation}
\frac{M^2_\rho}{\Lambda_b^2} = 2.16 \sqrt{\frac{\pi}{g_sN}} \,.
\end{equation}
As expected it is massive, reflecting the dynamical generation of a
quark mass.

\begin{figure}[!h]
\begin{center}
\includegraphics[height=4.3cm,clip=true,keepaspectratio=true]{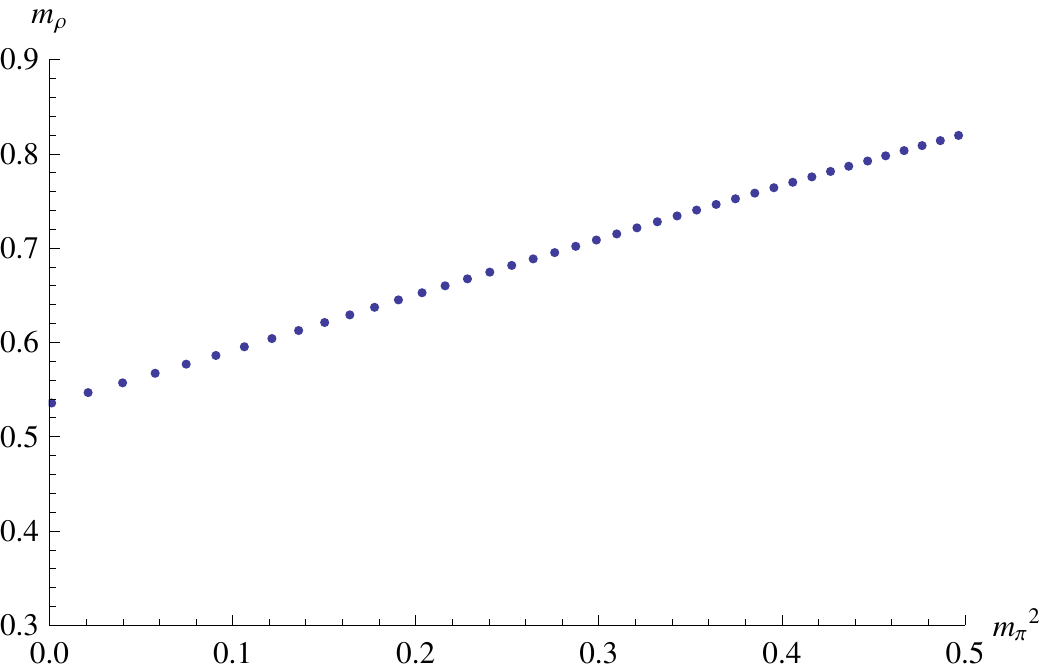}
\hspace{1cm}
\includegraphics[height=4.5cm,clip=true,keepaspectratio=true]{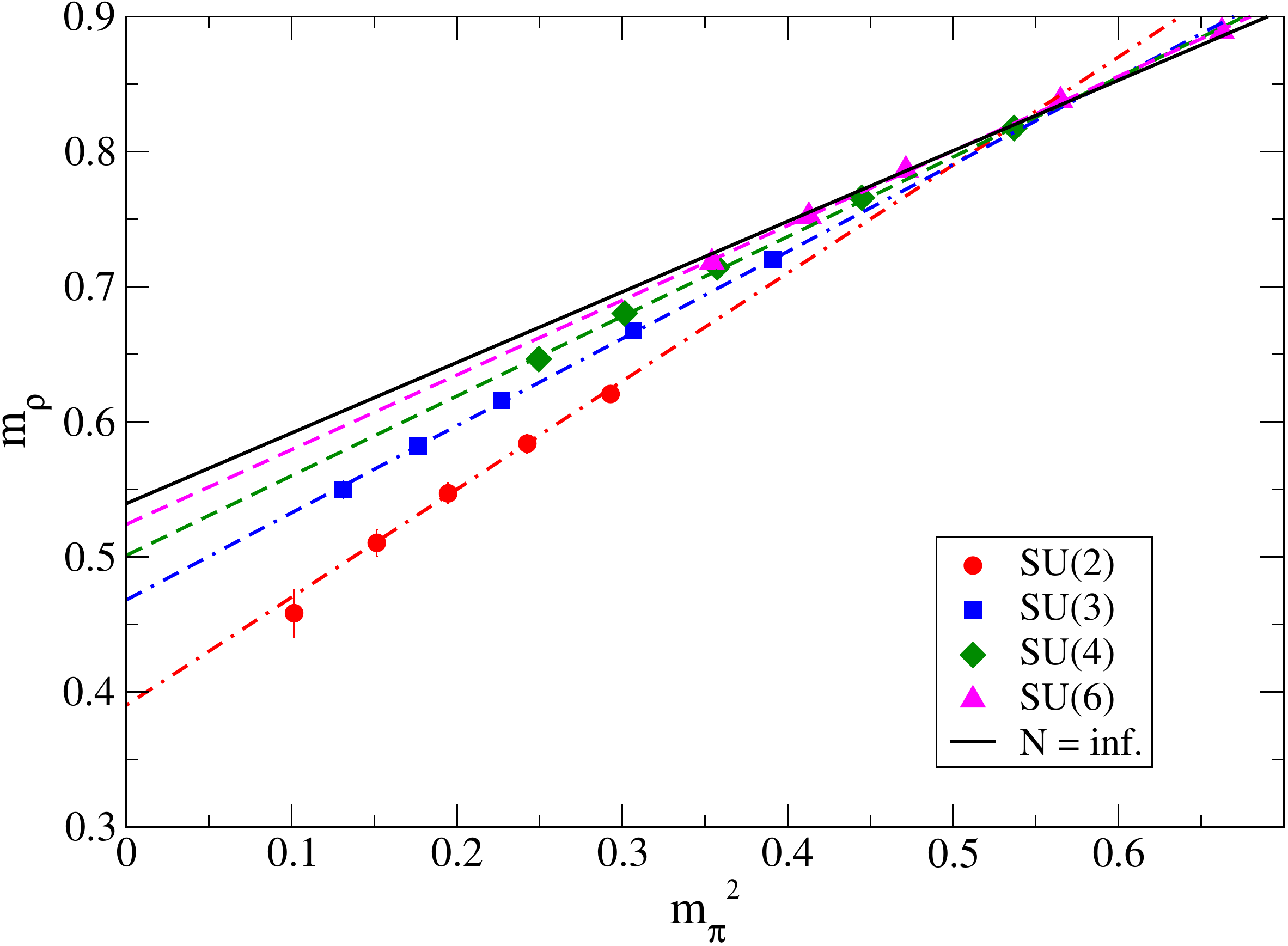}
\caption{A plot of $m_\rho$ vs $m_\pi^2$ in the Constable-Myers
background on the left (we thank Andrew Tedder for generating
this plot). Lattice data \cite{lattice}
(preliminary, quenched and at finite spacing) for
the same quantity is also shown on the right.}
\label{mpivsmrho}
\end{center}
\end{figure}

In figure~\ref{mpivsmrho} we plot the dependence of the rho meson mass
on the pion mass squared in this model, in dimensionless units fixed
by the choice of the supergravity scale $b$. The rho mass as function
of the pion mass squared has recently also been computed for large $N$
within lattice gauge theory \cite{Bali:2007kt,lattice}, and a direct
comparison of gauge/gravity and lattice results is possible. In the
lattice computations, the scale is set by the lattice spacing $a$. We
choose our units such as to be able to compare directly with the
lattice results of \cite{lattice}, which are shown on the right hand
side of figure~\ref{mpivsmrho}.  In units such that the offset at
$m_\pi =0$ coincides with \cite{lattice}, we find a linear dependence
of the rho mass on the pion mass squared, with slope 0.57.

For the lattice results of \cite{lattice}, the simulations are
performed in the quenched approximation.  This is appropriate for the
large $N$ limit, if not for smaller $N$. The lattice data of
\cite{lattice} is preliminary and at a fixed, finite lattice spacing.
Nevertheless it is striking that not only does the lattice data
display the same linearity as the gauge/gravity model, but also the
slope in the large $N$ limit is 0.52 and therefore is very close to
the gauge/gravity dual result.
The fact that the numbers  agree at
the level of the first digit is surprising. We will see in the
sections that follow that generically AdS-meson predictions match QCD
data better than one would naively expect. --
For large values of the quark mass,
we expect $m_\rho \propto m_\pi$ due to the onset of supersymmetry.

\subsection{Gauge theory in AdS$_4$ space}

Another clean example of a gravity dual description of chiral symmetry
breaking has been provided in \cite{Ghoroku:2006nh}.  There the
duality is adapted to look at the $\N=4$ gauge theory in a
four-dimensional anti-de-Sitter space. The gravity dual they provide
has a constant dilaton and is given by
\begin{equation}
ds^2_{10}=
\left\{
{r^2 \over R^2}A^2\left(-dt^2+a(t)^2\gamma(x)^2 (dx^i)^2\right)+
\frac{R^2}{r^2} dr^2+R^2 d\Omega_5^2 \right\} \ ,
\label{finite-c-sol}
\end{equation}
\begin{equation}
  A=1+({r_0\over r})^2, \quad a(t)={R^2\over 2r_0}\sin(2{r_0\over R^2} t),
\quad \gamma(x)={1\over 1-x_ix^i/(4\tilde{r}_0^2)}\ ,
\end{equation}
where  $R=\sqrt{\Lambda}/2=(4 \pi N )^{1/4}$
and $\tilde{r}_0$ is an arbitrary scale factor which sets the 4d
cosmological constant
\begin{equation}
\lambda = - 4 \frac{r_0^2}{R^4}\,.
\end{equation}

The presence of a cosmological constant breaks both conformal
invariance and supersymmetry.
\begin{figure}[htbp]
\vspace{.3cm}
\begin{center}
\includegraphics[angle=0,width=0.45\textwidth]{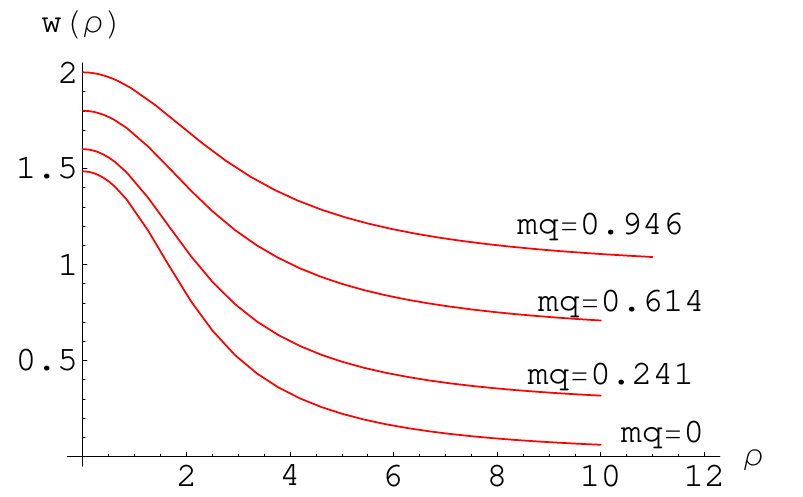}
\caption{D7 brane embeddings in the geometry with an AdS$_4$ subspace
  showing chiral symmetry breaking. 
Figure provided by K.~Ghoroku.}\label{ads4embed}
\end{center}
\end{figure}

Quarks are included in the geometry through probe D7 branes. The
computation follows those already seen and we display the embeddings
in figure~\ref{ads4embed}.  Chiral symmetry breaking is clearly
manifest.

\begin{figure}[htbp]
\vspace{.3cm}
\begin{center}
\includegraphics[angle=0,width=0.45\textwidth]{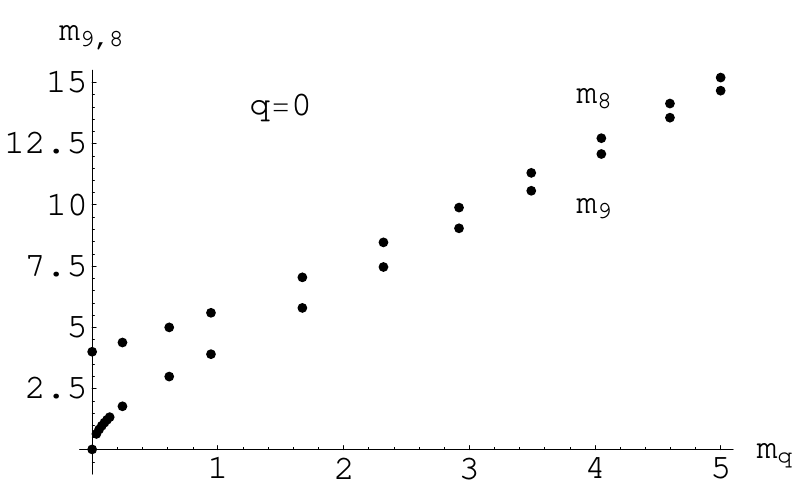}
            \hspace{1cm}
\includegraphics[angle=0,width=0.45\textwidth]{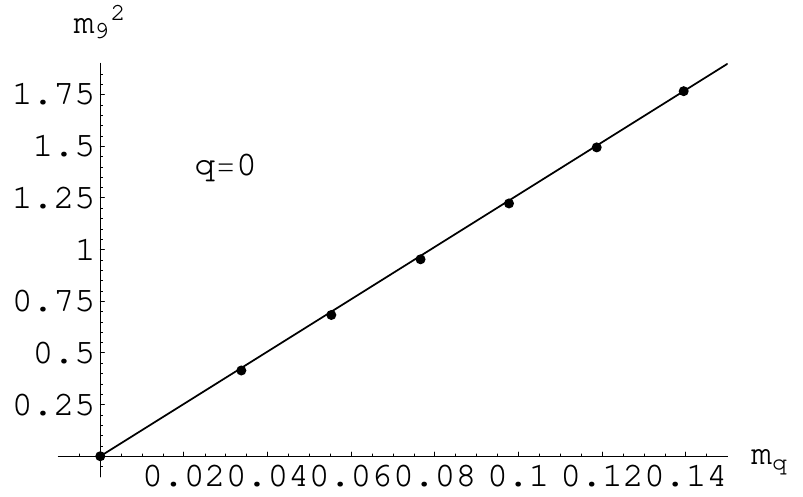}

\caption{Plot of the scalar ($m_8$) and pseudo-scalar ($m_9$) masses
vs the quark mass $m_q$ for $l=n=0$, $r_0=1.0$ and $R=1$ in the AdS$_4$
theory \cite{Ghoroku:2006nh}. Plots provided by K.~Ghoroku.}\label{adsmass}
\end{center}
\end{figure}

The meson spectra associated with fluctuations of the D7 brane in the
$w_5, w_6$ directions have also been computed and are shown in figure
\ref{adsmass}. There is a pion mode, whose mass fits the expected
Gell-Mann-Oakes-Renner relation for small quark mass, and a massive
sigma mode.

D7 embeddings have also been studied when the gauge theory lives in
de-Sitter space \cite{Ghoroku:2006af,Hirayama:2006jn} but there is no
spontaneous chiral symmetry breaking in that case (the behaviour is
like that of the $\N=4$ gauge theory at finite temperature described
in section~\ref{finiteT}).

\subsection{Chiral symmetry breaking in the D4/D6 system} \label{d4d6myers}

A similar model based on a D4 brane background in which one of the
space directions wrapped by the D4 branes is compactified on a circle
was studied in \cite{Kruczenski:2003uq}. There the flavour degrees of
freedom are provided by D6 brane probes. Spontaneous chiral symmetry
breaking of the $U(1)_A$ symmetry is seen in this model too. It has
the advantage of not displaying a singularity in the interior of the
curved space. On the other hand, the dual gauge theory becomes
five-dimensional in the UV.

The authors of \cite{Kruczenski:2003uq} consider the D4/D6 system with the
branes oriented as described by the following array:
\begin{center}
\begin{tabular}{|r|c|c|c|c|c|c|c|c|c|c|}
\hline
& 0 & 1 & 2 & 3 & 4 & 5 & 6 & 7 & 8 & 9 \\
\hline
$N$ \,\,D4    & X & X & X & X & X  &   &   &   &   &   \\
\hline
$N_f$ \,\,D6 & X & X & X   & X   &   & X & X  & X   &   &   \\
\hline
\end{tabular}
\label{intersectionKMMW}
\end{center}
The D4- and the D6-branes may be separated from each other along the
89-directions.  On the gauge theory side one has a supersymmetric,
five-dimensional $SU(N)$ gauge theory coupled to a four-dimensional
defect. The entire system is invariant under eight supercharges, {\em
  i.e.}~there is $\N =2$ supersymmetry in four-dimensional language.
The degrees of freedom localized on the defect are $N_f$
hypermultiplets in the fundamental representation of $SU(N)$, which
arise from the open strings connecting the D4- and the D6-branes. Each
hypermultiplet consists of two Weyl fermions of opposite chiralities,
$\psi_{\rm L}$ and $\psi_{\rm R}$, and two complex scalars.

Identifying the 4-direction with period $2\pi/M_{\rm KK}$, and with
anti-periodic boundary conditions for the D4-brane fermions, breaks
all of the supersymmetries and renders the theory effectively
four-dimensional at energies $E \ll M_{\rm KK}$. Further, the adjoint
fermions and scalars become massive. Now, the bare mass of each
hypermultiplet, $m_q$, is proportional to the distance between the
corresponding D6-brane and the D4-branes. Even if these bare masses
are zero, we expect loop effects to induce a mass for the scalars in
the fundamental representation. Generation of a mass for the
fundamental fermions is, however, forbidden by the chiral $U(1)_A$
symmetry.  At low energies, one is therefore left with a
four-dimensional $SU(N)$ gauge theory coupled to $N_f$ flavours of
fundamental quark.

\subsubsection{D4 brane background}

The type IIA supergravity background dual to $N$ D4-branes
compactified on a circle with anti-periodic boundary conditions for
the fermions takes the form
\begin{eqnarray}
ds^{2} &=& \left(\frac{U}{R}\right)^{3/2} \left( \eta_{\mu \nu} \,
dx^\mu dx^\nu + f(U) d\tau^{2} \right) + \left(
\frac{R}{U}\right)^{3/2} \frac{dU^{2}}{f(U)} +
R^{3/2} U^{1/2} \, d\Omega_{ 4}^{2} \,, \label{metric} \\
e^{\phi} &=& g_s \left( \frac{U}{R}\right)^{3/4}\,,
\sac F_{4} = \frac{N}{V_{ 4}} \, \varepsilon_{ 4}\,, \sac
f(U) = 1-\frac{\Ukk^{3}}{U^{3}} \,.
\label{metric1}
\end{eqnarray}
The coordinates $x^\mu=\{ x^0, \ldots , x^3\}$ parameterize the
four non-compact directions along the D4-branes whereas $\tau$
parameterizes the circular 4-direction on which the branes are
compactified. $d\Omega_{ 4}^2$ and $\varepsilon_{ 4}$ are the
$SO(5)$-invariant line element and volume form on a unit
four-sphere, respectively, and $V_{
  4}=8\pi^2/3$ is its volume. $U$ has dimensions of length and may be
thought of as a radial coordinate in the 56789-directions transverse
to the D4-branes. To avoid a conical singularity at $U=\Ukk$, $\tau$
must be identified with period
\begin{equation}
\delta \tau = \frac{4 \pi}{3} \, \frac{R^{3/2}}{\Ukk^{1/2}} \,.
\label{deltatau}
\end{equation}
This supergravity solution above is regular everywhere and is
completely specified by the string coupling constant, $g_s$, the
number of D4-branes $N$, and the constant $\Ukk$. The remaining
parameter, $R$, similar to the AdS radius, is given in terms of these
quantities and the string length, $\ell_s$, by
\begin{equation}
R^3 =  \pi g_s N \,\ell_s^3\,. \label{R}
\end{equation}
The $SU(N)$ field theory dual to (\ref{metric}, \ref{metric1}) is
defined by the compactification scale, $\mkk$, below which the theory
is effectively four-dimensional, and the four-dimensional coupling
constant {at} the compactification scale, $g_{\rm YM}$. These are
related to the string parameters by
\beq
\mkk =
\frac{3}{2}\frac{\Ukk^{1/2}}{R^{3/2}} = \frac{3}{2
  \sqrt{\pi}}\frac{\Ukk^{1/2}}{{( g_s N)}^{1/2} \ell_s^{3/2}} \sac
g_{\rm YM}^2 =3 \sqrt{\pi} \left( \frac{g_s\Ukk}{ N \ell_s}
\right)^{1/2} \,.
\label{parameters}
\eeq
The string length cancels in any calculation of a physical
quantity in the field theory. For example, the QCD string tension
is
\beq
\sigma =  \frac{1}{2\pi\ell_s^2} \left. \sqrt{-G_{tt} G_{xx}}
\right|_{U=\Ukk} = \frac{1}{2\pi \ell_s^2}
\left(\frac{\Ukk}{R}\right)^{3/2} = {2\over27\pi} g_{\rm YM}^2 \nc \,
\mkk^2 \,.
\eeq

\subsubsection{Probe D6 branes}

Flavour degrees of freedom are introduced into this model by adding D6
brane probes. Asymptotically (as $U\rightarrow \infty$), the D6-brane
is embedded as described by (\ref{intersectionKMMW}).  The analysis is
simplified by introducing isotropic coordinates in the
56789-directions. A new radial coordinate $\rho$ is related to $U$ by
\beq
U(\rho) = \left(\rho^{3/2} +
\frac{\ut^3}{4\rho^{3/2}}\right)^{2/3} \,.
\label{isomer}
\eeq
Moreover
five coordinates $\vec{z}=(z^5, \ldots, z^9)$ are introduced, such that
$\rho = |\vec{z}|$ and $d\vec{z} \cdot d\vec{z} = d\rho^2 + \rho^2
\, d\Omega_{\it 4}^2$. In terms of these coordinates the metric
(\ref{metric}) becomes
\beq
ds^{2} = \left(\frac{U}{R}\right)^{3/2}
\left( \eta_{\mu \nu} \, dx^\mu dx^\nu + f(U) d\tau^{2} \right) +
K(\rho) \, d\vec{z} \cdot d\vec{z} \,,
\eeq
where
\beq K(\rho) \equiv \frac{R^{3/2} U^{1/2}}{\rho^2}\,.
\eeq
Here $U$ is now thought of as a function of $\rho$. Finally, to
exploit the symmetries of the D6-brane embedding, it is useful to
introduce spherical coordinates $\l, \Omega_{2}$ for the
$z^{5,6,7}$-space and polar coordinates $r, \phi$ for the
$z^{8,9}$-space. The final form of the D4-brane metric is then
\beq
ds^{2} = \left(\frac{U}{R}\right)^{3/2} \left( \eta_{\mu \nu} \,
dx^\mu dx^\nu + f(U) d\tau^{2} \right) + K(\rho) \, \left(
d\lambda^2 + \lambda^2 \, d\Omega_{\it 2}^2 + dr^2 + r^2 \,
d\phi^2 \right) \,, \label{isometric}
\eeq
where $\rho^2 = \l^2 + r^2$.  In these coordinates the D6-brane
embedding takes a simple form, using $x^\mu$, $\l$ and $\Omega_{2}$ as
worldvolume coordinates (or $\xi^a,a=1, \dots 6$ collectively).  The
D6-brane's position in the 89-plane is specified as $r=r(\l)$, $\phi =
\phi_0$, where $\phi_0$ is a constant. Note that $\l$ is the only
variable on which $r$ is allowed to depend, by translational and
rotational symmetry in the 0123- and 567-directions, respectively.
Embeddings with $\tau=\mbox{constant}$ correspond to a single D6-brane
localized in the circle direction.

With this ansatz for the embedding, the induced metric on the
D6-brane, $g_{ab}$, takes the form
\beq
ds^2(g) = \left(\frac{U}{R}\right)^{3/2} \, \eta_{\mu \nu} \,
dx^\mu dx^\nu + K(\rho) \, \left[ \left( 1+\dot{r}^2 \right)
d\lambda^2 + \lambda^2 \, d\Omega_{\it 2}^2 \right] \,,
\label{induced}
\eeq
where $\dot{r}\equiv\partial_\lambda r$. The D6-brane action
becomes
\beq
S_{D6} = -\frac{1}{(2\pi)^6 \ell_s^7}
\int d^7\xi\, e^{-\phi} \sqrt{-\det g} =
-T_{\rm D6} \int d^7\xi\, \sqrt{h}
\left( 1 + \frac{\ut^3}{4\rho^3} \right)^2 \,
\l^2 \sqrt{1 + \dot{r}^2} \,,
\label{eq:D6action}
\eeq
where $T_{\rm D6}= 2\pi/g_s(2\pi \ell_s)^7$ is the six-brane
tension and $h$ is the determinant of the metric on the round unit
two-sphere.  The
equation of motion for $r(\l)$ is
\beq
\frac{d}{d\lambda} \left[\left(1+\frac{\ut^3}{4\rho^3}\right)^2
  \lambda^2 \frac{\dot{r}}{\sqrt{1+\dot{r}^2}} \right]
= -\frac{3}{2} \frac{\ut^3}{\rho^5}
\left(1+\frac{\ut^3}{4\rho^3}\right)
\lambda^2 \, r \, \sqrt{1+\dot{r}^2} \,.
\label{embedeq}
\eeq
Note that $r(\l)=r_0$, where $r_0$ is a constant, is a solution in the
supersymmetric limit ($\ut=0$), as in \cite{Karch:2002sh,
  Kruczenski:2003be}. This implies that there is no force on the
D6-brane, regardless of its position in the 89-plane. The solution
with $r_0 =0$ preserves the $\ua$ rotational symmetry in the
89-directions. If $\ut\neq 0$, the force on the D6-brane no longer
vanishes and causes it to bend, as dictated by the equation of motion
above. In this case the $U(1)_A$ symmetry is broken. In particular
there is spontaneous symmetry breaking by a quark condensate, exactly
in the same way as described in section~\ref{sectionConstableMyers}
above for the D3/D7 system.  Moreover, exactly as described in
\ref{sec:Goldstone}, there is a pseudo-Goldstone mode similar to the
$\eta'$.  Mesons with large spin in this model are studied in the
spirit of section~\ref{largespin} in \cite{Kruczenski:2004me,
  Paredes:2004is} and their decays are analyzed in
\cite{Peeters:2005fq}. Heavy-light mesons have been analyzed in
\cite{Bando:2006sp}.

For the case of multiple flavours, $N_f >1$, the authors of
\cite{Kruczenski:2003uq} present a holographic version of the
Vafa--Witten theorem \cite{Vafa:1984xg}, which states that the
$U(N_f)$ flavour symmetry cannot be spontaneously broken if $m_q
>0$. In the holographic description this is realized by the fact
that the $N_f$ D6-branes must be coincident in order to minimize
their energy.

A novel feature of \cite{Kruczenski:2003uq} is that there is a
discussion of the case when both D6 and $\rm \bar{\rm D6}$ brane
probes are present. This leads to a defect field theory in which the
fundamental degrees of freedom are confined to a $2+1$-dimensional
subspace. Nevertheless this is an important step towards the physical
model described in the next section.

\subsection{Non-abelian chiral symmetries}\label{sectionss}

The holographic models we have reviewed to date are intrinsically
supersymmetric gauge theories in the ultra-violet. This gives more
control but means only a U(1)$_A$ chiral symmetry can be realized
because of couplings between the quarks and the adjoint, scalar
super-partner of the gluons (there is a superpotential term
$\tilde{Q} \Phi Q$). To realize a more realistic non-abelian
chiral symmetry requires an analysis of an intrinsically
non-supersymmetric brane configuration - Sakai and Sugimoto
\cite{Sakai:2004cn,
  Sakai:2005yt} have provided such a candidate.

\begin{figure}
\begin{center}
\includegraphics[width=60mm]{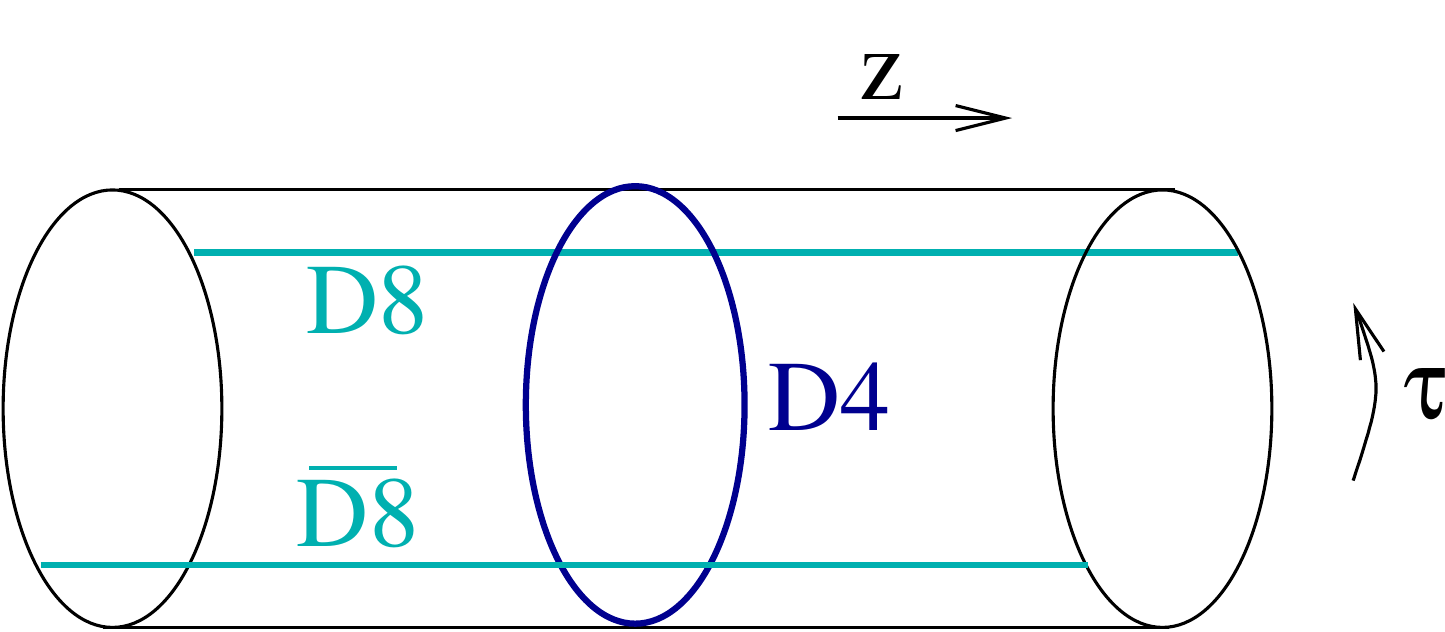} \hspace{2cm}
\includegraphics[width=60mm]{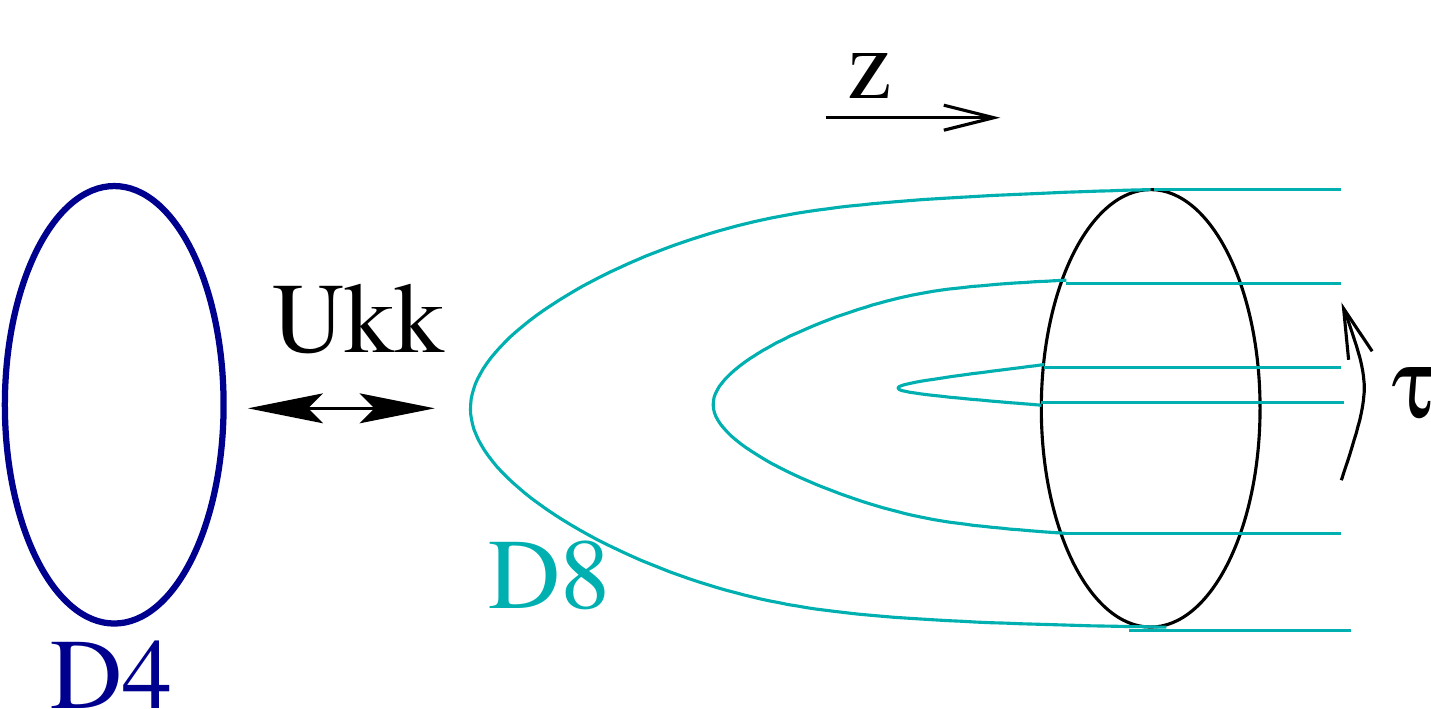}
\vspace{0.5cm}

\begin{tabular}{|r|c|c|c|c|c|c|c|c|c|c|}
\hline
& 0 & 1 & 2 & 3 & (4) & 5 & 6 & 7 & 8 & 9 \\
\hline
D4 & X & X & X & X & X & & & & & \\
\hline \raisebox{-0.5mm}{ D8 - $\rm \overline{D8}$} &
\raisebox{-0.5mm}{X} & \raisebox{-0.5mm}{X} &
  \raisebox{-0.5mm}{X} & \raisebox{-0.5mm}{X} & & \raisebox{-0.5mm}{X} &
        \raisebox{-0.5mm}{X} & \raisebox{-0.5mm}{X } & \raisebox{-0.5mm}{X
 } & \raisebox{-0.5mm}{X} \\
\hline
\end{tabular}
\end{center}
\caption{Sketch of the underlying D4-D8-$\overline{\rm D8}$-brane
construction and the chiral symmetry breaking embeddings in the
gravity dual.}  \label{d4d8per}
\end{figure}

The gauge degrees of freedom are provided, as in the model of
section~\ref{d4d6myers}, by a D4 brane with one direction wrapped on a
circle.  Quarks are included by including separated D8 branes and
anti-D8 branes.  These D8 branes fill the whole space except one
direction which is taken to be the circular direction. The model is
displayed in figure~\ref{d4d8per} - the compactified direction
is~$x^4$.

The D4-D8 (${\overline D8}$) strings generate chiral (anti-chiral) quark
fields in the gauge theory \cite{Sakai:2004cn}. The two $U(N_f)$ gauge
symmetries on the surfaces of the D8 and $\overline{\rm D8}$ branes are
interpreted as the chiral non-abelian flavour symmetries.

We should stress that the model describes a five-dimensional theory
with chiral quarks living on defects in the UV. The compactified
dimension renders a four-dimensional IR but we will not be able to
drive the compactification scale smaller than the typical scale of the
strong dynamics as one would like. Hopefully some universal features
of this class of model can teach us about the four-dimensional gauge
theory though.

We will see that the key feature of this model, when we take the
strong coupling limit to render a gravity dual, is chiral symmetry
breaking. In particular the D8 and $\overline{\rm D8}$ branes will prefer
to join into a single curved D8 brane as shown in
figure~\ref{d4d8per}. There is only one surviving SU($N$) gauge
symmetry corresponding to the chiral symmetries being broken to the
vector.  Furthermore there is a minimum D4-D8 separation so the quark
strings stretched between them have some minimal dynamically
determined mass.

\subsubsection{Gravitational background (D4-D8-$\overline{\rm {D8}}$)}

We can now consider the holographic dual of this D4-D8-$\overline{\rm D8}$
system by taking the near horizon limit of the geometry of a large
$N$ D4 brane stack wrapped on a circle (note $\alpha'$ corrections
to the metric are considered in \cite{Basu:2007yn}). We have

\beq ds^2 = \left ( \frac{u}{R} \right )^{\frac{3}{2}} \left  (
dx_4^2 +f(u) d\tau^2 \right ) + \left ( \frac{R}{u} \right
)^{\frac{3}{2}} \left ( \frac{du^2}{f(u)} +u^2 d \Omega_4^2 \right
) \eeq
with $f(u) \equiv 1- \left ( \frac{u_{KK}}{u} \right )^3$. Note here
$u$ is the holographic direction. There is a nonzero four-form flux
(not important for this discussion) and a dilaton $e^{-\phi}=g_s \left
(\frac{u}{R} \right )^{-\frac{3}{4}}$.

Note the coordinate $\tau$ is periodic with the period given by
$\delta \tau= \frac{4 \pi}{3}
\frac{R^{\frac{3}{2}}}{u_{KK}^{\frac{1}{2}}}$ forming a $S^1$ which is
wrapped by the D4 branes.  This compactification is necessary in order
to make the spacetime smooth and complete.  There is a horizon at
$u=u_{KK}$ (where the radius of the $S^1$ $\rightarrow 0$) which means
the co-ordinate $u$ is restricted to the range $[u_{KK}, \infty]$.
This scale represents the mass gap of the pure glue theory and the
block to smaller $u$ shows the theory is confining.

We will change variables to the radial coordinate $z$ where $1+z^2 =
\left ( \frac{u}{u_{KK}} \right )^3$ so the geometry becomes

\beq \begin{array}{ccl} ds^2 &  = & \left ( \frac{u_{KK}}{R}
\right )^{\frac{3}{2}} \left (\sqrt{1+z^2} dx_4^2 +
\frac{z^2}{\sqrt{1+z^2}} \; d \tau^2 \right )\\
&& \left. \right. \hspace{2cm} +\left ( \frac{R}{u_{KK}} \right
)^{\frac{3}{2}} u_{KK}^2 \left (\frac{4}{9} (1+z^2)^{-\frac{5}{6}}
\; dz^2+ (1+z^2)^{\frac{1}{6}} \; d \Omega_4^2 \right ) \,.
\end{array}\eeq

\subsubsection{Probe D8 branes}

As usual finding the full back-reacted geometry when D8 branes are
introduced is difficult so we will work in the probe limit
corresponding to quenching in the gauge theory - a good approximation
when $N_f \ll N$. The back-reaction has been addressed as an expansion
in the number of D8 branes in \cite{Burrington:2007qd} and the probe
embeddings below remain stable.

We can find the embeddings of a probe D8 brane in the above
background. These form a family of curves in the ($z$,$\;\tau$)-plane
which we parameterize as $\tau(z)$.  The Dirac-Born-Infeld (DBI)
action for the embedding is

\beq {S}_{DBI} = \int_{D8} d^8 \zeta \; e^{-\phi}  \sqrt{-
\det[\mathcal{P}\left ( g_{ab} \right )]} \,.
\eeq
This gives

\beq \label{taction} \begin{array}{ccl} {S}_{DBI}&  = &
Vol(S^4) \; \int d^4 x \; \int dz \; \frac{2}{3} g_s u_{KK}^5
\left( \frac{R}{u_{KK}} \right )^{\frac{3}{2}}
(1+z^2)^{\frac{2}{3}} \\
&& \left. \right. \hspace{4cm}\times \sqrt{1+\frac{9}{4 u_{KK}^2}
\left ( \frac{u_{KK}}{R} \right )^3 z^2(1+z^2)^{\frac{1}{3}}
\tau'(z)^2} \end{array} \,.\eeq

One finds the extremal configurations $\tau(z)$ for the D8 obey

\begin{equation}
\tau'(z)=\frac{2}{3} \left ( \frac{R}{u_{KK}} \right
)^{\frac{3}{2}} \frac{J}{\sqrt{u_{KK}^6 g_s^2 z^4(1+z^2)^2-J^2
u_{KK}^{-2} z^2(1+z^2)^{\frac{1}{3}}}} \,.
\end{equation}

Here $J=g_s u_{KK}^4 z_0(1+z_0^2)^{\frac{5}{6}}$ is chosen effectively
to make the gradient infinite at $z=z_0$.  This point is the point of
closest approach of the D8 to the horizon at $u=u_{KK}$.

This gives us a one-parameter family of embeddings where choosing a
particular value of $z_0$ specifies one particular curve.  Some
examples are shown in figure~\ref{embedders} for $z_0$ increasing in
factors of $\sqrt{10}$.  Note the curve for $z_0=0$ consists of two
horizontal pieces at $\tau= \pm \frac{\pi}{3}
\frac{R^{\frac{3}{2}}}{u_{KK}^{\frac{1}{2}}}$ plus a vertical piece at
$z=0$ connecting the two.  The vertical piece lies on the horizon.

\begin{centering}
\begin{figure*}
\begin{centering}
\includegraphics[width=80mm]{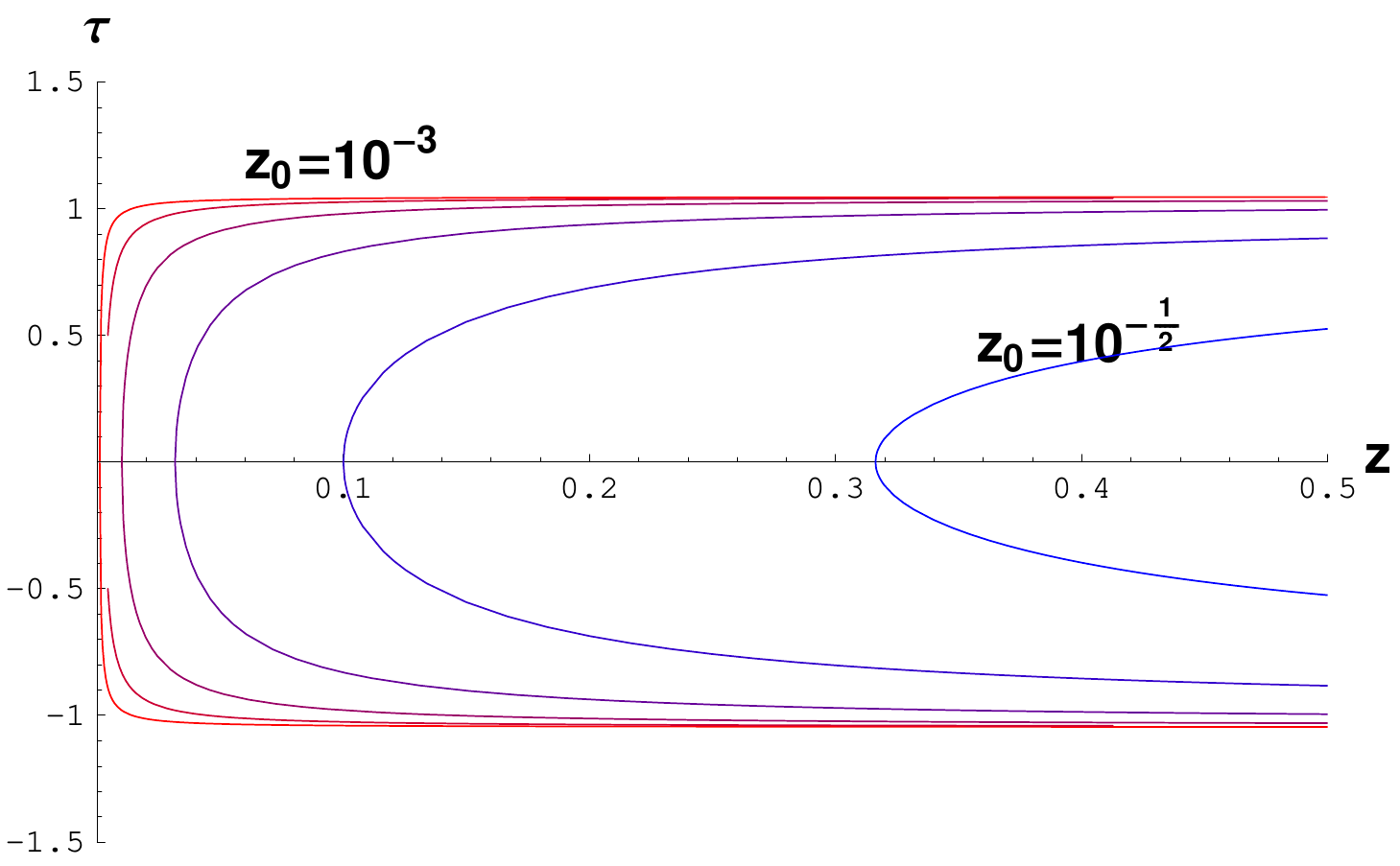}
\par
\end{centering}
\caption{Some regular D8-brane embeddings in the $z,\tau$-plane.
We have set $R=1$ and $u_{KK}=1$ for the numerical plot.}\label{embedders}
\end{figure*}
\par
\end{centering}

The large $z$ (UV) asymptotic behaviour of the solutions takes the
form

\beq  \label{asymptotic2} \tau = {\alpha} - {\beta \over z^3} \eeq
with $\alpha, \beta$ free parameters.

\subsubsection{The pion}\label{sspion}

For the moment let us restrict to discussing the $\beta=0$ solution
(we will return to the other embeddings in subsection~\ref{nonpodal}).
For this solution the D8 and $\overline{\rm D8}$ lie at anti-podal points
on the circle until the connection along the horizon at $u=u_{KK}$.
This configuration is interpreted as the theory with massless quarks
and chiral symmetry breaking on the same scale as the mass gap of the
glue. The chiral symmetry breaking should imply the existence of a
Goldstone boson - in the one flavour case this will be the equivalent
of the $\eta'$ in QCD although since we are at large $N$ the anomaly
is suppressed and the $\eta'$ behaves more like a pion.

If chiral symmetry is broken there should be a vacuum manifold with
different points corresponding to the different possible phases on the
quark condensate. In \cite{Sakai:2004cn} the phase of the quark
condensate was identified with the value of the gauge field $A_z$
living on the D8 world volume. To identify the vacuum manifold we
should find background solutions (that is, independent of the $x_4$
co-ordinates) for $A_z(z,x_4)$ which correspond to different global
choices of the phase $\pi$. $A_z$ is described by the DBI action
including a U(1) gauge field, which at low energy has the Lagrangian
density on the D8 world-volume
\beq {\cal L}= e^{-\phi} \sqrt{- \det[\mathcal{P}\left ( g_{ab}
\right )]} \; \left ( -1-\frac{1}{4} F^{ab} F_{ab} \right ) \,. \eeq

For the massless  D8-brane embedding we can take $\tau(z) = \pm
\frac{\delta \tau}{4}$ which evaluates to $\pm \frac{\pi}{3}
\frac{R^{\frac{3}{2}}}{u_{KK}^{\frac{1}{2}}}$ .  Physically, the
vertical part of the D8-brane in this case can be neglected because it
lies along the horizon where points separated in $\tau$ are
degenerate. Working on the upper branch of the D8-brane
($\tau(z)=+\frac{\pi}{3} \frac{R^{\frac{3}{2}}}
{u_{KK}^{\frac{1}{2}}}$) the action then takes the simple form
(neglecting the volume factor coming from the four-sphere angular
coordinates - we are working with states of zero spin on the $S^4$
here)
\begin{equation} \label{ssgaugeaction}
{S} = \frac{1}{2} \int_0^{\infty} dz \int d^4 x \; \left  (
e^{-\phi} \sqrt{-g} g^{zz} g^{11} \right ) \; \left (-(\partial_0
A_z)^2+(\partial_1 A_z)^2+(\partial_2 A_z)^2+(\partial_3 A_z)^2
\right ) \,.
\end{equation}

It is apparent that $F^{ab}$ and hence the action vanishes if $A_z$ is
the only non-zero field and if it is only a function of $z$. Any
function of $z$ is allowed. This is an artifact of gauge freedom in
the model and one should pick a gauge. For example one could gauge fix
by including a term

\beq \delta {\cal L} = {1 \over \xi} \; e^{-\phi} \sqrt{-
\det[\mathcal{P}\left ( g_{ab} \right )]} \; \left ( \nabla_a A^a -
\kappa(z)\right )^2 \,,  \eeq where $\kappa(z)$ is any arbitrary
function. Writing $A_z(z,x_4) \equiv g(z) \pi(x_4)$, there is
sufficient freedom to pick any functional form of $g(z)$. We will
follow the choice of Sakai and Sugimoto and pick
\beq g(z) = \frac{\rm \cal{C}}{1+z^2} \,. \eeq

The solution contains the arbitrary multiplicative factor ${\cal C}$
since the action is only quadratic in $A^z$. The freedom to pick the
constant ${\cal C}$ in this solution is the freedom to move on the
vacuum manifold.

We can now identify the pion field. It should correspond to space-time
($x^\mu$) dependent fluctuations around the vacuum manifold. In other
words we look at solutions of the form
\beq A_z(z,x) = \pi(x_4) \times \frac{2}{\sqrt{3 \pi}} \;
\frac{1}{1+z^2} \,. \eeq

Substituting this into the action (\ref{ssgaugeaction}) we find a canonically
normalized kinetic term for a massless field,
\beq {S} = \int d^4 x {1 \over 2} (\partial^\mu \pi)^2 \,. \eeq

This is the pion - the Goldstone mode of the chiral symmetry breaking.
The non-abelian partners of this state are discussed in
\cite{Gao:2007sh}. Note that interchanging the D8 and $\overline{\rm D8}$
branes corresponds to interchanging left and right handed quarks and
is therefore a manifestation of parity in the model. This state has
negative parity and is hence a pseudo-scalar.

\subsubsection{Meson spectrum and interactions}

Fluctuations of the D8 branes about the embeddings discussed
correspond to mesons of the gauge theory. Generically one looks for
solutions of the linearized field equations coming from the DBI action
of the form $f(u) e^{ikx}$.  Even and odd
functions $f(r)$ describe even and odd parity states.

Fluctuations of the vector field in the DBI action generate vector
and axial vector mesons (note that the links to the ideas of a
hidden local symmetry in QCD are made in \cite{Harada:2006di}). In
addition there is a scalar field corresponding to fluctuations of
the embedding. If we restrict these fluctuations to the trivial
harmonic of the four-sphere on the D8 transverse to the $x$
directions, we obtain QCD-like states. It is important to realize
there are additional states with higher harmonics that effectively
have R-charge indicating that there are light non-degenerate
``super-partners'' of the QCD fields in the field theory. There
are in addition fermionic fields in the DBI action that would
describe mesinos if supersymmetry were restored
\cite{Heise:2007rp}. Finally there are also Kaluza Klein modes of
the glueballs and gluino balls from the gauge sector. The typical
scale for the masses of all of these bound states is \beq M_{KK} =
\frac{3}{2} \frac{U_{KK}^{1/2}}{R^{3/2}} \,, \qquad R^{3/2} =
\sqrt{\pi g_s N l_s^3} \,. \eeq

Note that as in previous examples, the mesons are tightly bound in
the limit $g_s N \rightarrow \infty$, and hence rather
un-QCD-like. The values of the masses for states that can be
mapped to QCD have been computed in \cite{Sakai:2004cn,Sakai:2005yt}.
They find

\begin{center}
\begin{tabular}{ccccc}
$m_\rho$ &  0.67 $M_{KK}$,&&$m_{a_1}$ &  1.58 $M_{KK}$,\\
$m_\rho^*$ &  1.89 $M_{KK}$,&&$m_{a_1^*}$ & 2.11 $M_{KK}$,\\
$m_\rho^{**}$ & 2.21 $M_{KK}$.&&&\end{tabular}
\end{center}

The interactions between mesons can also be computed by inserting
the functions $f(r)$ back into the DBI action and integrating over
the four non-spatial directions on the D8. Some example values are

\begin{center}
\begin{tabular}{l}
$f_\pi^2 = \frac{1}{54 \pi^4} g^2_{YM}N^2 M^2_{KK} $\,,\\
\\
$g_{\rho \rho \rho} = 0.45 \frac{ (6 \pi)^{3/2}}{g_{YM} N}$\,,\\
\\
$g_{\rho \pi \pi} = 0.42 \sqrt{ \frac{216 \pi^3 }{g_{YM}^2 N^2}}$\,.
\end{tabular}
\end{center}

If one forces this model onto the QCD spectrum by fitting the
scale $M_{KK}$ and $g^2_{YM}N$,  then these results match the data
at the 20-30$\%$ level.

Purely pionic interaction terms exist as well which reproduce a
Skyrme style model of baryons. There has been considerable interest in
the baryonic sector of the model recently - see the  papers
\cite{Nawa:2006gv,Hata:2007mb,Nawa:2007gh,Hong:2007kx,Hong:2007tf,
Hong:2007ay,Domokos:2007kt,Imaanpur:2007qq}. See also work on introducing a
Fermi surface at high density \cite{Kim:2007xi,Rozali:2007rx}.

\subsubsection{Non-anti-podal embeddings}\label{nonpodal}

Finally we return to the non-anti-podal embeddings of figure
\ref{embedders}. Clearly these embeddings have a larger D4-D8
separation and hence a larger quark mass - there remains debate in the
literature about whether this mass is a hard mass \cite{Evans:2007jr}
or entirely dynamically generated \cite{Antonyan:2006vw,Antonyan:2006pg,
Basu:2006eb}.

The configurations differ in their asymptotic positions of the D8
branes. Perturbatively the configuration of a D8 and $\overline{\rm D8}$ on
a circle is not generically stable due to their attraction and they
would be expected to join, suggesting chiral symmetry breaking is
present at weak coupling.  Fluctuations of the D8s in the UV though
correspond to strings with both ends on one D8 and are hence dual to
operators in the adjoint of the U($N_f$) chiral flavour symmetries.
For example they could correspond to the coupling and source of a
(possibly higher dimension) operator of the form $\bar{q}_L \gamma^\mu
D_\mu q_L$.  Clearly dialing this coupling in the UV Lagrangian would
enhance the gluon exchange diagram between quarks and might well
increase the dynamically generated mass.  On the other hand if the
true parameter that is being changed is the quark mass then
dynamically that could feed through to set a different value for the
same operator's coupling and vev. The change in position of the D8 may
be an indirect signal of the presence of a quark mass.

One would think that the difference between hard and explicit chiral
symmetry breaking should be evident from the existence or otherwise of
a flat direction in the potential. For all of these configurations the
gauge freedom discussed in section~\ref{sspion} above remains so the
analysis there shows there is a flat direction.  However, in the full
string construction the quark mass is a field vev and there should be
a larger spurious symmetry of the form
\beq
q_l \rightarrow e^{i \alpha} q_L,\hspace{0.5cm} q_R
\rightarrow e^{-i \alpha} q_R, \hspace{0.5cm} m \rightarrow e^{-2
i \alpha}m \,.
\eeq
If the flat direction corresponds to this symmetry then fluctuations
in this direction are not physical modes in the gauge theory in which
the phase of the mass is fixed.

Another approach taken has been to include an explicit tachyonic mode
connecting the D8 and $\overline{\rm D8}$ in \cite{Casero:2007ae,
  Dhar:2007bz, Bergman:2007pm}. This field should directly describe
the quark mass and condensate and its vacuum solution does indeed set
the shape of the linked D8-$\overline{\rm D8}$ pair.  The precise tachyon
potential is not known though.

In practice for the spectrum (the pion apart) it is not too important
whether the mass is dynamical or hard. All the meson masses rise as
the D8s are brought closer together asymptotically.

An alternative attempt to introduce a quark mass by the introduction
of an instanton on the D8 world-volume can be found in
\cite{Hashimoto:2007fa}.

\subsection{More chiral symmetry breaking}

A number of other examples of holographic chiral symmetry breaking
exist in the literature. In \cite{Evans:2005ti} probe D7 branes were
numerically embedded in the non-supersymmetric Yang-Mills$^*$
deformation \cite{Babington:2002qt} of the $\N=4$ theory providing
evidence for chiral symmetry breaking. Quarks have also been added to
the beta-deformed $\N=4$ theory in \cite{Penati:2007vj} - chiral
symmetry breaking is again observed.  It appears to be generically
true that breaking supersymmetry in gravity duals leads to chiral
symmetry breaking.

Attempts have been made to construct gravitational duals of QCD in
non-critical string theories.  These gravity theories in less than ten
dimensions risk the presence of order one curvature and so are not
completely controlled. Nevertheless, the AdS$_6$ Schwarzschild black
hole \cite{Kuperstein:2004yk} is a possible candidate and shows
confining behaviour.  In \cite{Casero:2005se} quarks are introduced in
the spirit of section~\ref{sectionss} above via both a D4-$\overline{\rm
  D4}$ and D5-$\overline{\rm D5}$. Chiral symmetry breaking is observed in
the pattern of the Sakai-Sugimoto model and the vector meson masses
have been computed. A non-critical D3-D4-$\overline{\rm D4}$ configuration
is discussed in \cite{Cotrone:2007gs} and again displays chiral
symmetry breaking.

\subsection{Summary}

We have reviewed a number of holographic descriptions of chiral
symmetry breaking. Quarks are introduced via probe branes in
non-supersymmetric geometries. In each case, the repulsion of the core
geometry acting on the probes causes the spontaneous symmetry
breaking, which appears as a manifest breaking of a symmetry in the
geometrical set-up. The quarks acquire a dynamical mass resulting in a
non-zero vector meson mass even at zero quark mass. Goldstone bosons
of the symmetry breaking play the role of the pions of QCD. These are
all crucial dynamical results in view of holographically describing
QCD.

\newpage

\section{Mesons at finite temperature } \label{finiteT}
\setcounter{equation}{0}\setcounter{figure}{0}\setcounter{table}{0}

In previous sections we have focused on strongly-coupled gauge theories
at zero temperature. Considerable progress has also been made on
understanding the thermal properties of gauge theories using holography.

The gravitational dual of placing the $\N=4$ gauge theory at finite
temperature is to replace the AdS space with an AdS-Schwarzschild
black hole \cite{Witten:1998qj,Witten:1998zw}.  The black hole has all
the correct thermodynamic properties to describe the thermodynamics of
the gauge theory. Further, the horizon cuts off the holographic radial
direction corresponding to cutting off energy scales below that of the
temperature in the field theory.  In the infinite volume limit, the
free energy of the black hole solutions is lower than that of AdS with
a compact time direction for any temperature. If the spatial
directions of the theory are also compact, the transition between AdS
space and the AdS-Schwarzschild black hole can be shifted to higher
temperatures of order the inverse compactification scale. This
gravitational tunnelling transition was first described by Hawking and
Page \cite{Hawking:1982dh}.  Witten has interpreted it, within the
gauge-gravity duality, as the dual of the deconfinement transition.
The free energy of AdS scales as order one relative to the black hole
geometry's free energy which scales as $N^2$ - the high temperature
phase has deconfined gluons (and superpartners).

A considerable amount of work has been done on the holographic
description of the transport properties of the quark-gluon plasma.
Amongst these is the famous ratio of shear viscosity to volume density
of entropy which takes the value $\hbar/4\pi k_B$
\cite{Policastro:2001yc}.  This corresponds to the ``fluid'' with the
lowest known value of this ratio. There has been considerable interest
in this quantity, since the value of this ratio deduced from RHIC
heavy-ion collisions suggests that the quark-gluon plasma is an almost
perfect fluid of this type \cite{Shuryak:2006se}.

Here we will constrain ourselves to reviewing results on the thermal
properties of mesons using the AdS/CFT correspondence
\cite{Babington:2003vm, Mateos:2007vn, Ghoroku:2005tf,Albash:2006ew}.

\subsection{First order phase transition in the quark-gluon plasma}

An interesting new first order phase transition has been found which
occurs as the temperature increases and passes through the meson's
mass scale. At this scale the meson disassociates, or melts, into the
background plasma. This is an additional transition at energy scales
above the deconfinement scale.  The sharp transition is probably a
consequence of being at large $N$ and is not believed to be present in
QCD on the basis of lattice results
\cite{Fodor:2007sy,Karsch:2007dp,Karsch:2007dt}. Nevertheless, the gauge/gravity
dual description does allow the study of meson melting.

\label{adsbhsection}

\subsubsection{AdS-Schwarzschild solution}

The high temperature, deconfined, phase of the $\N=4$ gauge theory is
described by the AdS-Schwarzschild solution, given by
\begin{equation}
\label{bh}ds^2 = \frac{K(r)}{R^2} d\tau^2 + R^2 {\frac{dr^2}{ K(r)}} +
\frac{r^2}{R^2} d\vec x^2 + R^2 d\Omega_5^2 \, ,
\end{equation}
where
\begin{equation} K(r) = r^2 - {\frac{r_H^4}
{r^2}} \,.
\end{equation}
Asymptotically for $r \gg r_H$, the black hole solution approaches
$AdS_5 \times S^5$ whose radius is related to the 't~Hooft coupling of
the dual gauge theory by $R^4= 4 \pi \lambda \alpha'^2$. This spacetime is
smooth and complete if $\tau$ is periodic with period $\pi r_H$. Note
that the $S^1$ parameterized by $\tau$ collapses at the horizon
$r=r_H$.  The fact that the geometry ``ends'' at $r=r_H$ is
responsible for the existence of an area law for the Wilson loop and a
mass gap in the dual field theory (see \cite{Witten:1998zw}). For
convenience, in the numerical work below we shall set both $R$ and
$r_H$ equal to 1.

The temperature of the field theory corresponds to the Hawking
temperature of the black hole which is given by the radius of the
horizon, $T=r_H/(R^2 \pi)$. At finite temperature the fermions have
anti-periodic boundary conditions in the Euclidean time direction
\cite{Witten:1998zw} and supersymmetry is broken.  The black hole
solution thus describes a strongly interacting quark-gluon plasma
which is non-supersymmetric and non-conformal.  It is therefore
believed that, despite the presence of other fields not contained in
QCD, this plasma shares some properties with the quark-gluon plasma of
QCD.

As in the previous sections, we now introduce a D7-brane into this
background, which corresponds to the addition of matter in the
fundamental representation.  The dual field theory is the $\N=2$
gauge theory discussed in section~\ref{sec21}, but now at
 finite temperature.

\subsubsection{Embedding of a D7-brane}

To embed a D7-brane in the AdS black-hole background it is
useful to recast the metric (\ref{bh}) to a form with an explicit
flat 6-plane. To this end, we change variables from $r$ to $w$,
such that
\begin{equation}
{\frac{dw}{ w}}\equiv {\frac{r dr}{ (r^4-r_H^4)^{1/2}}} \, ,
\end{equation}
which is  solved by
\begin{equation}
2 w^2=r^2+\sqrt{r^4-r_H^4} \,.
\end{equation}
The metric is then
\begin{equation} \label{BHmetric}
ds^2=\left(w^2+\frac{w_H^4}{w^2}\right)d\vec
x^2+\frac{(w^4-w_H^4)^2}{w^2(w^4 +w_H^4)}dt^2+
\frac{1}{w^2}(\sum_{i=1}^6 dw_i^2) \, ,
\end{equation}
where $\sqrt{2}w_H=r_H$, $\sum_i dw_i^2 = dw^2 + w^2 d\Omega_5^2$,
which for reasons of convenience will also be written as $d\rho^2 +
\rho^2d\Omega_3^2 + dw_5^2 + dw_6^2$ where $d\Omega_3^2$ is the unit
three-sphere metric.  The AdS black hole geometry asymptotically
approaches $AdS_5 \times S^5$ at large $w$.  Here the background
becomes supersymmetric, and the D7 embedding should approach the
constant solutions $w_6 = m = const., w_5 = 0$ found in
section~\ref{flavour}.  To take into account the deformation, we will
consider a more general ansatz for the embedding of the form $w_6 =
w_6(\rho), w_5=0$, with the function $w_6(\rho)$ to be determined
numerically.  The DBI action for the orthogonal directions $w_5, w_6$
is
\begin{align} \label{DBIBH}
S_{D7} &=-\mu_7\int d^8\xi ~\epsilon_3 ~ {\cal G}(\rho,w_5,w_6) \\
&\times \left( 1 + { g^{ab} \over (\rho^2 + w_5^2 + w_6^2)} \partial_a w_5
\partial_b
w_5 + { g^{ab} \over (\rho^2 + w_5^2 + w_6^2)} \partial_a w_6 \partial_b
w_6 \right)^{1/2} \,,\nonumber
\end{align}
where the determinant of the metric is given by
\begin{align}
{\cal G}(\rho,w_5,w_6)
&= \sqrt{ g_{tt} g_{xx}^3 \rho^6 \over (\rho^2 + w_5^2
+ w_6^2)^4} \nonumber\\
&= \rho^3 { ( (\rho^2 + w_5^2 + w_6^2)^2 + w_H^4) ( ( \rho^2
+ w_5^2 + w_6^2)^2 - w_H^4) \over (\rho^2 + w_5^2 + w_6^2)^4} \,.
\end{align}

With the ansatz $w_5=0$ and $w_6 = w_6(\rho)$, the equation of
motion becomes
\begin{equation}
\label{eqnmot} {d \over d\rho} \left[ {\cal G}(\rho,w_6)
\sqrt{ 1 \over 1 + \left({  d w_6 \over d \rho}\right)^2} { d w_6
\over d \rho} \right] - \sqrt{1 + \left({  d w_6 \over d
\rho}\right)^2} {8 w_H^8 \rho^3 w_6 \over ( \rho^2 + w_6^2)^5} = 0
\,.
\end{equation}
The solutions of this equation determine the induced metric on the D7
brane which is given by
\begin{equation} ds^2 = \left( \tilde w^2 +
  \frac{w_H^4}{\tilde w^2} \right) d\vec x^2 + \frac{(\tilde w^4 -
  w_H^4)^2}{\tilde w^2 (\tilde w^4 +w_H^4)} dt^2 + \frac{1+(\partial_\rho
w_6)^2}{\tilde w^2} d
\rho^2
+ \frac{\rho^2}{\tilde w^2} d\Omega_3^2 \, , \label{D7metric}
\end{equation}
with $\tilde w^2=\rho^2+w_6^2(\rho)$. The D7-brane metric becomes
$AdS_5 \times S^3$ for $\rho \gg w_H,m$.

\subsubsection{First order phase transition at finite temperature}

We now compute the explicit D7-brane solutions.
The UV asymptotic (large $\rho$) solution, where the geometry returns
to $AdS_5\times S^5$,  is of the form
\begin{equation}
\label{asymp} w_6(\rho) \sim m + \frac{c}{\rho^2} \,.
\end{equation}
The parameters $m$ and $c$ have the interpretation as a quark mass and
bilinear quark condensate $\langle \bar \psi \psi \rangle$,
respectively, as discussed in section~\ref{sectioncsb}.  These parameters
can be taken as the boundary conditions for the second order
differential equation (\ref{eqnmot}), which are solved using a numerical
shooting technique.  Of course the physical solutions should not have
arbitrary $m$ and $c$. For a given value of $m$, $c$ is fixed by requiring
regularity throughout the space.

\begin{figure}[!ht]
\begin{center}
\includegraphics[height=6cm,clip=true,keepaspectratio=true]{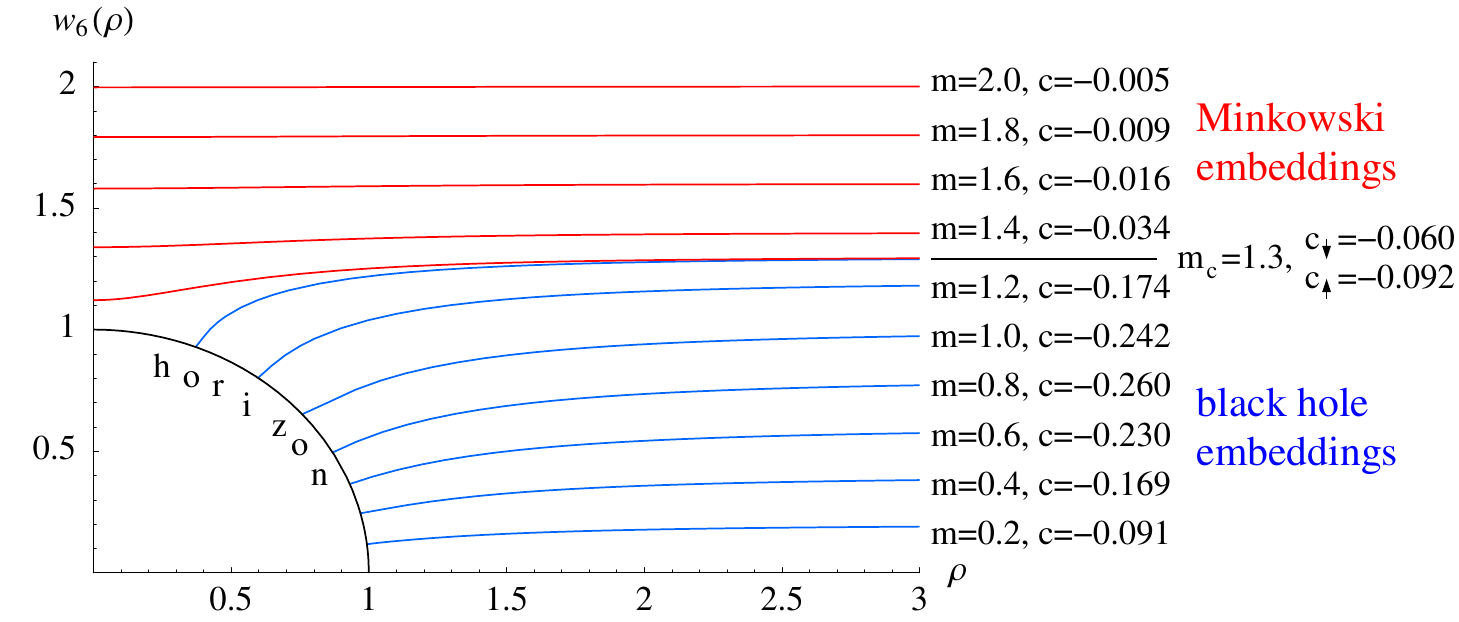}
\caption{Two classes of regular solutions in the AdS black hole
  background. The quark mass $m_q$ is the parameter $m$ in units of
  $\Lambda \equiv \frac{w_H}{2\pi\alpha'}$: $m_q=m \Lambda$. We set
  $\Lambda=w_H=1$. }\label{quiv2}
\end{center}
\end{figure}

The numerical solutions are illustrated in figure~\ref{quiv2} for
several choices of $m$.  We choose units such that the horizon is
represented as a quarter circle with radius $w_H=1$.

As can be seen from the figure, there are two qualitatively different
D7-brane embeddings.  At large quark masses the D7-brane tension is
stronger than the attractive force of the black hole. The D7-brane
ends at a point outside the horizon, $\rho =0$, $w_6 \ge w_H$, at
which the $S^3$ wrapped by the D7-brane collapses (see
(\ref{D7metric})). Such a D7-brane solution is called a {\em
Minkowski} embedding. They behave very similarly to the supersymmetric
solutions in $AdS_5 \times S^5$. As the mass decreases, there exists a
critical value of the mass $m=m_{\rm crit}\approx 0.92$ such that
$w_6(\rho=0)=w_H$. For smaller masses the D7-brane is forced to fall
into the black hole horizon, {\em i.e.}\ the D7-brane ends at the
horizon $w=w_H$ at which the $S^1$ of the black hole geometry
collapses. This is a so-called {\em black hole} embedding.

From a geometrical point of view the two classes of embeddings differ
by their topology: The D7-topology is $R^3 \times B^4 \times
S^1$ for Minkowski and $R^3 \times S^3 \times B^2$ for black hole
solutions. The appearance of a change in the topology of the embedding
at $m_{\rm crit}$ points to a phase transition in the dual field
theory at exactly this critical value of the quark mass.

In fact, this embedding behaviour is a specific example of the more
general problem of embedding a brane of arbitrary dimension in a black
hole geometry, as studied in \cite{Frolov:1998td}.  Expanding the
embedding equation near the horizon, it was shown that the equations
have a self-similarity which implies that for a given range of $m$, there are
an infinite number of embeddings. 

\begin{figure}[!ht]
\begin{center}

\includegraphics[height=4.5cm,clip=true,keepaspectratio=true]{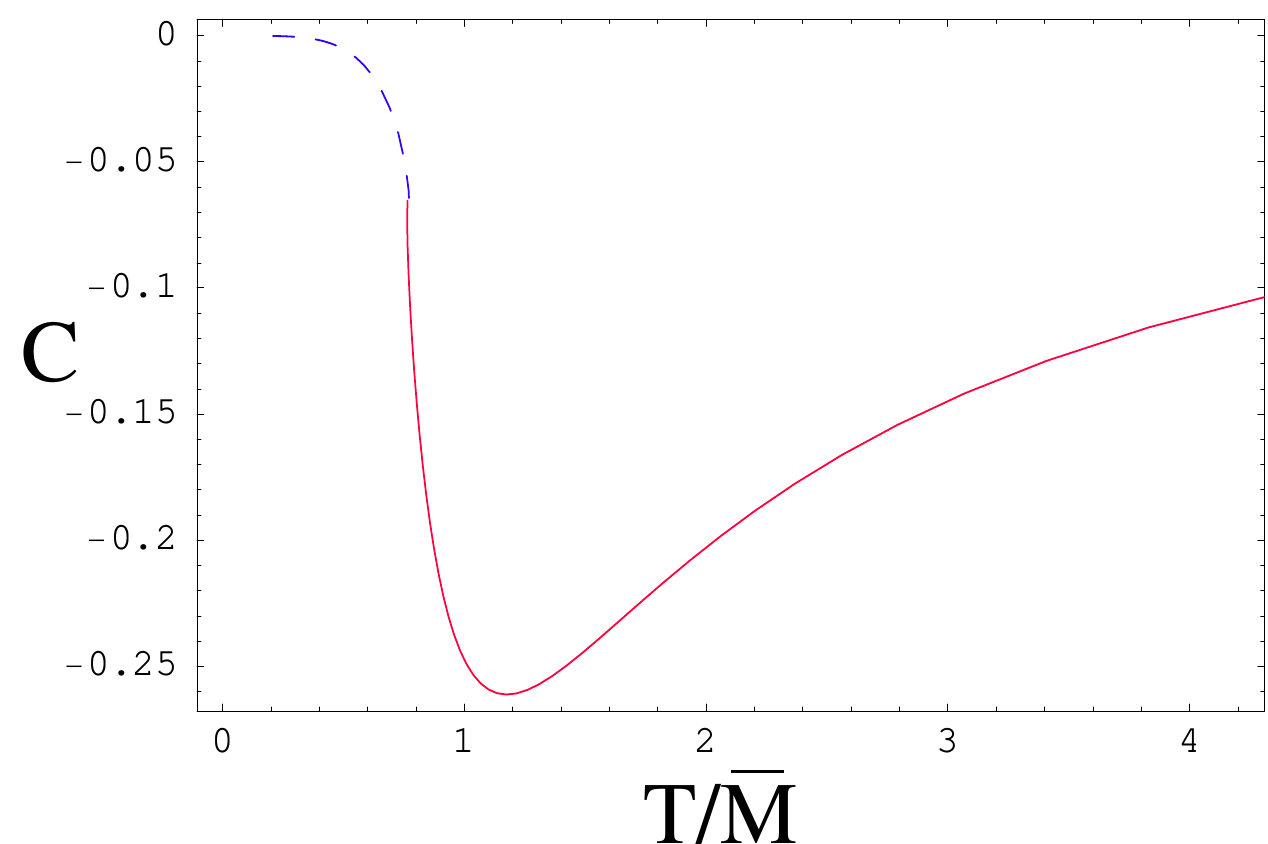}
\hspace{1cm}
\includegraphics[height=4.5cm,clip=true,keepaspectratio=true]{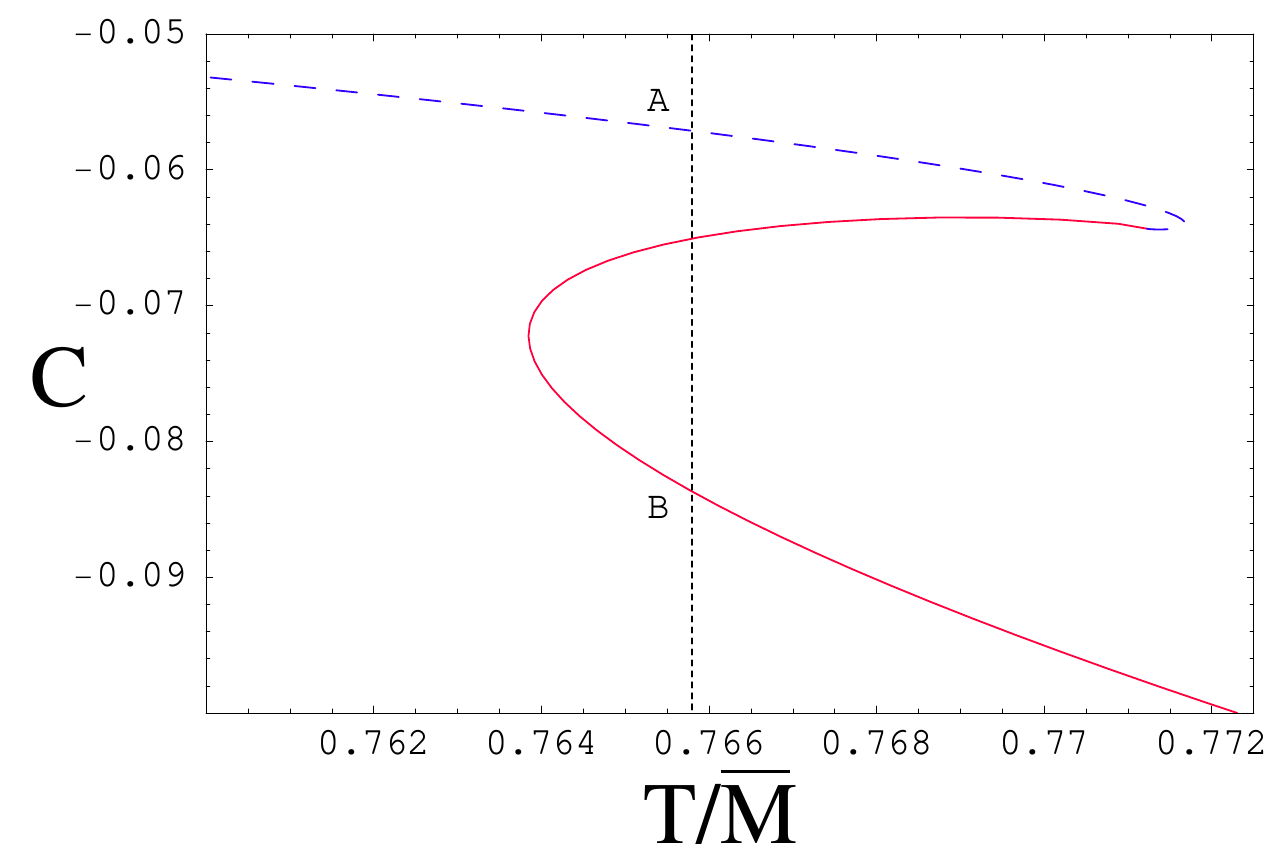}

\includegraphics[height=4.3cm,clip=true,keepaspectratio=true]{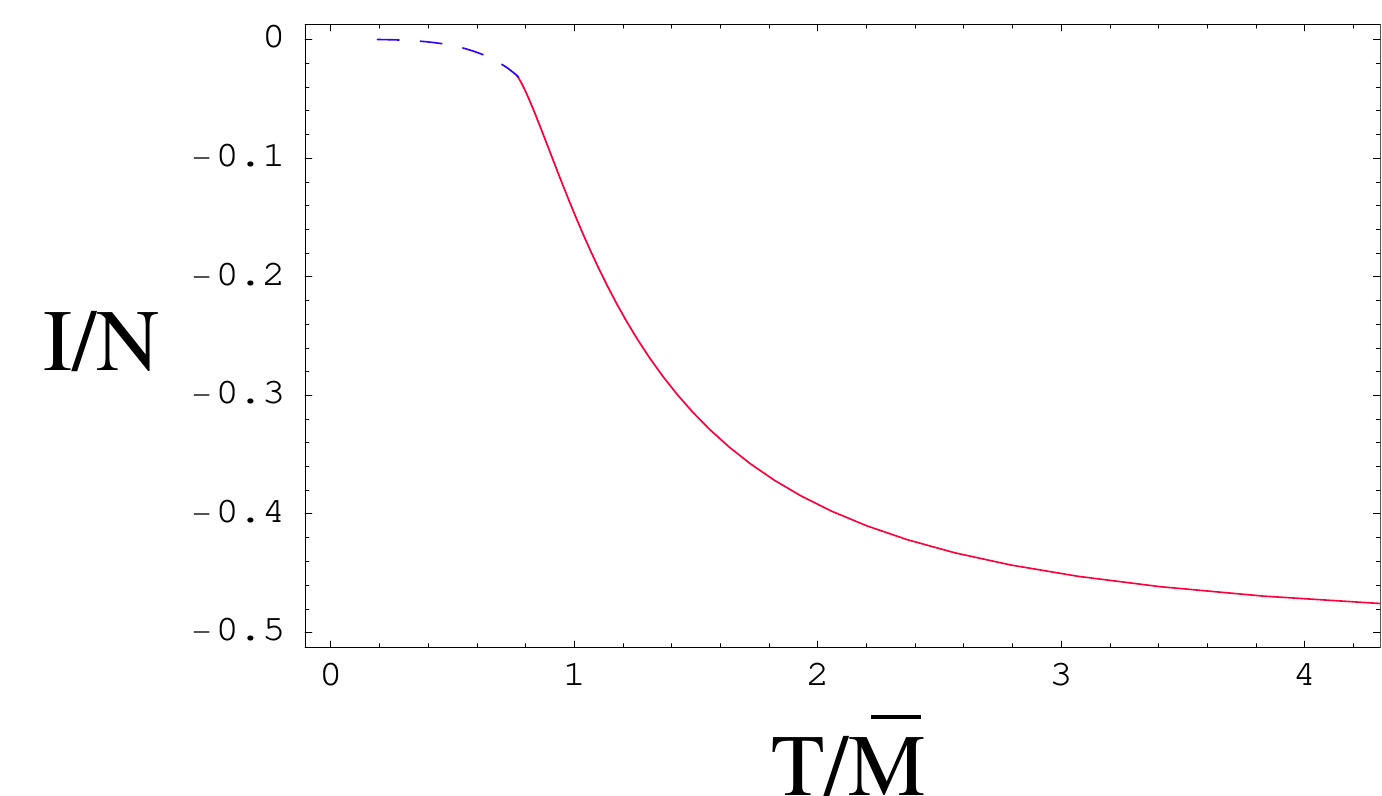}
\hspace{0.8cm}
\includegraphics[height=4.5cm,clip=true,keepaspectratio=true]{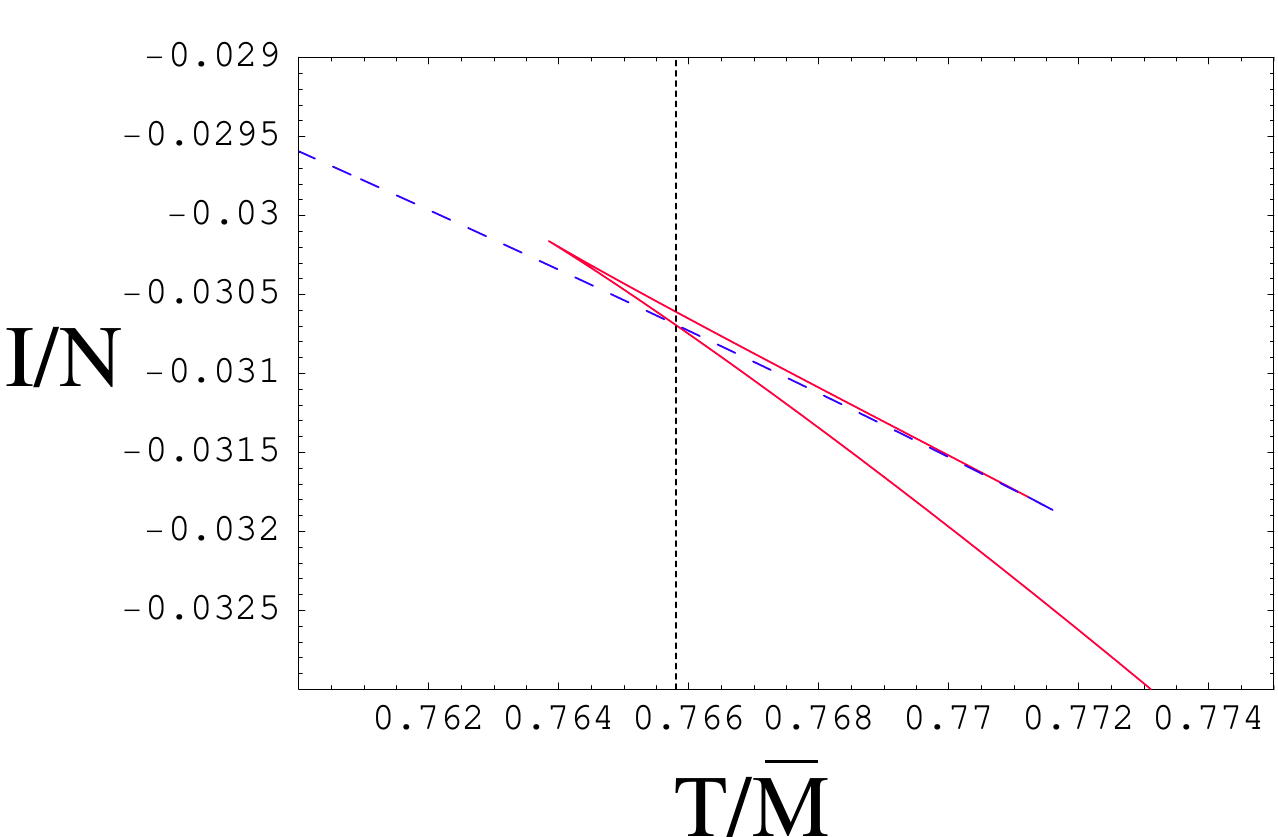}
\hspace{0.6cm} \mbox{}
\end{center}

\caption{Plots of the parameter $c$ vs $1/m \equiv T/\bar{M}$ for the
  regular solutions in AdS-Schwarzschild as given in
  \cite{Mateos:2007vn}. A close-up of the transition point is also
  shown displaying the first order transition. The action of the
  solutions is also plotted to determine the transition point. Note
  $\cal{N}$ is a normalization coefficient - see
  \cite{Mateos:2007vn}. Figures kindly provided by D.~Mateos, R.~Myers and
  R.~Thomson.  }\label{quivit}
\end{figure}

The dependence of the condensate on the mass is illustrated in
figure~\ref{quivit}.  At $m=0$ the condensate $c$ is zero (the brane
lies flat), so there is no chiral symmetry breaking in this gauge
theory. As $m$ increases, the condensate $c$ initially increases and
then decreases again. At sufficiently large $m$, the condensate
becomes negligible, which is to be expected as the D7-brane ends in
the region where the deformation of AdS is small.  Recall that there
is no condensate in the Yang-Mills theory with unbroken ${\mathcal N}
=2$ supersymmetry described by D7-branes in un-deformed AdS. Once
supersymmetry is broken by the temperature and the chiral symmetry is
broken by the quark mass, it would be surprising if a condensate were
not present though.

Since the D7-brane topology changes as $m_{\rm crit}$ is crossed, one
might expect a phase transition to occur at this point.  Zooming in
around $m_{\rm crit}$, we see in figure~\ref{quivit} that $c$ is
multi-valued around the critical mass $m_{\rm crit}$ as expected.
This means that for a given quark mass in the regime $1.295 \leq m
\leq 1.308$ there exist both Minkowski and black hole embeddings.
These solutions have the same quark mass $m$ but a different value of
the quark condensate $c$.

The $c$ vs.~$1/m$ plot can also be considered as a plot of the
condensate $c$ versus the temperature, since all dimensionful
quantities are normalized by the temperature by setting $r_H=1$. For
this we keep the quark mass $m$ fixed and vary the horizon $w_H \sim
T$.  Then, for small temperatures we recover the Minkowski embeddings,
while for high temperatures we have black hole embeddings. Heating up
the system from zero temperature, we eventually reach a critical
temperature $T_{\rm crit}$ at which further supply of external energy
does not increase the temperature of the system. It rather leads to
the formation of a quark condensate.  The jump in the quark condensate
shows that the phase transition is discontinuous and thus of first
order. The phase transition occurs in the deconfined phase of the
field theory at a temperature $T_{crit} > T_{deconf}$.

\subsection{Mesons in the AdS black-hole background}

\begin{figure}[!ht]
\begin{center}
\includegraphics[height=9cm,clip=true,keepaspectratio=true]{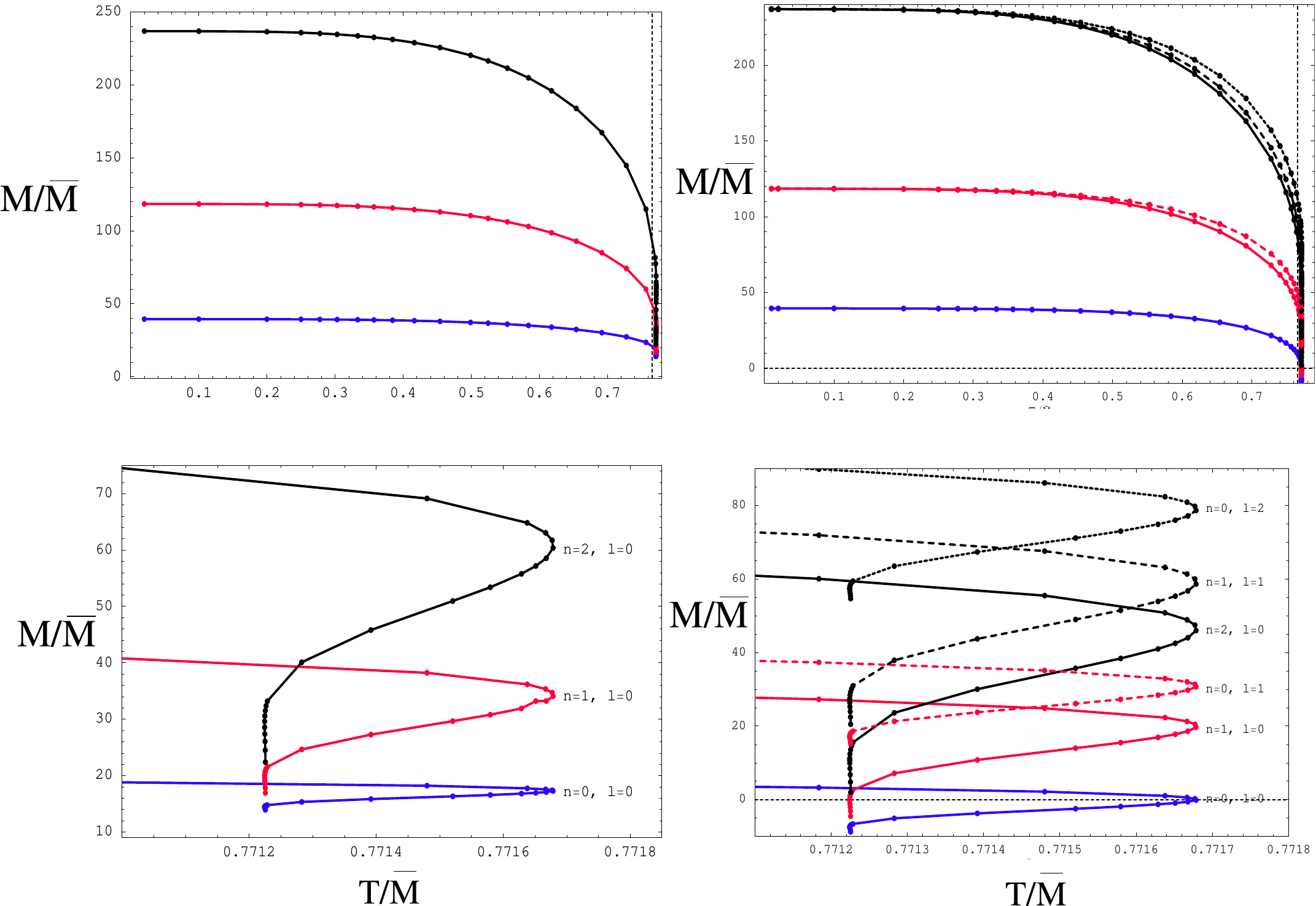}
\caption{Plots of meson masses in units of the quark mass vs $1/m
  \equiv T/\bar{M}$ for the Minkowski embeddings of the D7 brane 
  as found in \cite{Mateos:2007vn}.  The figures on the left are for
  fluctuations of the D7 brane in the angular direction in the
  $w_5-w_6$ plane. Those on the right for radial fluctuations. In each
  case the lower figure is a close-up of the transition
  region. Figure kindly provided by D.~Mateos, R.~Myers and 
R.~Thomson.}\label{thermalmass}
\end{center}
\end{figure}

The true physical nature of the phase transition corresponding to the
D7 branes switching from a Minkowski to a black hole embedding is
revealed through the behaviour of the mesons.

In the Minkowski phase, ({\em i.e.}\ when the D7 brane probe does not
reach the black hole horizon) the meson spectrum is similar to that in
the zero temperature theory. One can study perturbations of the D7
brane about the background embedding of the form $f(r) e^{-iwt},
w^2=M^2$ corresponding to stationary mesons. Requiring regularity for
$f(r)$ determines the allowed meson masses $M$. Plots (taken from
\cite{Mateos:2007vn}) of the masses of the mesons associated with
angular fluctuations in the $w_5-w_6$ plane and radial fluctuations in
that plane are shown in figure~\ref{thermalmass}. As the mass
approaches the critical value of $m$ the meson masses fall and the
lowest radial mode becomes tachyonic.

If we now move to the other side of the transition, in the black hole
phase, when the D7 brane probe terminates on the horizon, the mesons
become unstable and decay. In this case, there are no regular mesonic
fluctuations with real masses. Instead the black hole supports
quasi-normal modes - fluctuations of the D7 that are purely infalling
waves at the horizon.  The mass that is extracted from these solutions
is complex.  The interpretation is that the mesons are not stable in
the thermal plasma, and `melt' into it with a characteristic decay
width given by the imaginary part of the quasi-normal eigenfrequency.

This is nicely described in \cite{Hoyos:2006gb}.  An ansatz for D7
fluctuations of the form $f(\rho) e^{-( i \omega t + k.x)} $ is again
used. For the quasi-normal modes, the frequency $\omega$ develops a
negative imaginary part, which provides a damping and corresponds to
the decay width of the meson. The quasi-normal modes are eigenmodes
with infalling boundary condition at the black hole horizon.

In \cite{Hoyos:2006gb} the spectrum of scalar fluctuations of the D7
brane around its minimal-energy embedding was analyzed for a range of
quark masses for fluctuations with zero spatial momentum.  We
linearize the equation of motion obtained from the DBI action for
fluctuations of the D7 brane around the equilibrium configuration.

Consider embedding the D7 on the three-sphere within the five-sphere
parameterized as
\begin{equation}
d\Omega_5^2 = d\theta^2 +
\sin^2 \theta d \psi^2 + \cos^2 \theta d\Omega_3 ^2 \,.
\end{equation}
In the special case of zero quark mass the D7 embedding is trivial,
lying at $\theta=0$ for all $r$.

We consider fluctuations of the embedding in the $\theta$ direction of
the form $ \theta(r) e^{-i \omega t +i k \cdot x}$. Expanding the DBI
action to quadratic order in $\theta$ leads to the eigenvalue equation
in the variable $z \equiv \frac{r_{\rm H} }{r}$,
\begin{equation}
\theta''-\frac{3+z^2}{z(1-z^4)} \theta'+\frac{3}{z^2(1-z^4)} \theta
+ \frac{\Omega^2}{(1-z^4)^2} \theta -\frac{k^2}{(1-z^4)} \theta=0 \, ,
\end{equation}
where $\Omega = \omega R^2/r_H$.
In the UV (asymptotically AdS) limit ($z \rightarrow 0$) the solution
is a linear combination of $z^1$ and~$z^3$.  The latter is the
normalizable mode and corresponds to a field theory quark bilinear via
the AdS/CFT dictionary.

In the IR (near-horizon) limit ($z \rightarrow 1$) the solution is a
linear combination of $(1-z)^{+\frac{i \; \Omega}{4}}$ and
$(1-z)^{-\frac{i \; \Omega}{4}}$.  The solution with the negative
exponent corresponds to a purely infalling wave.

In this case the eigenvalue problem can be solved using a method known
in the GR literature as Leaver's method \cite{Leaver:1985ax} to obtain
the quasi-normal spectrum for $k=0$ (figure~\ref{quasis}).

\begin{centering}
\begin{figure}
\begin{centering}
\includegraphics[width=60mm]{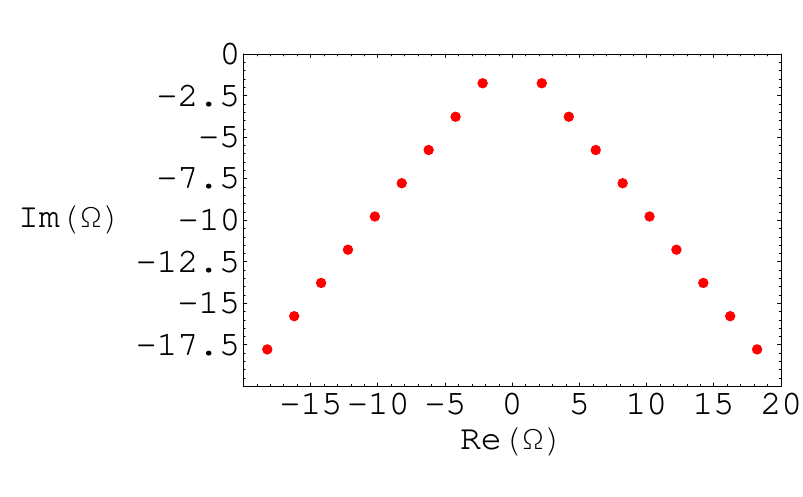} \hspace{1cm}
\includegraphics[width=60mm]{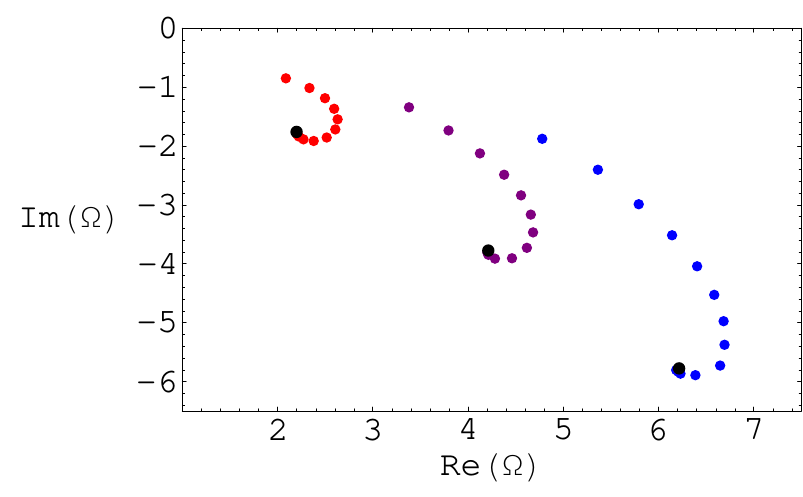}
\par
\end{centering}
\caption{The lowest quasi-normal modes for $m_q=0$ on the left and
the three lowest quasi-normal modes for increasing $m_q$ on the right.
The black points on the right show the limiting values for $m_q=0$.}
\label{quasis}
\end{figure}
\par
\end{centering}

For a nonzero quark mass the D7 embeddings are only known numerically
and the analysis is much more involved.  We need the solution to
behave like a purely ingoing wave at the horizon and to be
normalizable at infinity.  The technique is to perform numerical
integration of purely ingoing solutions outward from the horizon
surface and normalizable solutions inward from infinity and attempt to
match them smoothly at an intermediate value of the radial coordinate.
This matching is only possible for a discrete set of frequencies which
are the quasi-normal frequencies.  An interesting picture is obtained
for the temperature dependence of quasi-normal modes, shown in figure
\ref{quasis}.  As the embedding approaches the critical embedding the
imaginary part of the quasi-normal frequencies is becoming smaller as
one would expect - we are moving closer to the mesons being stable.
The evolution of the quasi-normal modes at large $T$ into the stable
mesons at small $T$ has been explicitly followed through the
computation of the theory's spectral function in \cite{Myers:2007we}.

Computations involving semi-classical strings in the D3-D7 system have
also been made. The properties of heavy light mesons at finite temperature
are determined in \cite{Peeters:2006iu}. A long D3-D7 string describes a heavy deconfined quark
and the energy loss and wake produced by such a string dragged through the plasma has been
studied in \cite{gubser:2006bz,Herzog:2006gh,Chesler:2007an,Gubser:2007zr}.

Thus the main physical characteristic of the phase transition is the
mesons melting into the background thermal plasma. Note that since the
temperature $T = r_H/(R^2 \pi)$ with $R=\lambda \alpha^{'2}$ and the
transition occurs when $m \sim r_H$, the temperature scale of the
transition is
\begin{equation}
T_c \sim \frac{m_q (2 \pi \alpha') }{\sqrt{\lambda} \alpha' \pi}
\sim \frac{2 m_q}{\sqrt{\lambda}} \,.
\end{equation}
The transition occurs at a temperature of roughly the meson mass.

\subsection{More thermodynamics}

The thermal transitions in the D3/D7 system compactified on an $S^3$
has been studied in \cite{Karch:2006bv}.  The meson spectra in the
presence of a black hole whose radius is growing with time has been
computed in \cite{Grosse:2007ty}.

In the presence of a finite quark or isospin density, introduced
through a vev for the time component of the gauge field on the probe
brane, the structure of the phase diagram becomes more involved. In
particular there are unstable regions in the phase diagrams. Studies
of finite chemical potential and finite density effects for D7 brane
probes in the AdS-Schwarzschild background may be found in
\cite{Apreda:2005yz,Kobayashi:2006sb,Albash:2006bs,Nakamura:2006xk,
  Kobayashi:2006sb,Mateos:2007vc,  Bergman:2007wp, 
  Karch:2007br,Ghoroku:2007re,Erdmenger:2007ap,Erdmenger:2007ja}.

We also note that equivalent phase transitions to those above occur in
the D4/D6 system of section~\ref{d4d6myers} (see
\cite{Kruczenski:2003uq} for the details and \cite{Ghoroku:2005tf} for
related work).

Considerable work has also been done on the Sakai-Sugimoto model (see
section~\ref{sectionss} above) at finite temperature. That model also
displays a first order meson melting transition as described in
\cite{Aharony:2006da,Peeters:2006iu}.  Additional finite density
studies can be found in \cite{Horigome:2006xu,
  Yamada:2007ys,Davis:2007ka,Kim:2007zm,Parnachev:2007bc,Kim:2007xi,
  Rozali:2007rx,Aharony:2007uu}.
\begin{figure}
\begin{center}
  \includegraphics[width=0.45\linewidth]{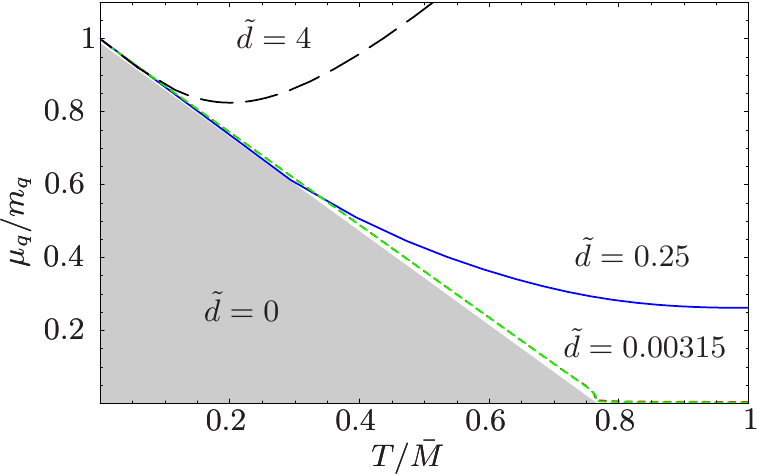}
  \caption{Phase diagram for D7 branes in AdS-Schwarzschild at finite
    baryon density: The quark chemical potential~$\mu_q$ divided by
    the quark mass is plotted versus the temperature~$T$ divided by
    $\bar M= 2 m_q/\sqrt{\lambda}$.  Two different regions are
    displayed: The shaded region with vanishing baryon density and the
    white region with finite baryon density. The multi-valued region at
    the lower tip of the transition line is not resolved here.  The
    curves are lines of equal baryon density.  The curve
    for the critical density $\tilde d^*=0.00315$ displays where
    the first order phase transition between two black hole
    embeddings disappears. Figure by
    M.~Kaminski and F.~Rust. }
  \label{fig:phaseDiagram}
\end{center}
\end{figure}
As an example we consider here spectral functions at finite
temperature and quark chemical potential as discussed in
\cite{Erdmenger:2007ap}. The phase diagram was found in
\cite{Kobayashi:2006sb,Mateos:2007vc} and is displayed in
figure~\ref{fig:phaseDiagram}. In the grey shaded area, the baryon
density $n_B$ is zero the first-order phase transition between
Minkowski and black hole embeddings occurs. In the white area, the
baryon density is non-zero. In this region, only black hole embeddings
are stable.  Lines of constant baryon density are displayed in colour.
For small non-zero values of the baryon density, a first-order
transition between two black hole embeddings occurs, which disappears
above a critical value for the quark density given by $\tilde d^*
\equiv 2^{5/2}n_B/(N_f\sqrt{\lambda}T^3) = 0.00315$, with $n_B$ the
baryon density.  Moreover there is a multi-valued region at the bottom
of the separation line between the grey-shaded and the white region,
which is not resolved here.  According to the phase diagram, within
the black hole phase ({\em i.e.}~in the white region) for fixed quark
mass, there is a temperature-dominated region for large temperatures
to the far right, and a potential-dominated region for small
temperatures to the left.  In the two regions, the spectral functions
show a qualitatively different behaviour.
\begin{figure}
  \includegraphics[width=.45\linewidth]{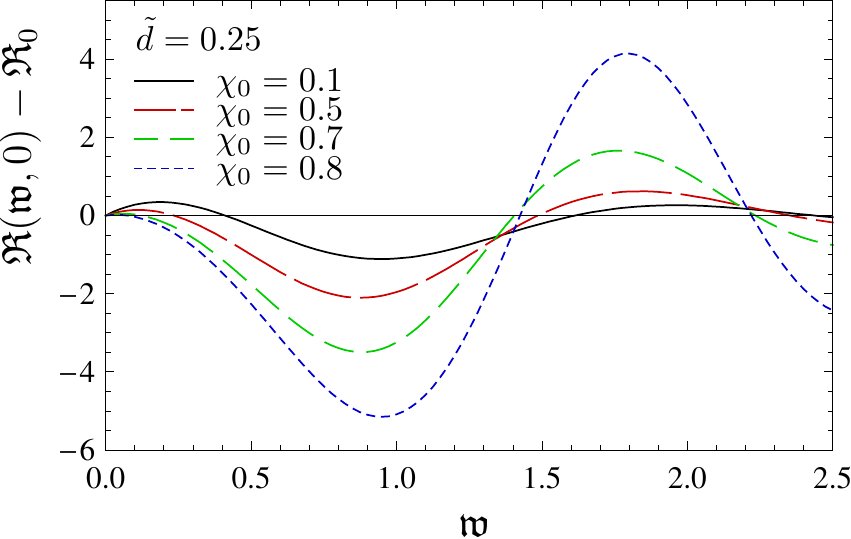}
  \hfill
  \includegraphics[width=.45\linewidth]{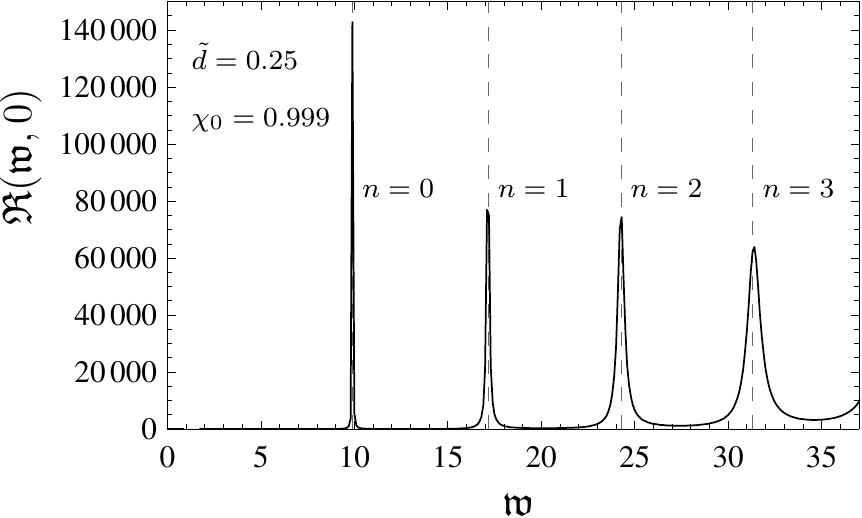}
  \caption{The finite temperature part of the spectral
    function~$\mathfrak{R}-\mathfrak{R}_0$ (in units of~$N_f N
    T^2/4$) in the temperature-dominated region (left plot) and in the
    potential-dominated region (right plot). $\tilde d$ parametrizes the 
quark density and $\chi_0$, introduced in \cite{Kobayashi:2006sb}, 
essentially  corresponds to $m_q/T$. Figures from
    \cite{Erdmenger:2007ap}. }
\label{fig:spectral}
\end{figure}
We consider the spectral functions for the current-current
correlator coupling to the gauge field on the D7 brane.  The
result is displayed in figure~\ref{fig:spectral}.  In the
temperature-dominated region, the spectral function, {\em
i.e.}~the imaginary part of the retarded Green function, displays
very broad peaks corresponding to unstable vector mesons. This is
shown on the left hand side of the figure.  In the
potential-dominated region however, the peaks become very narrow
and their location coincides exactly with the supersymmetric meson
spectrum discussed earlier in section~\ref{spectrum} (by
supersymmetry, the scalar and vector spectra coincide).

A further interesting point is that the location of the peaks first
moves to lower frequencies when the temperature is decreased until
they reach a minimum. When decreasing the temperature further, the
peaks move the larger frequencies again, while becoming narrower.
This corresponds to a movement of the poles similar to the one
displayed in figure~\ref{quasis}.

\subsection{Mesons from D7 branes with external B-fields} \label{secBfield}

Supersymmetric versions of embeddings in backgrounds with $B$ field
have been presented above in section~\ref{secps}. There are also
interesting effects in non-supersymmetric backgrounds with $B$ fields.

A Zeeman splitting is observed if a pure gauge-external $B$ field is
turned on in two spatial directions parallel to the $AdS$ boundary
\cite{Filev:2007gb,Filev:2007qu}.  Such a $B$ field breaks
supersymmetry completely.  As shown in
\cite{Filev:2007gb,Filev:2007qu}, it induces spontaneous chiral
symmetry breaking and Goldstone bosons by virtue of a similar
mechanism as discussed in section (\ref{sectioncsb}) above.

\begin{figure}[!ht]
\begin{center}
\includegraphics[width=17cm,clip=true,keepaspectratio=true]{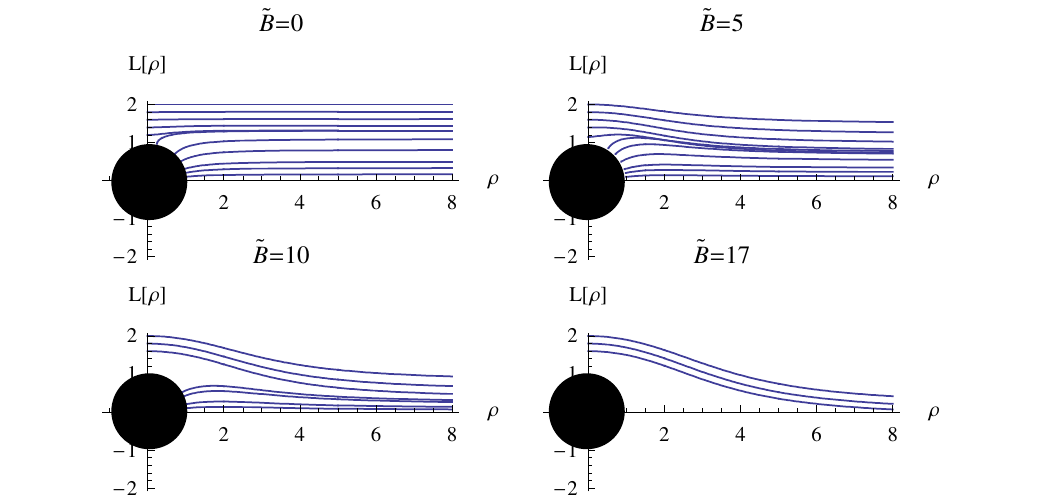}
\caption{Embedding function $L$ as function of the radial coordinate $\rho$
for D7 branes embedded in the black hole background with external magnetic
fields, for different values of the normalized (dimensionless)
external field $\tilde{B}$. Increasing values of $\tilde{B}$ for
fixed $T$ show the repulsive nature of the magnetic $B$ field, which is
switched on in two spatial directions parallel to the boundary.
We see that for large enough $\tilde{B}$, the black hole
phase is never reached, and spontaneous chiral symmetry breaking occurs.
From \cite{Erdmenger:2007bn}.}
\label{fig:BandTflowsv}
\end{center}
\end{figure}
As discussed in section~\ref{adsbhsection}, there is no spontaneous
chiral symmetry breaking in the finite temperature field theory dual
of the AdS-Schwarzschild black hole background. However, if a $B$
field of the form of \cite{Filev:2007gb} of sufficient strength is
switched on, the chiral symmetry breaking mechanism induced by this
$B$ field dominates and is present even in the black hole background
\cite{Erdmenger:2007bn,Albash:2007bk}.  This is shown in
figure~\ref{fig:BandTflowsv}.

With an external electric field, {\em i.e.}~a $B$ field turned on in
the temporal and one spatial direction parallel to the boundary, a
meson mass shift similar to the Stark effect arises
\cite{Erdmenger:2007bn}.  In this case, there is an attraction of the
D7 brane probes towards the origin and no chiral symmetry breaking
occurs \cite{Erdmenger:2007bn,Albash:2007bq}.

\subsection{Summary}

We have seen that the AdS/CFT Correspondence implies the existence of
a novel thermal phase transition in theories with quarks. As the
temperature passes through the scale of the meson mass, there is a
first order phase transition with a small jump in the value of the
quark condensate. The mesons of the theory melt into the thermal bath
at this scale. Note this transition is distinct from the deconfinement
transition of the glue-fields. Lattice calculations
\cite{Fodor:2007sy,Karsch:2007dp,Karsch:2007dt} do not reveal such
a first order transition in QCD, so it is probably that it is an
artifact of the large $N$ regime. Meson melting does occur in QCD
though and it is promising that we have a theoretical tool to address
that process.  The mesons of this theory are tightly bound and so
harder to dissociate than those in QCD - heavy-heavy mesons in QCD are
not expected to survive to as high a temperature as their mass scale
as we see here.

\newpage

\section{AdS/QCD} \label{secAdSQCD}
\setcounter{equation}{0}\setcounter{figure}{0}\setcounter{table}{0}

Inspired by holography a number of authors have proposed
phenomenological models of QCD generically called AdS/QCD. These
models consist of a gauge theory in a curved space (usually AdS) with
the field content picked to holographically match to certain QCD bound
states and operators. This sort of modelling is necessarily a leap in
the dark. From the string theory side one might expect that as one
approached QCD from the theories at infinite 't~Hooft coupling string
corrections would become large - one should be working in a string
theory and not a field theory.  Nevertheless, the string models
described above contain confinement and chiral symmetry breaking and
the ratio of meson masses do appear to match the QCD values to a few
10\%s (the absolute values do not match unless one extrapolates to
order one 't~Hooft coupling - in the string models these states are
tightly bound with mass $\sim m_{q} /\sqrt{g^2_{YM}N}$). It is
therefore interesting to try model building in the spirit of these
models.

\subsection{A simple model}

We will concentrate on the simplest example of this sort proposed in
\cite{Erlich:2005qh} and \cite{DaRold:2005zs} \cite{DaRold:2005vr}
which is closest in spirit to the string models in sections
\ref{D7nonsusy}-\ref{d4d6myers} (a phenomenological model in the
spirit of section~\ref{sectionss} can be found in \cite{Hirn:2005nr}).

The field theory will live in an AdS space in five dimensions
(discarding the extra five dimensions of the string theory removes the
SO(6) global symmetry of the ${\cal N}=4$ model as would happen were
the super-partners to be decoupled),
\beq ds^2 = r^2 dx_{4}^2 + \frac{dr^2}{r^2} \,. \eeq The radial
coordinate $r$ will be interpreted as the holographic energy scale of
the theory (see (\ref{energy})).  As written the metric has an SO(2,4)
symmetry and would appear to describe a conformal gauge background. To
break that symmetry and impose confinement a crude, hard wall is
imposed at $r=r_0$ - the theory will only live at $r \geq r_0$. One
can think of this scale as the mass gap of the gauge background.

We will choose to describe the quark mass and condensate and the pion
fields in the model. We introduce a scalar field
\beq X = X_0 e^{2i \pi^a t^a} \,. \eeq $X_0$ will be a background
field that describes the quark mass and condensate (these are both
assumed to be matrices in flavour space that are proportional to the
identity). As we saw in (\ref{asymptotic}) for a scalar to describe a
quark bilinear operator ($\Delta=3$) it must have mass squared
$m^2=\Delta(\Delta-4)=-3$ in AdS and then the solution is of the form
\beq X_0 = \frac{1}{2} \frac{m}{r} + \frac{1}{2}
\frac{\Sigma}{r^3} \,. \eeq
Remember $r$ has energy dimension so $m$ is the mass and $\Sigma$ the
condensate.  $\pi^a$ are then the $N_f^2-1$ pion fields.

In addition the model describes the vector and axial vector states
through two massless gauge fields dual to the operators $\bar{q}_L
\gamma^\mu q_L$ and $\bar{q}_R \gamma^\mu q_R$.\footnote{The
  mass-conformal dimension relation for vector operators is
  $m^2=(\Delta-1)(\Delta-3)$, thus $m^2=0$.} The action is
\beq
S = \int_{r_0}^\infty d^5x \sqrt{-g} Tr \left\{ |DX|^2 + 3 |X|^2 -
\frac{1}{4 g_5^2} (F_L^2 + F_R^2) \right\} \,,
\eeq
where $X$ transforms on the left under SU$(N_f)_L$ and on the right
under SU$(N_f)_R$.

It is of course completely ad hoc to only describe these states. In QCD
there are many other states with mass of order the $a_0$ and the
$\rho$ but we simply choose to ignore them.

The mass, condensate and the position of the hard wall will be
parameters of the theory that are fit. There is also $g_5$, which in
string theory duals is a prediction in terms of the gauge theory
't~Hooft coupling $g_{YM}^2 N$. In the phenomenological approach
though, this relation is abandoned and the value of $g_5$ is fitted to
the vector current correlator extracted from QCD,
\beq
  \int d^4x e^{iqx}\langle J^a_\mu(x)J^b_\nu(0)\rangle=\delta^{ab}
  (q_\mu q_\nu -q g_{\mu\nu})\Pi_V(-q^2) \,,
\eeq
where $J^a_\mu(x)=\bar q \gamma_\mu T^a q$. For QCD, the
leading order contribution to $\Pi_V(-q^2)$ is
\cite{Shifman:1978bx}
\beq\label{eq:piv-qcd}
  \Pi_V(-q^2)=-\frac{N}{24\pi^2}\ln(-q^2) \,.
\eeq
In order to calculate this quantity from the five-dimensional model,
we appeal to the AdS/CFT correspondence. The five-dimensional vector
field $V^a_\mu(x,r)= (A^a_{L\mu}(x,r)+A^a_{R\mu}(x,r))/2$ acts as a
source for the four-dimensional vector current $J^a_\mu(x)$ in the
limit $r\rightarrow\infty$. It obeys the equation of motion
\beq\label{eq:vec-eom}
  \partial_\mu\left(\frac{1}{g_5^2}~e^\phi\sqrt{-g}
  g^{\mu\alpha}g^{\nu\beta}(\partial_\alpha V^a_\beta
  - \partial_\beta V^a_\alpha)\right)=0 \,.
\eeq
We look for solutions of the form $V^\mu(x,r)=V_0^\mu(x)v(x,r)$, with
$\lim_{r\rightarrow\infty}v(x,r)=1$, so that $V^\mu_0(x)$ will act
as a dimension one source for $J^a_\mu(x)$. Solving the equation
of motion (\ref{eq:vec-eom}) in the $V^r(x,r)=0$ gauge gives \beq
  v(q,r)=-\frac{\pi}{2}\mathcal{Y}_1(q/r)\sim 1-\frac{q^2}{4r^2}\ln
  \left(\frac{-q^2}{r^2}\right)\,,\qquad {\textrm as}~\,
  r\rightarrow\infty\,,
\eeq where $\mathcal{Y}_1$ is a Bessel function of the second
kind. Substituting the solution back into the action and
differentiating twice with respect to the source $V^\mu_0$ gives
the vector current correlator \beq\label{eq:piv-ads}
  \Pi_V(-q^2)=\left[\frac{1}{g_5^2q^2}r^3\partial_rv(q,r)\right]_{r=\infty},
\eeq which (up to contact terms) yields
\beq
  \Pi_V(-q^2)=-\frac{1}{2g_5^2}\ln(-q^2)\,.
\eeq
Finally, comparing this to the perturbative QCD result
(\ref{eq:piv-qcd}) determines the 5d coupling as
\beq
\label{match}
g_5^2 = \frac{12 \pi^2}{N} \,.
\eeq
It may appear rather surprising to be fitting to the asymptotic
perturbative result when a gravity dual is inherently a description of
a strongly coupled gauge theory.  The argument that is usually made is
that perturbative QCD is conformal in the UV and so it is natural to
match to the UV behaviour in AdS which is also conformal. One captures
this conformality in the model if not the asymptotically free running
of the coupling.

Now as usual one can solve (\ref{eq:vec-eom}) for solutions of the
form $V = V(r) e^{ip.x},\, p^2=-M^2$ with $V(r)$ falling to zero as $r
\rightarrow \infty$. One must choose a (necessarily arbitrary)
boundary condition at the hard wall and we can for example choose
$\partial_r V =0$. We can therefore extract the masses of the
$\rho$ and its excited states.

One can also extract the decay constant for a $\rho$ decaying to a
photon. One integrates the action by parts treating one field $V$ as a
solution of the equation of motion and one as a background external
field. The coupling is then
\beq F_\rho^2 = {1 \over g_5^2} V^{''}_\rho(r \rightarrow \infty) \,.
\eeq

Similarly one can study the axial vector gauge field and the pion to
determine the pion mass, $a_1$ mass and their decay constants.  The
best fit results to the QCD data are shown in
table~\ref{adsqcdresults}.  There is a good fit to the data.

\begin{table}
\begin{center}
  \begin{tabular}{|c|c|c|c|}\hline
  Observable       & Measured         &  AdS A      &  AdS B  \\
                   & (MeV)            &  (MeV)      &  (MeV)
                   \\\hline
  $m_\pi$          & $139.6\pm0.0004$ &  $139.6^*$  &  $141$  \\
   $m_\rho$        & $775.8\pm0.5$    &  $775.8^*$  &  $832$  \\
  $m_{a_1}$        & $1230\pm40$      &  $1363$     &  $1220$   \\
    $f_\pi$        & $92.4\pm0.35$    &  $92.4^*$   &  $84.0$  \\
  $F_\rho^{1/2}$   & $345\pm 8$        &  $329$      &  $353$  \\
  $F_{a_1}^{1/2}$  & $433\pm 13$       &  $486$      &  $440$  \\
  \hline
  \end{tabular}
\caption{Results for meson variables in AdS/QCD. AdS A is the best
fit to the starred variables. Model B is the best fit to all the
observables.}\label{adsqcdresults}
\end{center}
\end{table}

\subsection{Higher order pion interactions}

The chiral symmetry breaking pattern of AdS/QCD means that the pions
necessarily take the form of a chiral Lagrangian model. In this
formalism the coefficients are a prediction though. In
\cite{DaRold:2005zs} the order $p^4$ terms in the chiral Lagrangian
were estimated in the simplest AdS/QCD model (assuming the lightest
rho dominated these terms). These terms take the form
\begin{eqnarray}
{\cal L}_4 &=& L_1 \,{\rm Tr}^2\big[ D_\mu U^\dagger D^\mu U\big]
+ L_2 \,\Tr\big[ D_\mu U^\dagger D_\nu U\big]
   \Tr\big[ D^\mu U^\dagger D^\nu U\big]
+ L_3 \,\Tr\big[ D_\mu U^\dagger D^\mu U D_\nu U^\dagger D^\nu
U\big]\,
\nonumber\\
&+& L_4 \,\Tr\big[ D_\mu U^\dagger D^\mu U\big]
   \Tr\big[ U^\dagger\chi + \chi^\dagger U \big]
+ L_5 \,\Tr\big[ D_\mu U^\dagger D^\mu U \left( U^\dagger\chi +
\chi^\dagger U \right)\big]\,
\nonumber\\
&+& L_6 \,{\rm Tr}^2\big[ U^\dagger\chi + \chi^\dagger U \big] +
L_7 \,{\rm Tr}^2\big[ U^\dagger\chi - \chi^\dagger U \big] + L_8
\,\Tr\big[\chi^\dagger U \chi^\dagger U + U^\dagger\chi
U^\dagger\chi\big]
\nonumber \\
&-& i L_9 \,\Tr\big[ F_R^{\mu\nu} D_\mu U D_\nu U^\dagger +
     F_L^{\mu\nu} D_\mu U^\dagger D_\nu U\big]
+ L_{10} \,\Tr\big[ U^\dagger F_R^{\mu\nu} U F_{L\mu\nu} \big] \,
. \label{l4}
\end{eqnarray}
We reproduce the results from \cite{DaRold:2005zs} in following
table:

\begin{table}[htb] \label{pfour}
\begin{center}
\begin{tabular}{cccc}
\hline
  & {\rm Experiment}
    &{\rm AdS$_5$}
\\ \hline
 $L_1$ & $0.4\pm0.3$ &
  $0.4$
\\
 $L_2$ & $1.4\pm0.3$ &
       $0.9$
\\
 $L_3$ & $-3.5\pm1.1$ &
        $-2.6$
\\
 $L_4$ & $-0.3\pm0.5$ &
 $0.0$
\\
 $L_5$ & $1.4\pm0.5$ &
            $1.7$
\\
 $L_6$ & $-0.2\pm0.3$ &
 $0.0$
\\
 $L_9$ & $6.9\pm0.7$ &
      $5.4$
\\
 $L_{10}$ & $-5.5\pm0.7$
& $-5.5$
\\
\hline
\end{tabular}
\end{center}
\end{table}

\subsection{Glueballs}

It is also possible to include glueballs into AdS/QCD
\cite{BoschiFilho:2002ta,BoschiFilho:2002vd,BoschiFilho:2005yh}
through additional scalars in the bulk. We can associate the $0^{++}$
glueballs with the operator ${\rm Tr}\, F^2$ a dimension 4 operator -
the usual AdS dictionary teaches that the dual supergravity field
should be massless.  The equation of motion (for a solution of the
form $\phi = \phi(r) e^{ipx}$, $p^2 = -M^2$) is
\beq \left(\frac{1}{r} \partial_r r^5 \partial_r + M^2 \right)
\phi(r) = 0 \,. \eeq

If we again impose Neumann boundary conditions ($\partial_z \phi = 0$)
at the hard wall we find the glueball masses (normalizing to the
lattice gauge theory \cite{Morningstar:1997ff,Morningstar:1999rf}
value for the lightest mass state) $M_1= 1.63$~GeV, $M_2 =2.98$~GeV,
$M_3= 4.33$~GeV etc.

\subsection{A plethora of AdS/QCD phenomenology}

A considerable number of other aspects of QCD phenomenology have
been successfully addressed using AdS/QCD which we can not
completely review here.  The reader is referred to the following
references. Strange quarks are added in \cite{Shock:2006qy}.
Higher spin mesons are studied in \cite{Katz:2005ir}. Baryon
states are included in \cite{deTeramond:2005su}. Four-point
current-current correlators relevant to the $\Delta I= 1/2$ rule
and the $B_K$ parameter for K-meson mixing are analyzed
in~\cite{Hambye:2005up}.  Heavy quark potentials are computed
in~\cite{White:2007tu}.  The AdS/QCD model is related to light
cone QCD in \cite{Brodsky:2006uqa, Brodsky:2007hb} allowing form
factor computations.  Form factors for mesons are also in
\cite{Grigoryan:2007vg,Grigoryan:2007wn}.

Properties of QCD at high temperature and density and the
deconfinement transition have been analyzed in this context in
\cite{Ghoroku:2005kg,Ghoroku:2006cc,Nakano:2006js,
  Kajantie:2006hv,Kim:2006ut,Cai:2007zw,Kim:2007gq}.

Such models have also been adapted to describe walking
\cite{Holdom:1981rm} technicolour \cite{Weinberg:1979bn,
  Susskind:1978ms} dynamics for electroweak symmetry breaking in
\cite{Hong:2006si,Hirn:2006nt, Piai:2006hy, Carone:2006wj,
  Carone:2007md}. It is worth remarking that very similar ideas to
these models have inspired the field of Higgsless electroweak models
\cite{Csaki:2003dt,Csaki:2003zu} and their deconstructed
\cite{Hill:2000mu,ArkaniHamed:2001ca} partners for example in
\cite{Foadi:2003xa}.

\subsection{Regge behaviour and the soft wall}

The basic AdS/QCD model does not have the expected Regge behaviour for
the towers of radially excited states ($M_n^2 \sim n$)
\cite{Cata:2006ak,  Masjuan:2007ay, Shifman:2007xn, Huang:2007fv}. 
To see this, consider the action
for the gauge field in AdS describing the rho mesons,

\beq I \sim \int d^5x e^{-\Phi(z)} \sqrt{-g} F^2 \,.\eeq Here we have
included a dilaton field $\Phi$ that is a constant in the basic
AdS/QCD model. The equation of motion for a solution of the form
$A_x = f(z) e^{i kx}, k^2=M^2$ is
\beq  \left( r \partial_r r^3 \partial_r + M^2 \right) f = 0 \,. \eeq
Changing variables to $z=1/r$ and substituting
\beq f = e^{B/2} \psi, \hspace{1cm} B = \phi + \ln r \,,\eeq we find
\beq - \psi^{''} + V(z) \psi = M^2 \psi \,, \hspace{1cm} V =
\frac{1}{4} (B')^2 - \frac{1}{2} B^{''} \,, \eeq which is of a
Schr\"odinger equation form.

If we impose the IR boundary by putting in a hard cut off then the
Schr\"odinger potential in the IR is that of a square well. The
mass spectrum therefore grows as $M_n^2 \sim n^2$ in contradiction
with the physically observed Regge behaviour.

One might simply argue that this is a sign that the supergravity
approximation is breaking down when we try to apply these methods to
QCD - string theory naturally gives Regge behaviour, so a resolution
would be to work with a full string theory.  In \cite{Karch:2006pv} it
was pointed out though that if the dilaton grows as $\frac{1}{r^2}$ in
the IR the potential V will be of the form
\beq V = z^2 + \frac{3}{4z^2} \,. \eeq

The exact solution is known and $M_n^2 = 4(n+1)$. Regge style
behaviour is therefore accessible in principle in the supergravity
regime. None of this behaviour is derivable though merely posited.

\subsection{Improvement and perfection}

An obvious criticism of AdS/QCD is that it is a model rather than
being derived explicitly from the QCD Lagrangian. There is no
understanding of systematic errors. Can we hope to improve the model
then?

Presumably in reality the weakly coupled string theory model will only
be valid in the strong coupling regime of QCD at low energies. It is
therefore a low energy effective theory.  An obvious consequence of
this is that a UV cut off should be imposed \cite{Evans:2005ip,
  Shuryak} and the scaling dimension of operators, values of higher
dimension operator couplings and expectation values of operators
should all be matched at the cut off. In principle, this is possible
although there is no obvious truncation to a finite number of such
matchings and the resulting model need not be weakly coupled.

The introduction of expectation values for relevant operators are
discussed and introduced in \cite{Ghoroku:2005vt,
  Csaki:2006ji,Hirn:2005vk,Shock:2006gt}. Instanton effects are
included in \cite{Shuryak}. In \cite{Gursoy:2007cb,Gursoy:2007er}
back-reacted geometries in 5d non-critical string theory are generated
that have a dilaton profile set to match the QCD running coupling -
the models display confinement and chiral symmetry breaking (through
the addition of D4 and $\overline{\rm D4}$ branes). Meson properties have
not yet been computed there though.  The inclusion of higher dimension
operator couplings are discussed in \cite{Evans:2005ip,Evans:2006ea}.
Improvements to the phenomenological fit can be achieved by all these
methods although at the expense of additional free parameters.

The basic AdS/QCD model also inputs chiral symmetry breaking and the
quark mass through two independent parameters $c$ and $m$. In reality
the quark condensate should be a prediction of the background gauge
dynamics (the metric) and the value of $m$. The more complete string
models of chiral symmetry breaking discussed above in sections
\ref{D7nonsusy}-\ref{d4d6myers} do contain this explicit
dynamics. In \cite{Evans:2006dj} the dilaton flow model of chiral
symmetry breaking was adapted to an AdS/QCD model keeping that
dynamical behaviour. The computations are essentially those of section
\ref{D7nonsusy} but with $g_5$ fixed as in (\ref{match}).
The resulting model has one fewer free parameter and gives a match to
the data of similar quality to the basic AdS/QCD model.

Finally higher dimension operators in the gravity dual's action,
representing stringy corrections, have
been included in \cite{Grigoryan:2007iy}.

\subsection{Summary}

AdS/QCD is a tidy crystallization of the ideas of holographic chiral
symmetry breaking models applied to QCD. Generically such models do
well at reproducing QCD phenomenology at the 10\% level, suggesting that
their parent string theory models are capturing crucial aspects of QCD
dynamics. It remains a challenge though both to  understand how to
systematically move towards a complete description of QCD, and how to
precisely embed the AdS/QCD models into string theory.

\newpage

\section{Conclusion}
\setcounter{equation}{0}\setcounter{figure}{0}\setcounter{table}{0}

In this review we have seen how a new theoretical technique for
calculating in strongly coupled gauge theory has emerged from string
theory. The AdS/CFT Correspondence in its initial form described the
highly supersymmetric and conformal $\N=4$ Yang Mills theory at large
't~Hooft coupling.  Deformations of the gravity dual have since led to
understanding of confinement in non-conformal theories.  Here we have
concentrated on the next step necessary
for moving towards QCD, which consists  of adding quarks
in the fundamental representation. The strong dynamics of the gauge
fields bind the quarks into tightly bound mesonic states.  When quarks
are included in non-supersymmetric geometries, we have seen that chiral
symmetry breaking is generated  - the quarks acquire a dynamical mass
and there are Goldstone fields associated with the symmetry breaking,
the analogues of the pions. It is very pleasing that the examples
presented show a large number of  phenomena we observe in QCD.

The gravity dual description also works well for strongly coupled
finite-temperature field theories, for example for describing dynamical
processes such as diffusion and meson melting. These results are
potentially relevant for the quark-gluon plasma, for which standard
perturbative or lattice gauge theory methods are not easily available.

Given these qualitative successes, it has been tempting to make
quantitative comparisons to QCD. This necessarily involves ignoring
the absence of asymptotic freedom and the presence of massive, but not
decoupled, superpartners. The 't~Hooft coupling is also brought down
from the infinite coupling limit to make these comparisons.
Surprisingly though, such comparisons do hold up well (typically at the
10$\%$ level). This suggests that a wide range of gauge theories share
a number of even quantitative properties. There is considerable hope
that calculations relevant to QCD can be performed. This hope must be
tempered though by the difficulties of bringing systematic errors
under control.

We want to stress though that the value of the gravity dual approach
are of a more principal nature.  They provide an opportunity for new
exchanges between string theory and quantum field theory, which leads
to a fresh look at both fields. This has lead to progress in both
areas. Moreover in this context, string theory as a candidate for a
unified theory of fundamental interactions has made a significant step
towards a more applied approach of being applicable to experimentally
testable models.

\newpage

\section{Other reviews}

A number of other recent reviews may be of interest to our readers.
The ``classic'' review of the AdS/CFT Correspondence is
\cite{Aharony:1999ti}. \cite{Aharony:2002up} contains a description of
generalizations to theories with broken conformal invariance.
\cite{RodriguezGomez:2007za} covers material on D7 brane probes in
supersymmetric theories.  The Sakai-Sugimoto model is reviewed in
\cite{Peeters:2007ab}. Thermal properties of these theories are
reviewed in \cite{mateos:2007ay}.

\vspace{2cm}

\section{Acknowledgements}

We would like to thank our co-authors with whom we have worked on some
of the topics described over the last years - Riccardo Apreda, James
Babington, Johannes Gro\ss e, Kazuo Ghoroku, Zachary Guralnik,
Matthias Kaminski, Dieter L\"ust, Ren\'e Meyer, Felix Rust, Jonathan
Shock, Christoph Sieg, Andrew Tedder, Diana Vaman and Tom Waterson.

We would like to thank Martin Ammon,
Dietmar Ebert and Kasper Peeters for a critical reading
of the manuscript, as well as Daniel Arean, Gunnar Bali, Biagio Lucini and
Alfonso Ramallo for comments.

J.E.~is grateful to the Isaac Newton Institute, Cambridge, for
hospitality in August 2007, where part of this review was written.

\newpage



\providecommand{\href}[2]{#2}\begingroup\raggedright\endgroup

\end{document}